\documentclass[12pt,preprint]{aastex}

\begin{document}
\title{Radiation Mechanisms and Physical Properties of GeV-TeV BL Lac Objects}
\author{Jin Zhang\altaffilmark{1,2}, En-Wei Liang\altaffilmark{3}, Shuang-Nan Zhang\altaffilmark{4,1}, Jin-Ming Bai\altaffilmark{5,6}}
\altaffiltext{1}{National Astronomical Observatories, Chinese Academy of Sciences, Beijing, 100012, China;
zhang.jin@hotmail.com}\altaffiltext{2}{College of Physics and Electronic Engineering, Guangxi Teachers Education
University, Nanning, 530001, China} \altaffiltext{3}{Department of Physics and GXU-NAOC Center for Astrophysics
and Space Sciences, Guangxi University, Nanning, 530004, China} \altaffiltext{4}{Key Laboratory of Particle
Astrophysics, Institute of High Energy Physics, Chinese Academy of Sciences, Beijing 100049,
China}\altaffiltext{5}{National Astronomical Observatories/Yunnan Observatory, Chinese Academy of Sciences,
Kunming, 650011, China}\altaffiltext{6}{Key Laboratory for the Structure and Evolution of Celestial Objects, Chinese Academy of Sciences, Kunming 650011, China}

\begin{abstract}
Broadband spectral energy distributions (SEDs) simultaneously or quasi-simultaneously observed with Fermi/LAT and the other instruments are complied from literature for 24 TeV BL Lac objects. Two SEDs are available for each of 11
objects, and the state of the sources is identified as a low or high state according to its flux density at $1\ {\rm TeV}$. The well-sampled, clean SEDs without contaminations of the accretion disk and external Compton
process of these sources are the best candidates for investigating the radiation mechanisms and the physical properties of the jets. 
Assuming that the electron spectrum is a broken power-law with a break at $\gamma_{\rm
b}$ and using the peak frequencies and their corresponding lumminosities ($\nu_{\rm s}$, $\nu_{\rm c}$, $L_{\rm s}$, and $L_{\rm
c}$) of the SEDs, we fit the SEDs with the single-zone synchrotron + synchrotron-self-Compton (SSC) model and determine the physical parameters of the jets, including the Doppler boosting factor ($\delta$), the magnetic field strength ($B$), the size of radiating region ($R$), the bolometric luminosity ($L_{\rm bol}$),
and the jet total power ($P_{\rm jet}$). The model well represents the SEDs, and the observed relation between $\nu_{\rm s}$ and $\nu_{\rm c}$ also favors the model. In this scenario, we find that $\gamma_{\rm b}$ is
significantly different among sources and even among the low and high states of a given source, but $B$ is distributed narrowly within the range of $0.1-0.6 $ G, indicating that the shocks in the jets are significantly
different among sources and the magnetic field may be independent of the shocks. $\delta$ ranges from 14 to 30, and $R=(2.6\sim 100)\times 10^{15}$ cm, suggesting that flux variations with a minimum timescale from an hour to
one day may be observed for these sources. Prominent flux variations with a clear spectral shift are observed and the ratio of the flux density at 1 TeV is correlated with the ratio of the $\gamma_{\rm b}$ in the low and
high states, indicating that the relativistic shocks in the jets may be responsible for the flux variations and the spectral shift. $\delta$ of the high state is systematically larger than that of the low state, but the
ratios of $\delta$ and the flux density in the high and low states are not correlated. The ratio of $L_{\rm c}/L_{\rm s}$ is anti-correlated with $\nu_{\rm s}$ in the co-moving frame for the sources in both the high and
low states, but the slopes of the anti-correlations are significantly different. This anti-correlation is possibly due to the Klein-Nishina effect, but not the cooling effect of the photon fields outside the jet, as
proposed for explaining the blazar sequence. No excess in the GeV band due to the interaction between the TeV photons and the extragalactic background light is observed, implying that the strength of the intergalactic
magnetic field would be much larger than $ 10^{-16}$ G. The observed $L_{\rm bol}$ is not correlated with $P_{\rm jet}$. An anti-correlation between $P_{\rm jet}$ and the mass of the central black hole is observed,
i.e., $P_{\rm jet}\propto M^{-1}_{\rm BH}$, disfavoring the scenario of a pure accretion-driven jet. We suggest that the spin energy extraction may be significant for powering jets in these sources, implying that smaller
mass black holes may be spinning more rapidly in this sample.
\end{abstract}


\section{Introduction}           
\label{sect:intro} The broadband spectral energy distributions (SEDs) of blazars are bimodal. The underlying
radiation mechanisms that shape such a feature are still not well understood. The proposed models can be
classified into two kinds, i.e., leptonic and hadronic models (Maraschi et al.1992; Bloom \& Marscher 1996;
Mannheim \& Biermann 1992). The more accepted one is the leptonic model. In this model, the bump at the
IR-optical-UV band is explained with the synchrotron process of relativistic electrons accelerated in the jets,
and the bump at the GeV-TeV gamma-ray band is due to the inverse Compton (IC) scattering of the same electron
population (e.g., Ghisellini et al. 1996; Urry 1999). The seed photon field may be from the synchrotron
radiation themselves (the so-called SSC model; Maraschi et al. 1992, Ghisellini et al. 1996) or from external
radiation fields (EC), such as the broad line region (BLR; Sikora et al. 1994), accretion disk (Dermer et al.
1992), torus (BLazejowski et al. 2000), and cosmic microwave background (CMB; Tavecchio et al. 2000). Broadband
SEDs obtained simultaneously or quasi-simultaneously are critical to reveal the radiation mechanisms. The low
energy bump is well sampled with the data from multi-wavelength campaigns in the radio, optical, and X-ray
bands. The observations with CGRO/EGRET and ground-based TeV gamma-ray telescopes sketch the feature of the high
energy bump, which is however poorly constrained.  The second bump usually peaks at the GeV-TeV and can be
confined with observations in the MeV-GeV-TeV band with the Large Area Telescope (LAT) on board the Fermi
satellite, which covers an energy band from 20 MeV to $\sim 300$ GeV. The observations with Fermi/LAT, together
with the ground-based observations in the TeV gamma-ray band, now convincingly pin down both the luminosity and
the frequency of high energy peak in the observed SEDs. 

About 40 active galactic nuclei (AGNs) have been detected in the TeV gamma-ray band since the first detection of
TeV gamma-rays from Mkn 421 with the Whipple imaging atmospheric-Cherenkov telescopes in 1992 (Punch et al.
1992). Most of them are Blazars. Observations show that TeV AGNs have violent variability in multi-frequency,
especially in the X-ray and the TeV bands (Buckley et al. 1996; Takahashi et al. 2000; Sambruna et al. 2000).
The variations in the X-ray band and the TeV band for almost all sources, such as Mkn 421, are associated (Blazejowski
et al. 2005). Some sources detected by EGRET are not confirmed by Fermi/LAT, and some in the 1FGL (LAT
first-year catalog) with fluxes much over the EGRET threshold were not detected by EGRET (Abdo et al. 2010a).
Four TeV AGNs are still do not detected by Fermi/LAT so far. These facts indicate that AGNs should have
significant variability in the GeV-TeV band. In the high state of TeV emission, the peak frequency ($\nu_{\rm
s}$) of the synchrotron radiation bump in the SEDs moves to higher energies for some sources, showing a tendency
that a brighter TeV emission corresponds to a harder spectrum for the emission in the X-ray and gamma-ray bands,
as observed in Mkn 501 (Anderhub et al. 2009a). It is
unclear what may be responsible for the variability and the spectral shift in the low and high states of TeV
emission. The abundant data observed with Fermi/LAT in the MeV-GeV band, together with the TeV gamma-ray data,
now provide an excellent opportunity to reveal the physical origin of the temporal and spectral variations of
these TeV sources.

It is well known that the observed emission from BL Lac objects, a sub-sample of blazars, is
jet dominated. The jets of AGNs are powered by the central massive black hole (BH). However, the mechanism of the jet
production is still a mystery. The proposed models can be simplified as two types, i.e., accretion and rotation
of the BH. The connection between jet and accretion disk has been widely investigated (Ulvestad \& Ho 2001;
Marscher et al. 2002; Merloni et al. 2003; Falcke et al. 2004; King et al. 2011). With a sample of 23 blazars in
the three months of Fermi satellite survey, Ghisellini et al. (2009) found that the jet power is slightly larger
than the disk luminosity and proportional to the mass accretion rate. However, the accretion process cannot
explain the radio loudness of AGN, the BH spin is suggested to be the possible ingredient (Sikora et al. 2007;
Lagos et al. 2009). In fact, there is some evidences for rapidly spinning BH (Cao \& Li 2008; Wu et al. 2011).
The jets may be powered by both the accretion process and the spin of the central BH (Fanidakis et al. 2011).
The SEDs of BL Lac objects dominated by jet emission suffer less contamination of the emission from accretion
disk and EC process. Therefore, they are the best candidates for investigating the jet properties and the
relation to the cental BHs.

AGNs are the only confirmed extragalactic TeV sources so far. It is well known that TeV photons from the high
redshift universe would be absorbed through interaction with the extragalactic background light (EBL), producing
electron-positron pairs. The electron-positron pairs may scatter the CMB photons to the GeV band. Therefore, the
intrinsic spectrum in the TeV band should be harder than the observed one, especially for the sources at high
redshift, and an excess component in the GeV band may be observed in the observed SEDs as suggested by some
authors\footnote{It was also reported that the electron-positron pairs would be deflected by the IGMF from the
initial TeV photon direction and thus the secondary GeV photons produced by them would exhibit a halo around
the cental bright point-source instead of a GeV excess in the SEDs (Aharonian et al. 1994; Dolag et al. 2009;
Neronov \& Semikoz 2009).}. However, one should note that the detection or not of such a component depends on
the the EBL model and the intergalactic magnetic field (IGMF). Assuming that the intrinsic spectrum of TeV
emission is the same as that of the GeV emission, one can confine the EBL model and the intergalactic magnetic
field (IGMF) (Tavecchio et al. 2010a). Using the data of Mkn 501 during its high state, Dai et al. (2002)
calculated the spectrum under different IGMF strengths and reported that the cascade GeV emission is detectable
by Fermi/LAT if the IGMF strength is $\leq10^{-16}$ G. Comparing the predicted emission in the GeV band with the
upper limits measured by the Fermi/LAT for the blazar 1ES 0229+200, Tavecchio et al. (2010a) suggested that the
IGMF is larger than $10^{-15}$G. The GeV-TeV observations for a large sample of BL Lac objects can present solid
evidence for this cascade component, hence gives strong constraints on the IGMF.

The well-sampled, clean, broadband SEDs without
contaminations of the accretion disks and external inverse Compton processes are the best candidates to
investigate the radiation mechanisms and the physical properties of the jets, such as the electron acceleration,
the bulk motion, the magnetic field, and the jet power. In this work, we compile broadband SEDs simultaneously or quasi-simultaneously observed with Fermi/LAT and the
other instruments from literature for TeV BL Lac objects and investigate the radiation mechanisms and the physical
properties of the jets for these sources. With a large sample of TeV BL Lac objects, we explore the physical
reasons that result in the flux variation and the spectral shift in these sources and the relation of the jet
production to the accretion and spin of the cental BHs. We also discuss the cascade emission of the TeV photons
and the constraint on the IGMF. Our sample selection and observed SEDs are presented in \S 2. Modeling the SEDs
with the single-zone synchrotron + SSC model is presented in \S 3. The physical properties of the jets are
reported in $\S$4. Discussions on our results are presented in $\S$5. A summary for our results is in $\S$6.

\section{Sample Selection and Data}
\label{sect:data} In order to obtain well-sampled and clean SEDs without contaminations of the accretion disk
and external inverse Compton processes to investigate the radiation mechanisms and the physical properties of
the jets, we consider only the TeV BL Lac objects that have positive Fermi/LAT detections. The low energy bump
of the SEDs should be well determined with the observations in the radio-IR-optical-X-ray band and high energy
bump can be confined with the GeV and TeV observations. Twenty-four BL Lac objects are included in our sample,
as listed in Table 1. We compile their broadband SEDs that were simultaneously or quasi-simultaneously observed
with Fermi/LAT and the other instruments from literature. For 13 sources there is only one GeV-TeV detection and
thus only one SED is assembled. For the other 11 sources, two or more GeV-TeV detections are available. We select two well-sampled SEDs for each source in order to study the difference of the jet parameters in the high and low states, which are defined with the observed or
extrapolated flux density at 1 TeV. The observed broadband SEDs are shown in Figure \ref{Fig:1}. The description
of each source is summarized in the Appendix.

\section{Modeling the SEDs}
\begin{figure*}
\includegraphics[angle=0,scale=0.30]{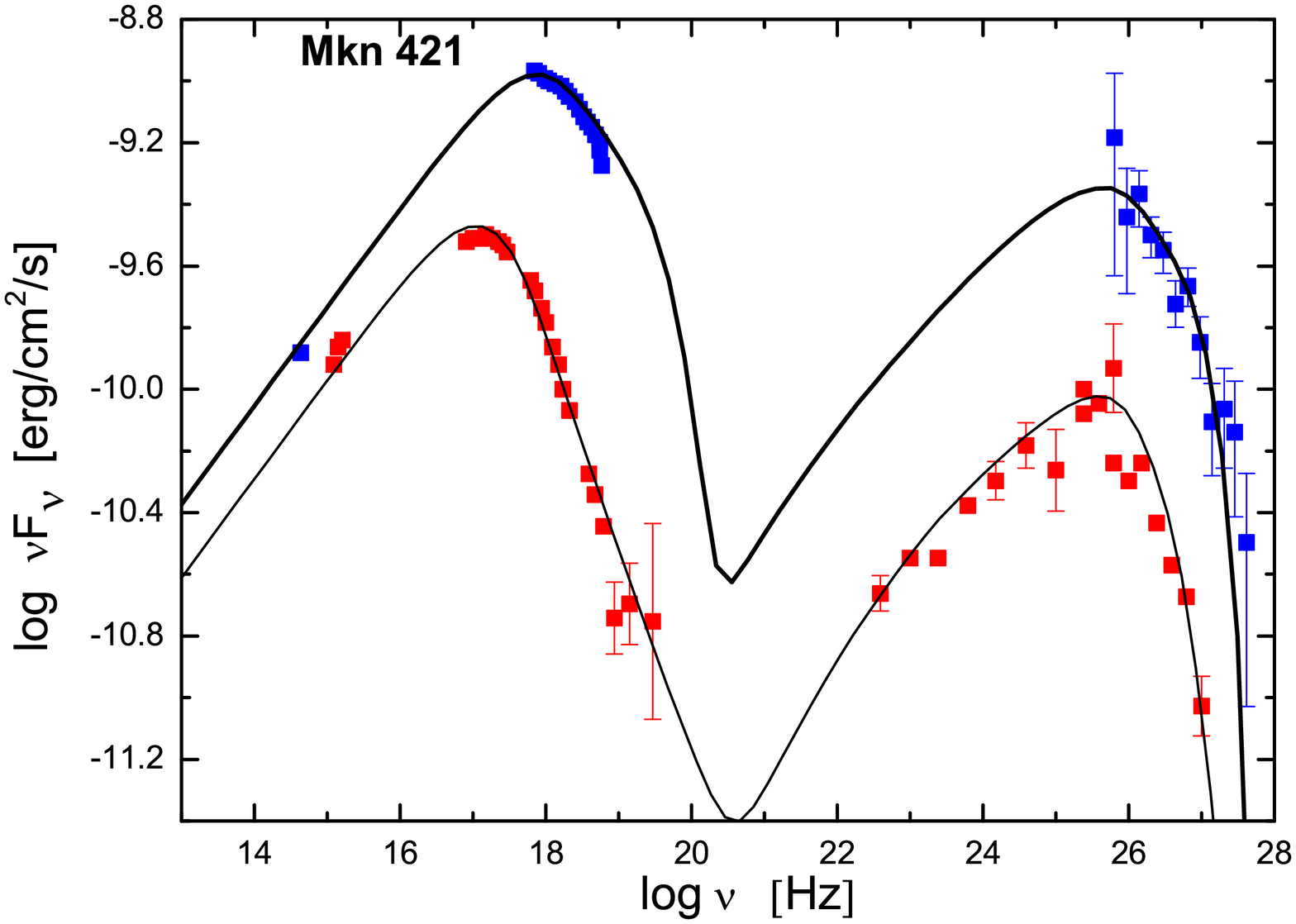}
\includegraphics[angle=0,scale=0.30]{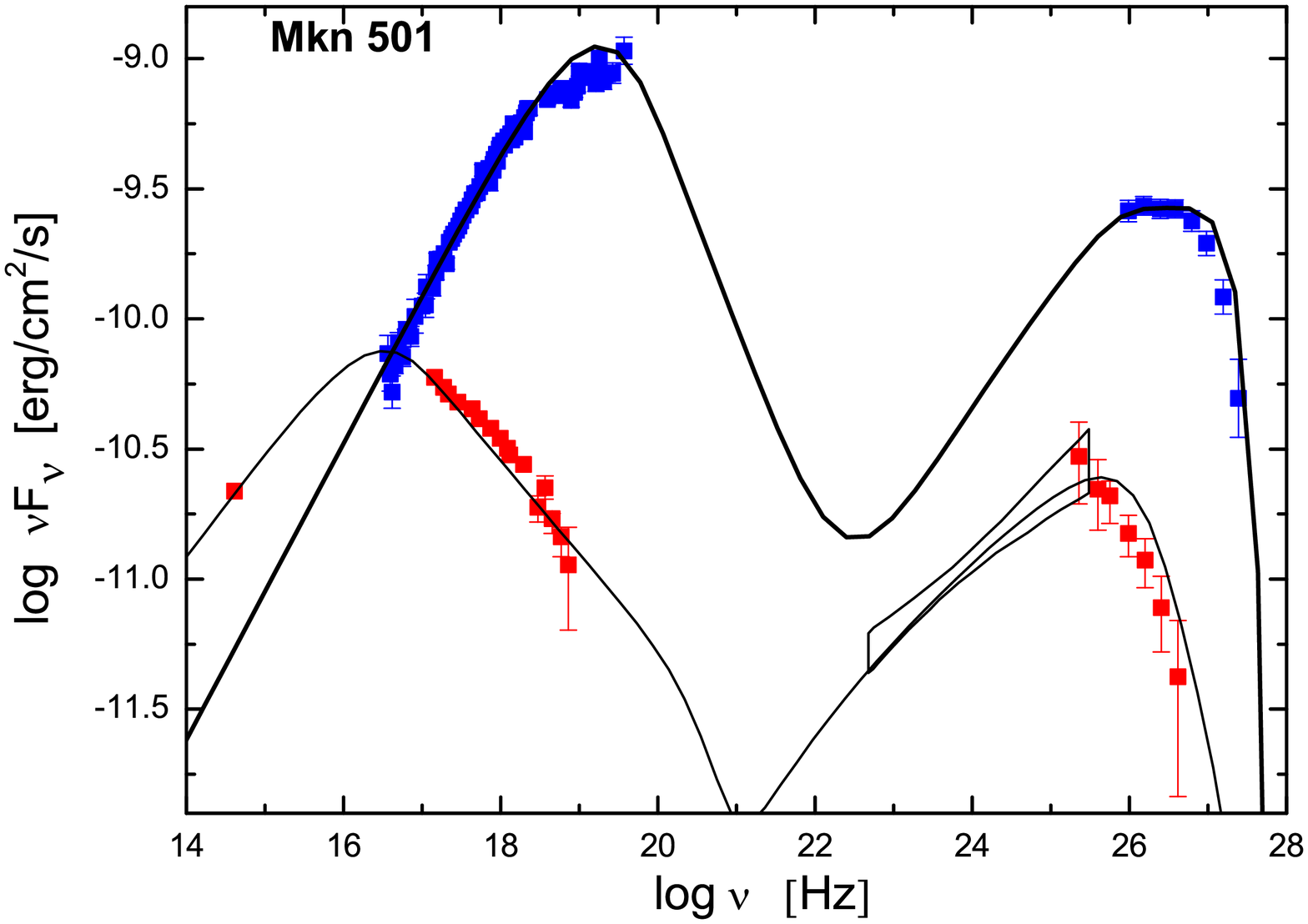}
\includegraphics[angle=0,scale=0.30]{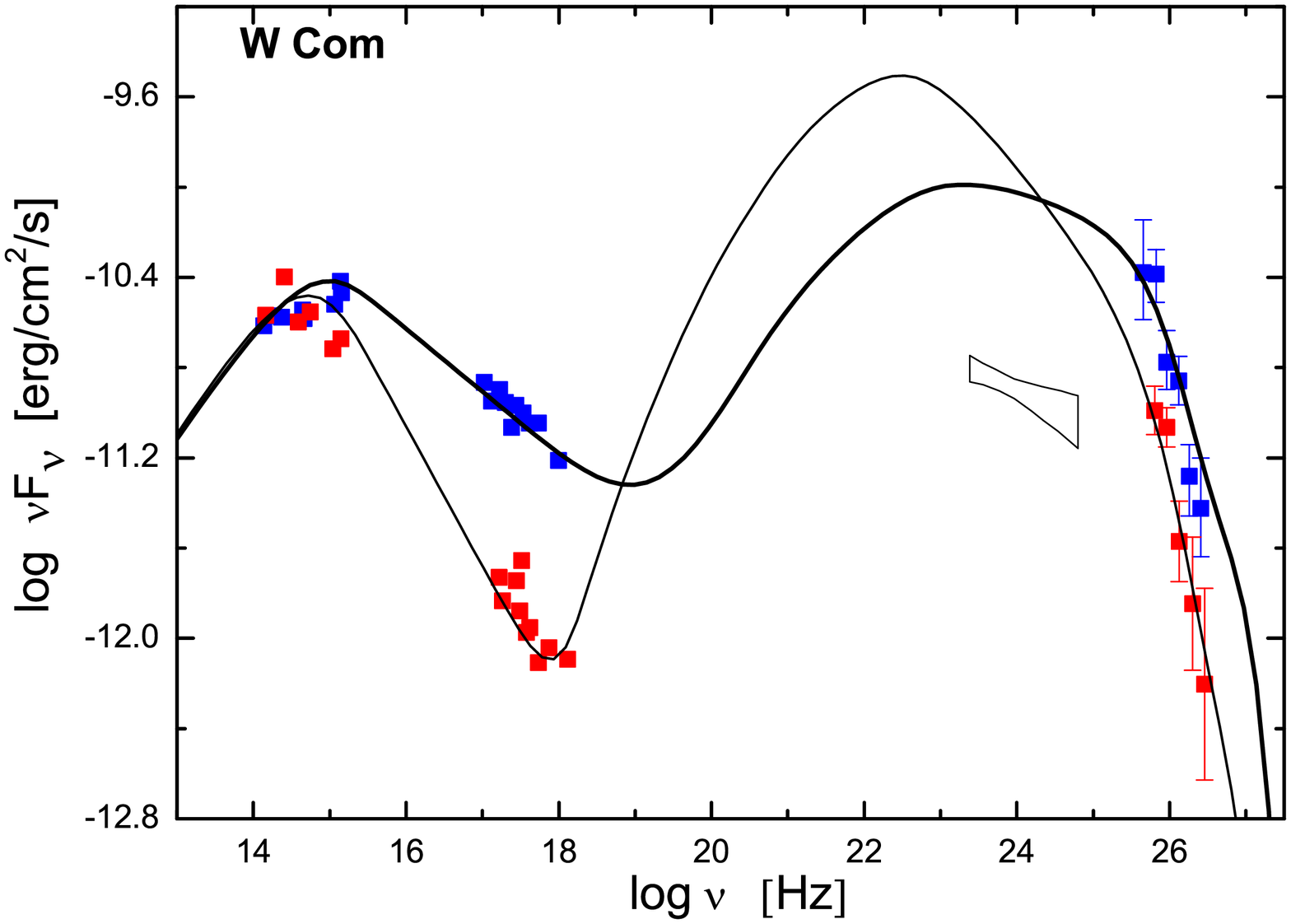}
\includegraphics[angle=0,scale=0.30]{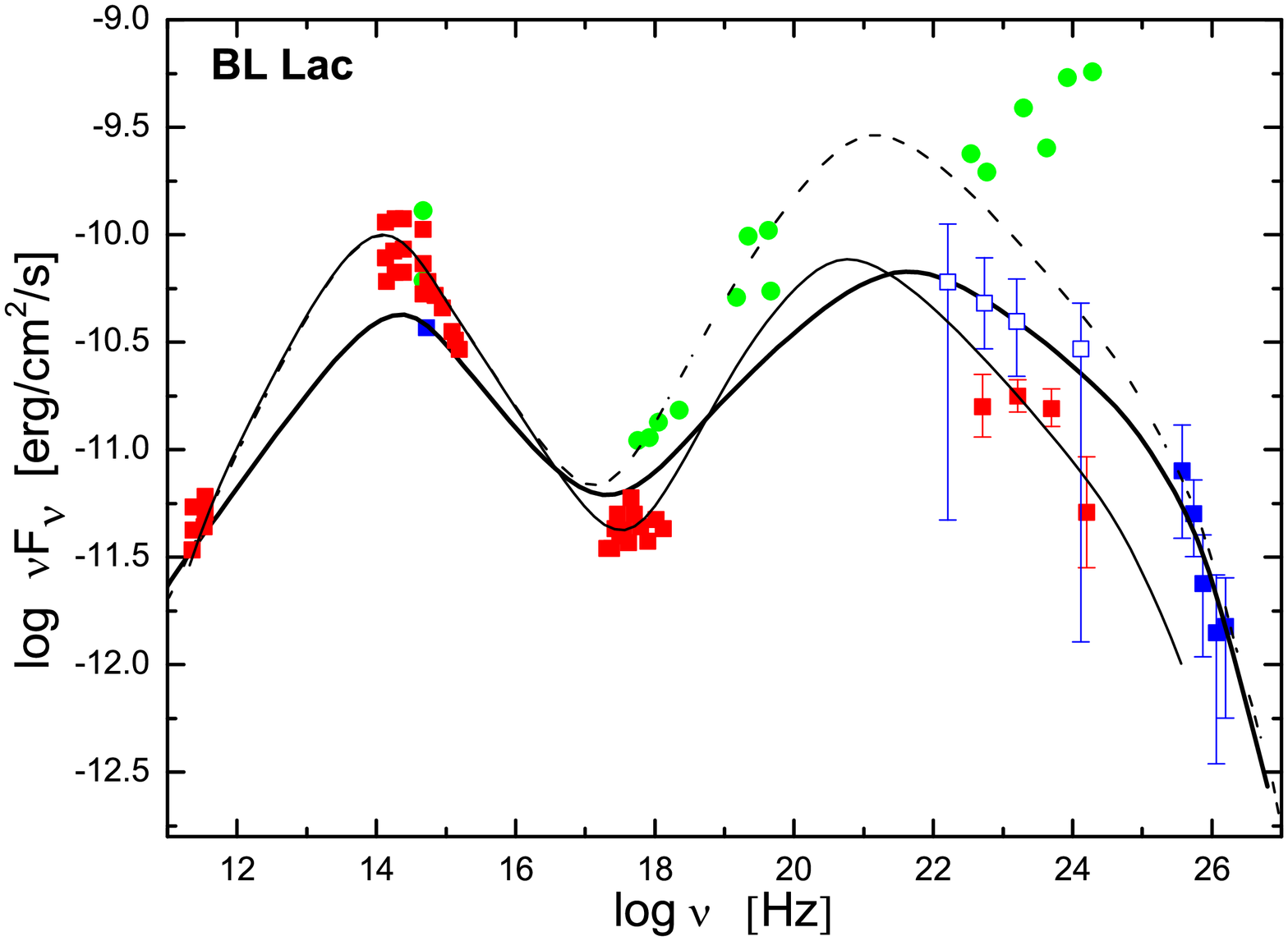}
\includegraphics[angle=0,scale=0.30]{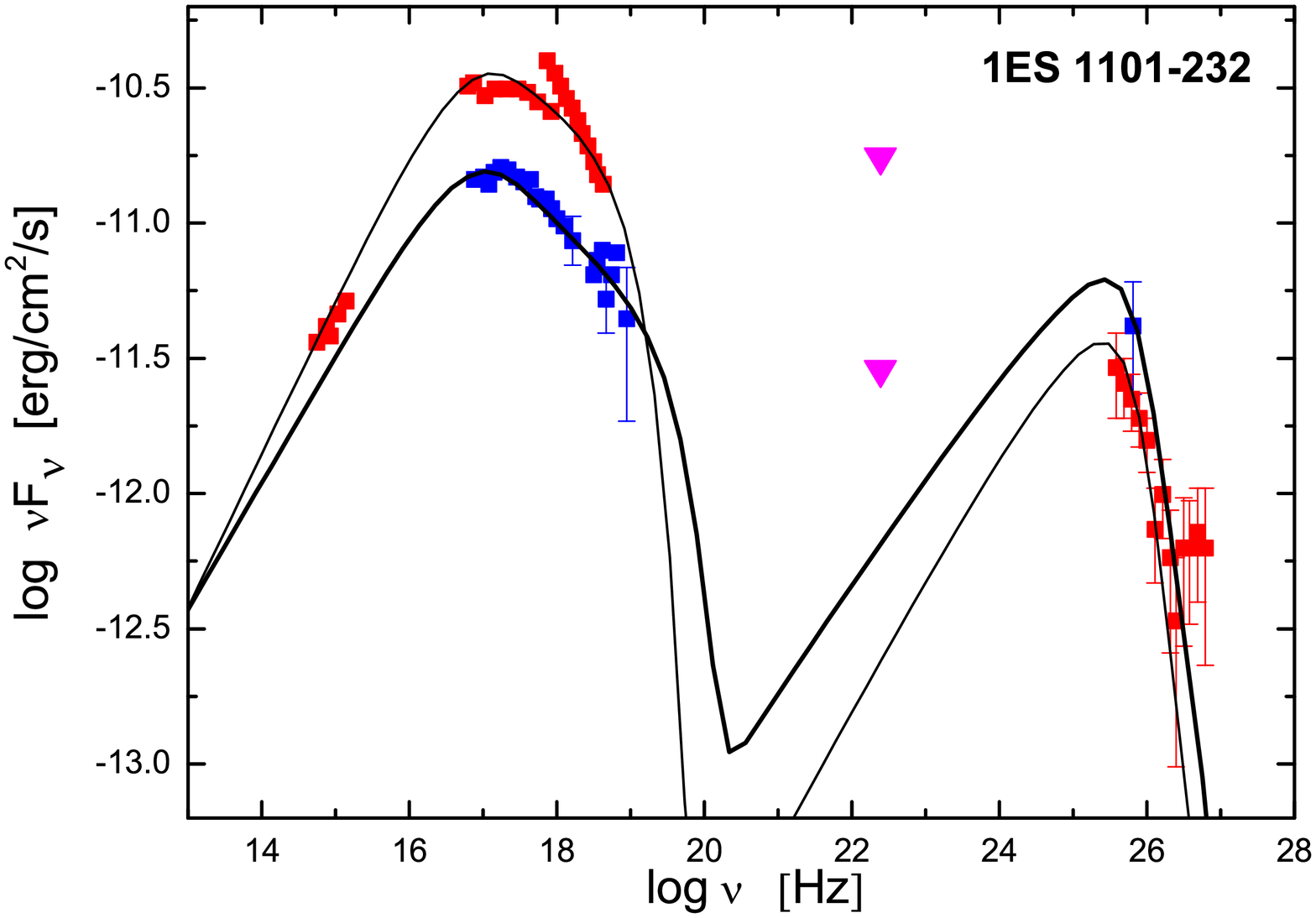}
\includegraphics[angle=0,scale=0.30]{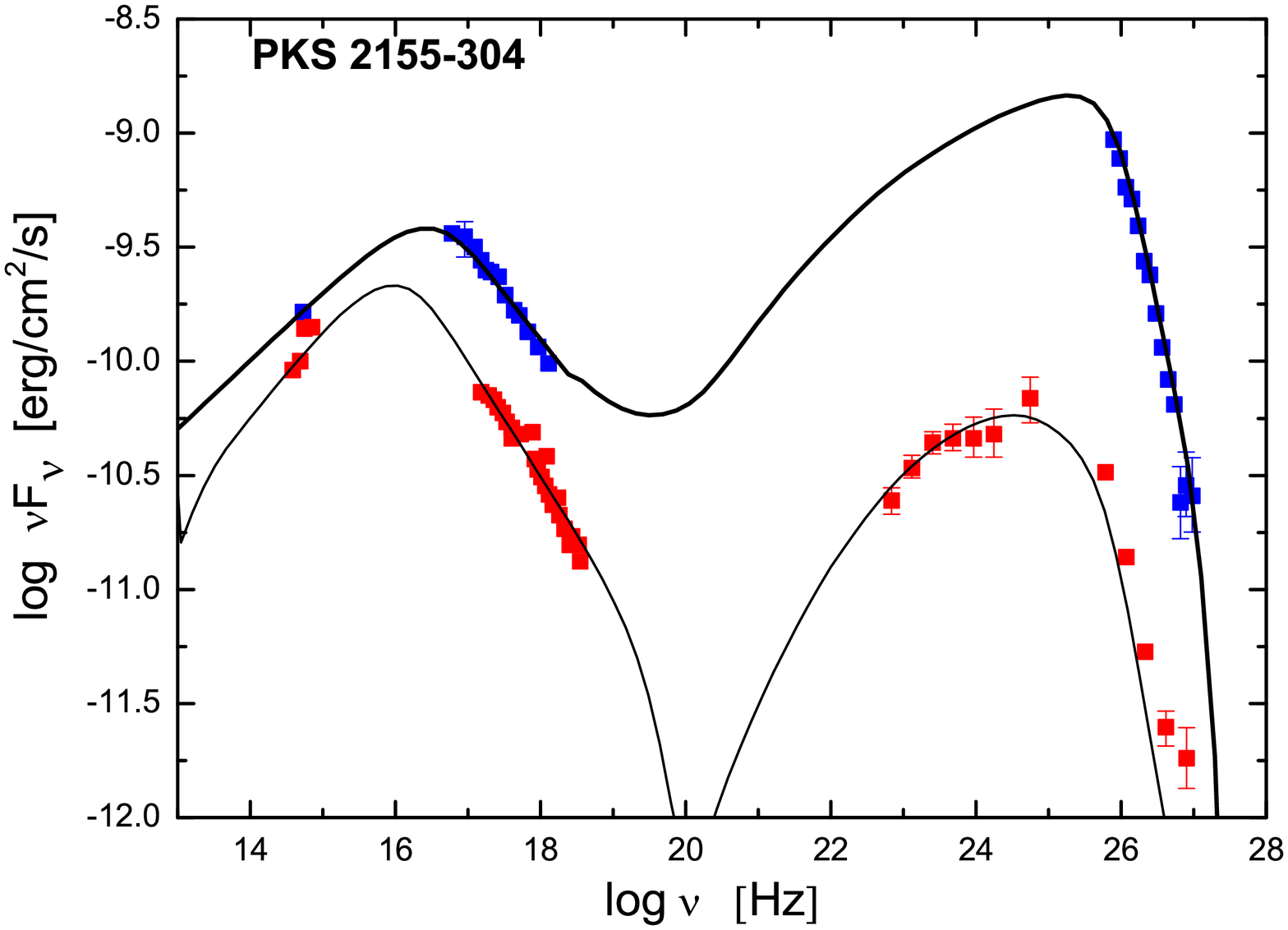}
\includegraphics[angle=0,scale=0.30]{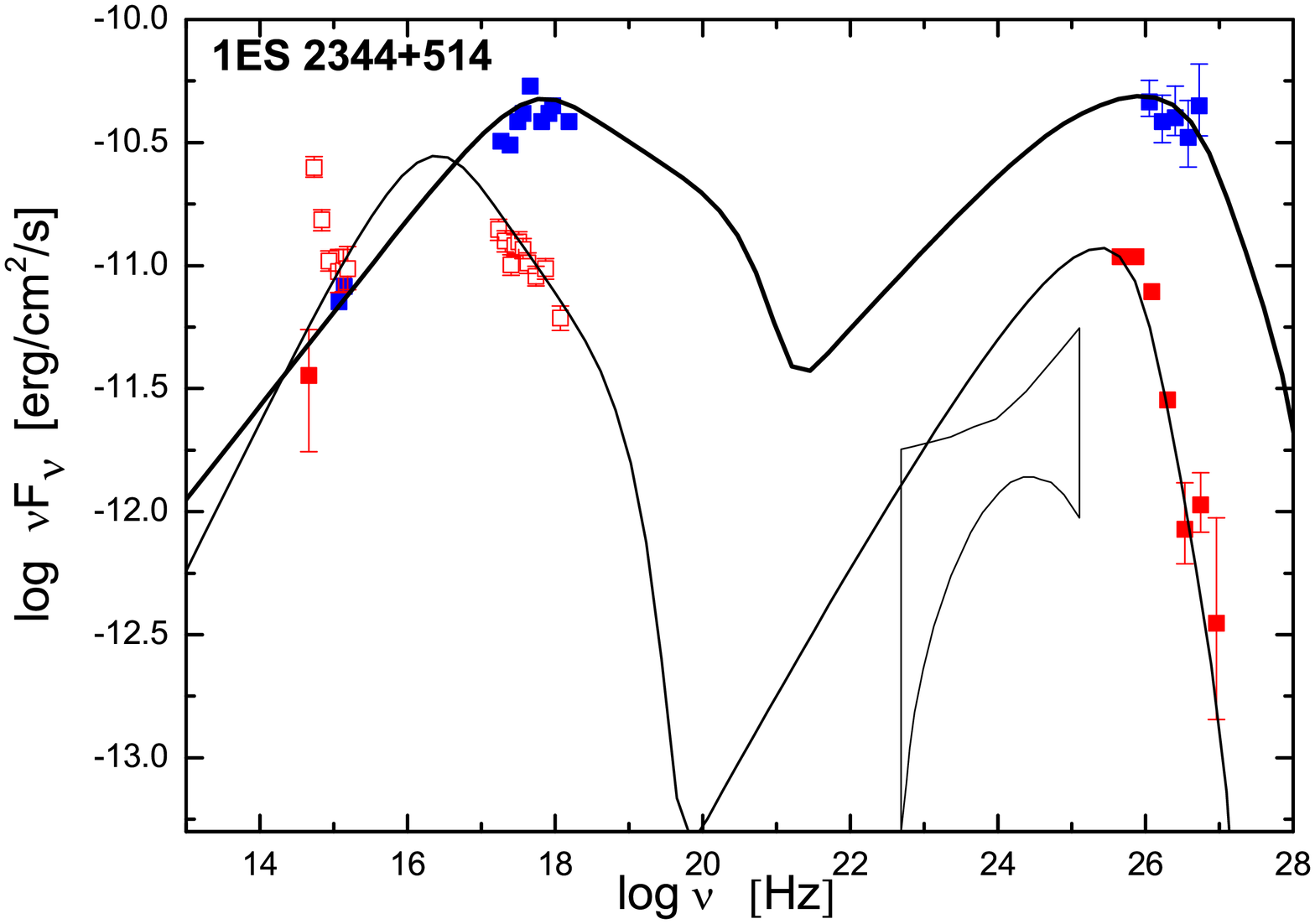}
\hfill
\includegraphics[angle=0,scale=0.30]{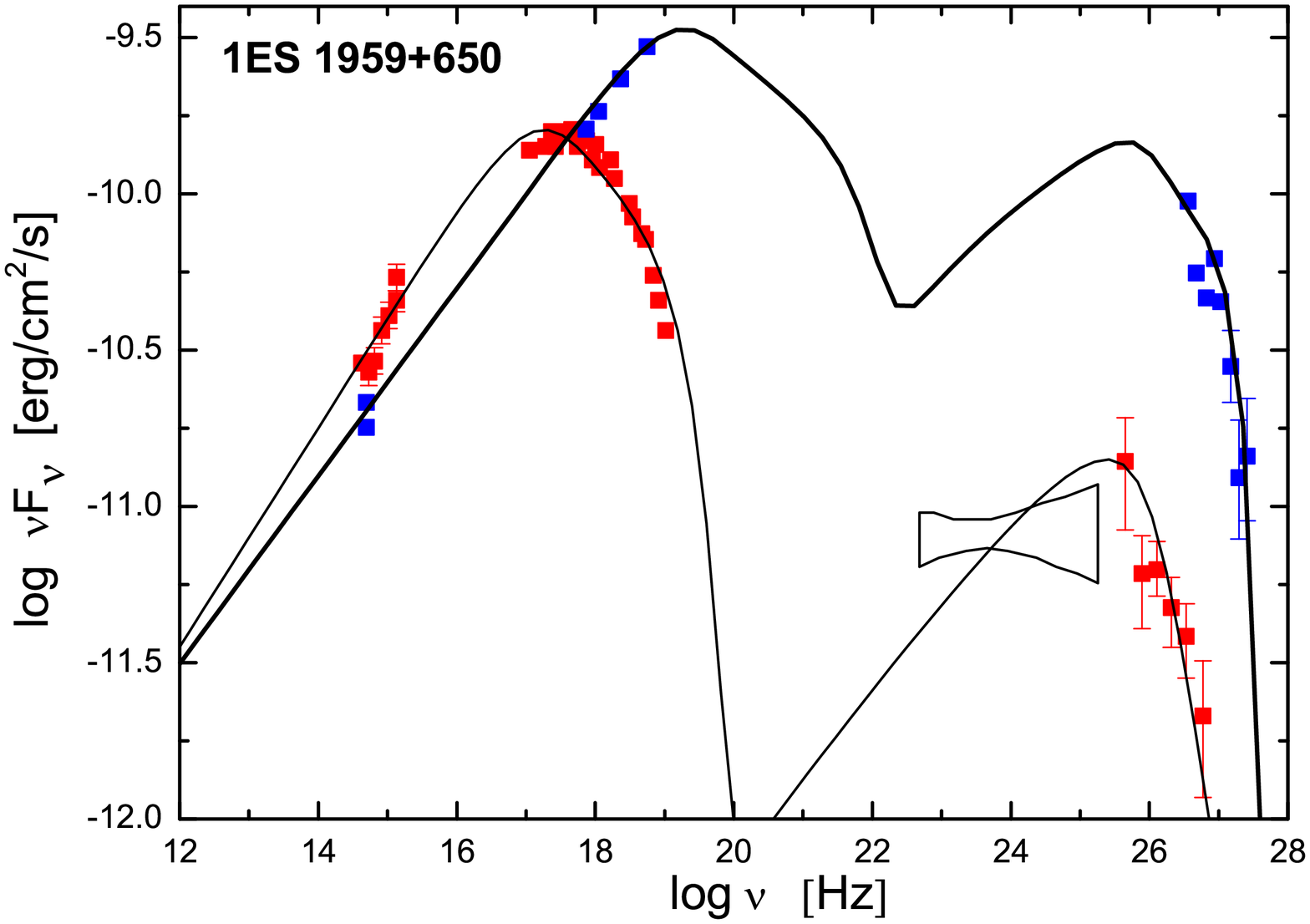}
\caption{Observed SEDs ({\em scattered data points}) with our model fits ({\em lines}). The data of high and low
states are marked with blue and red symbols, respectively. If only one broadband SED is obtained, the data are
shown with black symbols. Opened symbols are for the data that were not observed simultaneously. The Fermi/LAT
observations are presented with a bow-ties or a magenta triangle (upper-limits for five sources, namely, 1ES
1101-232, H 2356-309, RGB J0152+017, 1ES 0229+200 and PKS 0548-322).} \label{Fig:1}
\end{figure*}

The observed SEDs shown in Figure \ref{Fig:1} are double-peaked. We use the single-zone syn + SSC model to fit
the SEDs. Since the external photon fields are very weak compared to the synchrotron radiation photon field for
the BL Lac objects, we do not consider the contributions of the external photon fields. The radiation region is
taken as a homogeneous sphere with radius $R$. The electron distribution is assumed as a broken power law with
indices $p_1$ and $p_2$ below and above the break energy $\gamma_{\rm b} m_{\rm e}c^2$,
\begin{equation}
N(\gamma )= N_{0}\left\{ \begin{array}{ll}
\gamma ^{-p_1}  &  \mbox{ $\gamma \leq \gamma _{\rm b}$}, \\
\gamma _{\rm b}^{p_2-p_1} \gamma ^{-p_2}  &  \mbox{ $\gamma > \gamma _{\rm b}$,}
\end{array}
\right.
\end{equation}
where $p_{1,2}=2\alpha_{1,2}+1$ and $\alpha_{1,2}$ are the spectral indices. The spectrum predicted by the syn +
SSC model can be specified with seven parameters: the strength of magnetic field $B$, the size of the radiating
region $R$, the Doppler boosting factor $\delta$, $p_1$, $p_2$, $\gamma_{\rm b}$, and the electron density
parameter $N_0$ (Tavecchio et al. 1998). Generally, $R$ can be constrained by the minimum variation timescale
$\Delta t$, i.e., $R\sim \delta c\Delta t$. Considering the uncertainty of $\Delta t$ and $\delta$, we adjust
the value of $R$ in a range $ 2.6\times10^{15}-2.6\times10^{16}$ cm (corresponding to $\Delta t\sim 1$ day for
$\delta=1\sim 10$) in order to model the SEDs. $p_1$, $p_2$, $\gamma_{\rm b}$, and $N_0$ can be confined by the
observed parameters in the SEDs, i.e., $\alpha_{1}$, $\alpha_{2}$, $\nu_{\rm s}$, and $L_{\rm s}$. In the
GeV-TeV regime, the Klein-Nishina (KN) effect could be significant and make the IC spectrum have a high energy
cut off. We take this effect into account in our model calculations. The high energy gamma-ray photons would
also be absorbed by EBL, yielding electron-positron pairs, and the observed spectrum in VHE band must be steeper
than the intrinsic one. The absorption in the GeV-TeV band is considered with the EBL model as proposed by
Franceschini et al. (2008).

Note that both $B$ and $\delta$ are critical in modeling the observed SEDs. We constrain their values by the
transparency and KN effect for the TeV gamma-ray photons. A lower limit of $\delta$ can be obtained from the
condition of $\gamma$-ray transparency for pair-production absorption (Dondi \& Ghisellini 1995),
\begin{equation}
\delta> [\frac{\sigma_{\rm T}}{5hc^2}d_L^2(1+z)^{2\beta}\frac{F(\nu_0)}{t_{\rm var}}]^{1/(4+2\beta)},
\end{equation}
where $F(\nu_0)$ is the flux density at the target photon frequency $\nu_0=1.6\times10^{40}/\nu_{\gamma}$ and
$\beta$ is the spectral index of the target photons ($\beta=\alpha_1$ for $\nu_0<\nu_{\rm s}$ and
$\beta=\alpha_2$ for $\nu_0>\nu_{\rm s}$). Without considering the KN effect, The relation between $B$ and
$\delta$ is (Tavecchio et al. 1998)
\begin{equation} B\delta=(1+z)\frac{\nu_{\rm s}^2}{2.8\times10^6\nu_{\rm c}}.
\end{equation}
In the GeV-TeV regime, the spectrum will be significantly affected by the KN effect if $\delta < \delta_{\rm
KN}$,
\begin{equation}
\delta_{\rm KN}=[\frac{\nu_{\rm s}\nu_{\rm c}}{(3/4)(mc^2/h)^2}]^{1/2}.
\end{equation}
In this scenario, the relation between $B$ and $\delta$ is replaced with (Tavecchio et al. 1998)
\begin{equation}
\frac{B}{\delta}=\frac{\nu_{\rm s}}{\nu_{\rm
c}^2}(\frac{mc^{2}}{h})^2\frac{g(\alpha_1,\alpha_2)^2}{3.7\times10^6}\frac{1}{1+z},
\end{equation}
where
\begin{equation}
g(\alpha_1,\alpha_2)=\exp[\frac{1}{\alpha_1-1}+\frac{1}{2(\alpha_2-\alpha_1)}].
\end{equation}
We constrain both $B$ and $\delta$ with these constraints in our SED modeling.

Our model fits to the SEDs are shown in Figure \ref{Fig:1}. It is found that all the SEDs can be well explained
with the syn + SSC model. The parameters for all the SED fits are reported in
Table 1.

\section{Physical Properties of TeV BL Lac Objects}
As shown above, our single-zone Syn + SSC model can well represent the observed SEDs. In this section we explore
the physical properties for the emitting regions in the framework of this model.

\subsection{Electron Spectrum and Magnetic Field of Radiating Region}
The electron energy spectrum and magnetic field of the radiation region are critical to understand the
acceleration and radiation mechanisms of the electrons in the jets. Figure \ref{Fig:2} shows the distributions
of $\gamma_{\rm b}$ and $B$ for the sources in our sample. It is found that $\gamma_{\rm b}$ ranges from
$10^{3}$ to $10^{6}$. It is generally believed that the electrons are accelerated by the violent shocks via the
Fermi acceleration mechanism. The diversity of $\gamma_{\rm b}$ among sources, even among states of a given
source, likely indicates that the intensity of shocks significantly varies among sources (or states).
$\gamma_{\rm b}$ could be an indicator of the intensity of shocks to some extent. Different from $\gamma_{\rm
b}$, the distribution of $B$ is narrowly clustered at $0.1\sim 0.6$ G.  We do not find any correlation between
$\gamma_{\rm b}$ and $B$. Therefore, the magnetic field strengths of the radiating region may not be produced
through the amplification of interstellar medium magnetic field by the shocks.

\subsection{Bulk Motion, Radiating Region Size and Variability of BL Lac Objects}
Since $\delta$ and $R$ are important parameters in our modeling of the SEDs, we initially take $R$ in the range
$2.6\times10^{15}-2.6\times10^{16}$ cm but not rigidly fix it in this range, and adjust the parameter $\delta$
in order to model the observed SEDs. We assume that the size of the radiating region in different states of a
given source does not change. We find that $R=(2\sim 40)\times 10^{15}$ cm for most sources.  $\delta$ is found
to range from 14 to 30. We show the distributions of the derived beaming factor $\delta$ and emitting region
scale $R$ for the sources in our sample in Figure \ref{Fig:3}(a)(b). The corresponding minimum timescale of
variability $\Delta t$ derived from the relation $R\sim \delta c\Delta t$ is $10^3-10^5$ seconds as shown in
Figure \ref{Fig:3}(c), indicating that variations with a minimum timescale from an hour to one day may be
observed for these sources (Xie et al. 2001; Bai et al. 1998; Liang \& Liu 2003).

\subsection {Jet Properties}
Jet power ($P_{\rm jet}$) is essential to understand the physics of jet production. It can be estimated with
$P_{\rm jet}=\pi R^2 \Gamma^2 c (U^{'}_{e}+U^{'}_{p}+U^{'}_{B})$, where $\Gamma$ is the bulk Lorenz factor of
the radiating region and $U^{'}_{i},\ (i={\rm e,\ p,}\ B)$ are the energy densities associated with the emitting electrons
$U^{'}_{e}$, the cold protons $U^{'}_{p}$ and magnetic field $U^{'}_{B}$ measured in the comoving frame
(Ghisellini et al. 2010), which are given by
\begin{eqnarray}
U_{\rm e}^{'}=m_{\rm e}c^2\int N(\gamma)\gamma d\gamma,\\
U_{\rm p}^{'}=m_{\rm p}c^2\int N(\gamma)d\gamma,\\
U_B^{'}=B^2/8\pi,
\end{eqnarray}
assuming that there is one cool proton per emitting electron. These powers can be calculated with the SED
fitting parameters. The radiation power is estimated with the observed luminosity,
\begin{equation}
P_{\rm r}=\pi R^2\Gamma^2 cU_{r}^{'}=L_{\rm obs}\frac{\Gamma^2}{4\delta^4}\approx\frac{L_{\rm obs}}{4\Gamma^2},
\end{equation}
where $L_{\rm obs}$ takes the bolometric luminosity ($\sim L_{\rm bol}$); $P_{\rm r}$ may serve as a robust
lower limit of $P_{\rm jet}$. We calculate $L_{\rm bol}$ in the band  $10^{11}-10^{27}$ Hz based on our SED
fits, then derive $P_{\rm r}$. The calculated $L_{\rm bol}$ and $P_{\rm jet}$ are reported in Table 1 and no
statistical correlation between them is found (as shown in Figure \ref{Fig:4}(a)). The distributions of the
powers and their ratios to the total power, i.e., $\epsilon_{\rm e}=P_{\rm e}/P_{\rm jet}$, $\epsilon_{\rm
p}=P_{\rm p}/P_{\rm jet}$, $\epsilon_B=P_B/P_{\rm jet}$, and $\epsilon_{\rm r}=P_{\rm r}/P_{\rm jet}$, are shown
in Figure \ref{Fig:5}. One can observe that the power in the jets is carried by cold protons for most sources,
with a ratio $\epsilon_{\rm p}=P_{\rm p}/P_{\rm jet}
> 0.5$ for most sources. The portions of the power carried by the electrons and magnetic field are
comparable with $\epsilon_{\rm r}$.

It is generally believed that the jets are fed by accretion of the central massive BHs. We investigate the
relation of $P_{\rm jet}$ to the mass of the central BH with a sub-sample of 18 sources in our sample. The
masses of the BHs of these sources are collected from literature and reported in table 1. The jet power $P_{\rm
jet}$ as a function of BH mass $M_{\rm BH}$ is shown in Figure \ref{Fig:4}(b). A tentative anti-correlation
between the two parameters is observed, i.e., $P_{\rm jet}\propto M_{\rm BH}^{-\alpha}$, $\alpha=1.06$ for high
state and $\alpha=1.03$ for low state, respectively. The correlation coefficients estimated with the Spearman
correlation analysis method are $r=-0.55$ for high state data and $r=-0.68$ for low state data. If the jet is
powered purely by accretion of BH and reflects the accretion rate to some extent, i.e., $P_{\rm
jet}=\eta\dot{M}c^2$, the negative correlation of jet power and BH mass would imply that the accretion rate is
correlated with $\frac{1}{M_{\rm BH}}$. That is clearly unreasonable, and thus the relationship between $P_{\rm
jet}$ and $M_{\rm BH}$ disfavors the pure accretion powered jet scenario. Therefore, we suggest that drawing the
spin energy of the central BH should play a significant role in producing these jets. On the other hand, Figure
\ref{Fig:4}(b) would imply a decrease in the BH spin with an increase in $M_{\rm BH}$. These suggest that the
growth of BH mass in our sample (all with small redshift $z<1$ ) is mostly through a series of random accretion
processes that decrease the BH spin statistically (King \& Pringle 2006; King et al. 2008, 2011). Wang et al.
(2009) and Li et al. (2010) also suggested that minor mergers are important in the BH growth at low redshift and
major mergers may dominate at high redshift, consistent with our results.

\subsection{Flux Variation of TeV Sources}
Figure \ref{Fig:6}(a) shows the distributions of $\nu_{\rm s}$ and $\nu_{\rm c}$ for all the SEDs. It is found
that $\nu_{\rm s}$ ranges from the infrared to the X-ray band (from $10^{13}$ Hz to $10^{19}$ Hz). The value of
$\nu_{\rm c}$ covers the range from $10^{19}$ Hz to $10^{26}$ Hz, but most of them are narrowly clustered at
$10^{24}\sim10^{26}$ Hz. Therefore, these sources may have significant flux variations from IR-Optical to
gamma-ray bands. Since $\nu_{\rm c}$ is narrowly clustered at $10^{24}\sim10^{26}$ Hz, the observations of
Fermi/LAT play an important role in pining down the SSC peak.

Significant flux variations are observed in the TeV AGNs, as shown in Figure 1. The SEDs of both the low and
high states are available for each of 11 sources in our sample. We compare $\nu_{\rm s}$ and $\nu_{\rm c}$ between the
high and low states in Figure \ref{Fig:7}. It is clear that the SEDs in the high states shift to a higher energy
band. Figure \ref{Fig:6}(b) displays the correlation between the peak luminosities of the synchrotron radiation
component [$L_{\rm s}=\nu_{\rm s} L_{\nu_{\rm s}}$] and the SSC component [$L_{\rm c}=\nu_{\rm c} L_{\nu_{\rm
c}}$]. The sources with only one SED available are also shown in Figure \ref{Fig:6}(b). One can see that $L_{\rm
c}$ and $L_{\rm s}$ are tightly correlated and share the same relation regardless the sources are in high or low
states. The Spearman correlation analysis gives the correlation coefficients and the chance probabilities of
$r=0.63$ and $p=0.04$ for the high state data, $r=0.49$ and $p=0.13$ for the low state data, $r=0.87$ and
$p=1.2\times10^{-4}$ for these one-time only observation data, and $r=0.71$ and $p < 10^{-4}$ for all the data,
respectively. The best linear fit obtains $\log L_{\rm c}=(0.55\pm5.75)+(1\pm0.13)\log L_{\rm s}$, suggesting that $U^{'}_{\rm syn}
\sim U_B^{'}$. In the $\nu_{\rm c}-\nu_{\rm s}$ plane, they are also strongly correlated and share the same
relation, as shown in Figure \ref{Fig:6}(c). The Spearman correlation analysis yields a correlation coefficient
of $r=0.93$ and a chance probability of $p < 10^{-4}$ for the high state SEDs, $r=0.86$ and $p=6\times10^{-4}$
for the low state data, and $r=0.92$ and $p < 10^{-4}$ for the one-time only observation data, and $r=0.94$ and
$p < 10^{-4}$ for all the data points, respectively. In the regime $\nu_{\rm s} <10^{16}$ Hz, $\nu_{\rm c}$ is
sensitive to $\nu_{\rm s}$. However, The dependence of $\nu_{\rm c}$ on $\nu_{\rm s}$ is much weaker from
$\nu_{\rm s}
> 10^{16}$ Hz. We fit the $\nu_{\rm c}$ as a function of $\nu_{\rm s}$ with a smoothly broken power law,
\begin{equation}
\nu_{\rm c}=\nu_{c,0}[{(\frac{\nu_{\rm s}}{\nu_k})}^{-s_1}+{(\frac{\nu_{\rm s}}{\nu_k})}^{-s_2}]^{-1},
\end{equation}
and obtain $s_1=2.0$, $s_2=0.5$, and $\nu_k=1.05\times 10^{16}$ Hz. Note that $\nu_{\rm c}=(4/3)\gamma_{\rm b}^{2} \nu_{\rm s}$
in the Thomson scattering regime and $\nu_{\rm c} \propto \gamma_{\rm b}$ in the KN regime.
 Since $\nu_{\rm s}\propto \gamma_{\rm b}^2$, we thus get $\nu_{\rm c}\propto \nu_{\rm s}^2$ in the Thomson regime and $\nu_{\rm c}\propto \nu_{\rm s}^{0.5}$
in the KN regime. Therefore the observed $\nu_{\rm s} - \nu_{\rm c}$ relation is consistent with the Syn +
SSC model\footnote{Although the $\nu_{\rm c}$ and $\nu_{\rm s}$ are derived from our SED fits,
observations of Fermi/LAT together with the instruments in the radio-IR-optical-X-ray and TeV gamma-ray
bands well pin down the two peaks for most AGNs in our sample. The peak frequencies $\nu_{\rm s}$ and $\nu_{\rm c}$
would be roughly regarded as observational parameters.}.

To investigate what may be responsible for the flux variations and the spectral shift in the low and high
states, we derive the ratios of the flux density at 1 TeV $ (R_{\rm 1\ TeV})$ and the physical parameters
($R_x$) in the high state to that in the low state for the 11 sources, where $x$ stands for $L_{\rm bol}$, $B$,
$\delta$, $\gamma_{\rm b}$, and $P_{\rm jet}$. Figure \ref{Fig:8} shows $R_x$ as a function of $R_{\rm 1\ TeV}$.
The values of $\gamma_{\rm b}$, and $P_{\rm jet}$ of the high states are systematically higher than that of the
low states. We measure the correlations between $R_{\rm 1\ TeV}$ and $R_x$ with the Spearman correlation
analysis method, and find a tentative correlation between $R_{\gamma_{\rm b}}$ and $R_{\rm 1\ TeV}$, with a
correlation coefficient $r=0.73$ and chance probability $p=0.01$. The best fit line and robust fit line between
$R_{\gamma_{\rm b}}$ and $R_{\rm 1\ TeV}$ are also shown in Figure \ref{Fig:8}(d). No statistically significant
correlations between $R_{\rm 1\ TeV}$ and other parameters are found. Since $\gamma_{\rm b}$ could be an
indicator of the intensity of shocks, it is possible that the flux variation is probably due to the difference of
the internal shocks that are produced by the collision of two relativistic blobs inside a jet in different states. The collision of two relativistic blobs with larger difference in velocity would result in stronger relativistic shocks (Ravasio et al. 2002). The difference of the blob velocity would be due to the instability of the accretion process.

\section{Discussion}
\subsection{Radiation Mechanism: Leptonic {\em vs.} Hadronic Models}
The underlying radiation mechanisms of gamma-ray emission for blazars are still not well understood. The
proposed models can be classified into two kinds, i.e., leptonic and hadronic models (Maraschi et al.1992; Bloom
\& Marscher 1996; Mannheim \& Biermann 1992). Prior to the launch of the Fermi satellite, the ground-based
observations in the Radio-IR-Optical band are well explained with the synchrotron radiation of relativistic
electrons. The observations with Fermi/LAT, together with the ground-based telescope observations in the TeV
gamma-ray band, now convincingly pin down both the luminosity and the frequency of the high energy peak in the
observed SEDs, placing tight constraints on the radiation mechanisms. Our simple single-zone leptonic model can
well represent the observed SEDs. We derive the mono-luminosity of the sources at $5\times10^{14}$ Hz, 1 keV and
1 GeV, i.e., $L_{\rm 5E14}$, $L_{\rm 1\ keV}$ and $L_{\rm 1\ GeV}$, respectively, and show their correlations
with the best fitting line in Figure \ref{Fig:9}. The Spearman correlation analysis between $L_{\rm 1\ GeV}$ and
$L_{\rm 5E14}$ yields a correlation coefficient $r=0.83$ and a chance probability $p < 10^{-4}$. But only a
tentative correlation between $L_{\rm 1\ GeV}$ and $L_{\rm 1\ keV}$ with a correlation coefficient $r=0.4$ and a
chance probability $p=0.02$ is present. These facts imply that the high energy gamma-ray emission is correlated
with the low energy emission and the photon field of the low energy could be served as the target photons for
the high energy emission. The derived $\nu_{\rm c}-\nu_{\rm s}$ correlation also strongly favors the leptonic
model. It is expected that $\nu_{\rm c}=(4/3)\gamma_{\rm b}^{2}\nu_{\rm s}$ in the Thomson scattering regime
and $\nu_{\rm c} \propto \gamma_{\rm b}$ in the KN regime. As shown in Figure \ref{Fig:6}(c), $\nu_{\rm c}$ is
proportional to $\nu_{\rm s}^{2}$ for $\nu_{\rm s}<10^{16}$ Hz and $\nu_{\rm s}^{0.5}$ for $\nu_{\rm s}>10^{16}$
Hz. This feature well supports the leptonic model.

\subsection{Origin of the Magnetic Field in AGN Jets}
The origin of magnetic field in the AGN jets is still uncertain. We find that the values of magnetic field $B$
are almost uniform among sources and in different states, ranging from $0.1\sim 0.6$ G. This result indicates
that the magnetic field may be independent of the jet properties. As reported in Zhang et al. (2010), the $B$
values in both the jet knots and hot spots are $\sim 10^{-5}-10^{-4}$ G, which may be due to the amplification
of the interstellar medium magnetic field by the violent shocks. The strong magnetic field in the core region
might have a different physical origin. It is possible that the magnetic field is directly from the vicinity
around the central BH or from the accretion disk. The strength of the magnetic field in the inner accretion disk
region can be estimated by comparing it with that of BH binaries (e.g., Zhang et al. 2000), e.g., ${B_{\rm
AGN}}/{B_{\rm Binary}}\propto({T_{\rm AGN}}/{T_{\rm Binary}})^{2} \propto ({{10\rm \ eV}/{1\rm \ keV}})^2$,
where $T$ is the typical peak temperature of the accretion disk, and then $B_{\rm AGN}$ is $\sim10^{-4} B_{\rm
Binary}\sim 10^{4}$ G, much larger than what we found in the jets. It is thus plausible that the magnetic fields
in the jets are carried from the accretion flow, but diluted significantly as the jets propagate and expand
outwards; this also explains why the magnetic fields further out, e.g., in the jet knots and hot spots, are even
much weaker.

\subsection{Implications for Blazar Sequence}
With a large sample of different types of blazars, Fossati et al. (1998) reported a spectral sequence of
FSRQ--LBL--HBL, i.e., along with this sequence, an increase in the peak frequencies corresponding to the
decreases of bolometric luminosity and the ratio of the powers for the high- and low-energy spectral components.
This spectral sequence was interpreted by Ghisellini et al. (1998) with the cosmic evolution of the external
photon fields (such as the BLRs) of the blazars. 
This spectral sequence was interpreted by Ghisellini et al. (1998) with the cosmic evolution of the external
photon fields (such as the BLRs) of the blazars, or more physically, the sequence is due to the mass and accretion rate of the BH ( Ghisellini \& Tavecchio (2008). More recently, Chen \& Bai (2011) extend this sequence to narrow line Seyfert 1 galaxies.
The significant cooling effect for the electrons by external
field photons from BLR of FSRQs may result in a low $\gamma_{\rm b}$ in the electron spectrum. The contribution
from the IC process of the external photon fields would significantly increase the radiative energy density
$U_{\rm r}$ and the ratio of the $L_{\rm c}/L_{\rm s}$. Therefore, observationally, one can expect that both
$L_{\rm bol}$ and $L_{\rm c}/L_{\rm s}$ are anti-correlated with $\nu_{\rm s}$ in the co-moving frame, and thus
an anti-correlation between $\gamma_{\rm b}$ and $U_{\rm r}$ is also expected (Ghisellini et al. 1998). We
calculate the luminosity and the peak frequency of synchrotron emission in the co-moving frame with $L^{'}_{\rm
bol}=L_{\rm bol}/\delta^4$ and $\nu_{\rm s}^{'}=\nu_{\rm s} (1+z)/\delta$. The $L^{'}_{\rm bol}$ and $L_{\rm
c}^{'}/L_{\rm s}^{'}$ as a function of $\nu_{\rm s}^{'}$ are shown in Figure \ref{Fig:10}. The $\gamma_{\rm b}$
as a function of $U_{\rm r}$ ($U_{\rm r}=U_{\rm syn}+U_B$ in this work) is also shown in Figure \ref{Fig:10}. No
any correlation is found for pairs $L_{\rm bol}^{'}-\nu_{\rm s}^{'}$ and $\gamma_{\rm b}-U_{\rm r}$. This lack
of correlation in our sample can be explained by the fact the EC process is negligible for the majority of the
BL Lac objects in our sample.

However, the ratio of $L^{'}_{\rm c}/L^{'}_{\rm s}$ is indeed anti-correlated with $\nu_{\rm s}^{'}$ in their
high and low states, except for PKS 2005-489 in the high state. The Spearman correlation analysis yields a correlation coefficient
$r=-0.48$ and a chance probability $p=0.13$ for the high state data, $r=-0.79$ and $p=0.004$ for low state data,
respectively. Excluding PKS 2005-489 in the high state, our best fits give $\log L_{\rm c}^{'}/L_{\rm s}^{'}=(6.54\pm1.84)-(0.47\pm0.13)\log {\nu_{\rm s}^{'}}$ and $\log L_{\rm c}^{'}/L_{\rm
s}^{'}=(2.64\pm0.75)-(0.17\pm0.05)\log {\nu_{\rm s}^{'}}$ for the sources in
the low and high states, respectively. The slopes in the high and low states are significantly different. If
this anti-correlation is due to the cooling of the external photons, the slopes in the low and high states would
be similar. Since the 11 sources are BL Lac objects, the external photon fields outside their jets are much
weaker than the synchrotron radiation photon field and are thus not considered in this work. Therefore, the
$L_{\rm c}^{'}/L_{\rm s}^{'}-\nu_{\rm s}^{'}$ anti-correlation may have a different physical origin. It is
possible that this anti-correlation is due to the KN effect. As $\nu_{\rm s}$ increases, the SEDs shift to the
higher frequency end and the KN effect should be more significant. Thus, the ratio of $L_{\rm c}/L_{\rm s}$
would decrease as $\nu_{\rm s}$ increases since $U_B$ is almost constant.

\subsection{Implications for IGMF}
The TeV gamma-ray photons may interact with the extragalactic background light, hence produce electron-positron
pairs. As a result, the observed spectrum in the TeV band would be steeper than the intrinsic one. On the other
hand, these $e^{\pm}$ may interact with cosmic microwave photons through IC scattering and yield a GeV emission
component in the observed SEDs (Dai et al. 2002; Yang et al. 2008; Tavecchio et al. 2010a). GeV emission
observations for the TeV sources thus may place constraints on the EBL model and the intergalactic magnetic
field (IGMF) strengths. However, as shown in Figure \ref{Fig:1}, all the SEDs are well modeled with the one-zone lepton model and no extra-excess components in GeV band are found. Note that BL Lacertae experienced a gamma-ray outburst in 1997. The SED observed during the outburst cannot be explained with the syn+SSC model. We show also the SED and our model fit in Figure 1 (green circles and dashed line), but the fitting parameters of this SED are not considered in above analysis and discussion. It is found that an extra emission component over the SSC process was detected. Since BL Lacertae has broad emission lines (Vermeulen et al. 1995), being different from other BL Lac objects. It is thought that the GeV flare in 1997 might be produced by the EC process as proposed by Ravasio et al. (2002). In addition, the cascade GeV emission would be a common feature for all TeV sources. However, we do not see a similar feature in the other sources. Therefore, The extra GeV emission of BL Lacertae in 1997 outburst would not be due to the cascade emission.

The flux and the spectral shape of this reproduced emission in the GeV band depend on the primary SSC component
(TeV spectrum) and the strength of the IGMF. Dai et al. (2002) reported that this component may be detectable in
Mkn 501 with Fermi/LAT if the IGMF strength is $\leq10^{-16}$ G. By comparing the observed flux upper-limit in
the GeV band from the sources with that from the cascade emission expected by the model, Tavecchio et al. (2010a)
suggested that the IGFM strength should be larger than $B\cong10^{-15}$ G. Our systematical analysis on the SEDs
of GeV-TeV sources clearly indicates that no excess emission over the SSC component is presented in the SEDs.
Therefore, the IGMF strength would be much higher than $10^{-16}$ G.

\section{Summary}
The peak frequency and corresponding luminosity of the high energy bump of SEDs of AGNs are well determined with
simultaneous or quasi-simultaneous observations in the TeV and GeV bands. We compile the broadband SEDs of 24 BL
Lac objects that were simultaneously or quasi-simultaneously observed with Fermi/LAT and the other instruments
from literature. The clean SEDs without contaminations of the accretion disks and external inverse Compton
processes of these sources are good candidates for investigating the radiation mechanisms and the physical
properties of the AGN jets. Our results are summarized as following:
\begin{itemize}
\item  We find that the one-zone synchrotron + SSC model can well represent the observed SEDs.
The observed positive correlation between $\nu_{\rm s}$ and $\nu_{\rm c}$ also favors this scenario.

\item The distribution of $\gamma_{\rm b}$ ranges from $10^{3}$ to $10^{6}$, but the magnetic
strength $B$ is distributed within a narrow range of 0.1-0.6 G. These results indicate that the intensity of the
shocks for electron acceleration violently varies among sources, and the magnetic field may not be due to the
amplification of the interstellar magnetic field by the shocks in the jet. We propose that the magnetic field
may be carried from the accretion flow.

\item The Doppler boosting factor $\delta$ of the jets ranges from 14 to 30, and the sizes of the
radiating regions are roughly $(2\sim 40)\times 10^{15}$ cm, suggesting that flux variations with a minimum
timescale from an hour to one day may be observed for these sources. Significant flux variations are observed
for the sources in our sample. The SEDs in the high state shift to higher frequencies. The ratio of the flux
density at 1 TeV is correlated with the ratio of the $\gamma_{\rm b}$ in the low and high states, indicating
that the relativistic shocks in the jets may be responsible for the spectral shift between the low and high
states. The $\delta$ value of the high state is systematically larger than that of the low state, but the ratios
of $\delta$ and flux density at 1 TeV in the high and low states are not correlated. No systematical difference
of $B$ is found between high and low states.

\item We calculate the bolometric luminosity and the jet power for the sources in our sample.
The jet power is dominated by the kinetic energy for most sources. No correlation between $L_{\rm bol}$ and
$P_{\rm jet}$ is found. An anti-correlation between the jet power and the mass of the central BH is observed,
i.e., $P_{\rm jet}\propto M^{-1}_{\rm BH}$. This disfavors the scenario of a pure accretion-driven jet. We
suggest that the energy injection from the spin of the central BHs would be significant for these sources, and
that thus BHs with smaller masses should have higher jet power efficiencies.

\item No correlation is found between $L_{\rm bol}^{'}-\nu_{\rm
s}^{'}$ or $\gamma_{\rm b}-U_{\rm r}$, indicating that EC processes are not important for the sources in our
sample. The ratio of $L^{'}_{\rm c}/L^{'}_{\rm s}$ is anti-correlated with $\nu_{\rm s}^{'}$ for the sources in
both high and low states; however, the slopes of the anti-correlations between $L_{\rm c}^{'}/L_{\rm s}^{'}$ and
$\nu^{'}$ for the sources in the low and high states are significant different. We suggest that this
anti-correlation is possibly due to the KN effect.

\item  It was suggested that TeV gamma-ray photons may interact with the extragalactic background light and produce an
excess component in the GeV band if the $\leq10^{-16}$ G (Dai et al. 2002). However, no such an excess component
is found for the sources in our sample, indicating the strength of the IGMF would much larger than $10^{-16}$ G.
\end{itemize}

\section{Appendix}
\emph{W Com.} The first TeV intermediate-frequency-peaked BL Lac (IBL) object, an evidently bright outburst in
the optical and X-ray bands was observed in 1998 (Tagliaferri et al. 2000). W Com is confirmed to be a TeV source by
VERITAS observation during a strong TeV flare in the middle of March 2008 with an integrated photon flux above
200 GeV of $\sim9\%$ crab (Acciari et al. 2008) and the data with quasi-simultaneous observation of Swift are presented as red squares
in Figure 1. Both the SEDs quasi-simultaneously obtained during the TeV flare and during
the optical/X-ray outburst can be well fit by the single-zone synchrotron + SSC model (Zhang 2009).
Subsequently, another outburst of very high energy gamma-ray emission was detected in 2008 June by VERITAS with
the flux of $(5.7\pm0.6)\times10^{-11}$cm$^{-2}$s$^{-1}$ (Acciari et al. 2009a), three times brighter than the observation of March
2008 (blue squares in Figure 1). The bow-tie of Fermi/LAT observation (Abdo et al.
2009) are also presented in Figure 1.

\emph{Mkn 421.} It is the first confirmed TeV AGN by Whipple (Punch et al. 1992). Violent variation of the flux
in the GeV-TeV regime was detected, and was associated with that observed in the X-ray band (BLazejowski et al.
2005). The SEDs in high and low states are shown in Figure 1. The data of high state are from BLazejowski et al.
(2005), when the source underwent an outburst in 2004 April with the peak flux $\sim135$ mcrab in the X-ray band
and $\sim3$ crab in the gamma ray band. Mkn 421 is found in a rather low/quiet state from August 2008 to August
2009. The data of the low state are taken from this period (Paneque et al. 2009).

\emph{Mkn 501.} It was detected by Whipple during an observation of 66 hr. An average flux of
$(8.1\pm1.4)\times10^{-12}$cm$^{-2}$s$^{-1}$ above 300 GeV and the variability in a timescale of days were
observed (Quinn et al. 1996). From 1997 April to 1999 June, the observations with BeppoSAX showed that the peak
frequency of synchrotron emission shifted from 100 keV back to 0.5 keV, and correspondingly the flux decreased
(Tavecchio et al. 2001). A multi-wavelength campaign in 2006 July with Suzaku and MAGIC was performed. During
this program the average VHE flux above 200 GeV is $\sim 20\%$ of crab flux with a photon index $2.8\pm0.1$ from
80 GeV to 2 TeV, indicating that the source was in a low state (Anderhub et al. 2009a). The data of the TeV high
state are from Tavecchio et al. (2001) with the BeppoSAX quasi-simultaneous observations in 1997 April 16th. The
Fermi/LAT data observation is from Abdo et al. (2009), but it is not simultaneous with the low and high states.

\emph{PKS 2155-304.} A high-frequency-peaked BL Lac (HBL) object, its strong VHE emission was first detected in
1997 November, at the same time the strongest X-ray emission ever observed and GeV gamma rays were also detected
by BeppoSAX and EGRET (Chadwick et al. 1999). A simultaneous observation with HESS, Chandra and the Bronberg
optical observatory was carried in the night of July 26-30 2006 during a high-activity state in gamma-ray band,
and the gamma ray flux reached $\sim11$ times the crab flux (Aharonian et al. 2009a). The emission between X-ray
and VHE gamma-ray is strongly correlated. The broadband SED of this campaign is shown as blue squares in Figure
1. The GeV-TeV observation with Fermi and HESS was performed between 25 August and 6 September 2008, and the
low-energy component was simultaneously covered by ATOM telescope, RXTE and Swift observations (red squares in
Figure 1, Aharonian et al. 2009b). During that period PKS 2155-304 was at a low-activity state in X-ray and
gamma-ray bands, whereas the optical was much higher. The optical emission was correlated with the VHE emission,
but no correlation between X-ray and VHE was found.

\emph{1ES 1101-232.} A HBL object hosted by an elliptical galaxy, and detected by HESS in March-June 2004 and
2005 with a very hard spectrum and no significant variation was found (Aharonian et al. 2007a). A
multiwavelength campaign data set, including observations of VHE by HESS and X-ray by RXTE satellite in 2005 and
XMM-Newton in 2004, were obtained (blue squares in Figure 1, Costamante 2007). The Suzaku observation
simultaneously covered with the HESS measurement was carried out in 2006 May (red squares in Figure 1, Reimer et
al. 2008). No significant X-ray or gamma-ray variability was detected during this program, and the object was in
a quiescent state with the lowest X-ray flux ever measured. 1ES 1101-232 was not detected by Fermi/LAT
(Tavecchio et al. 2010b), and only an upper-limit was given as shown in Figure 1.

\emph{BL Lacertae.} The first low-frequency-peaked BL Lac (LBL) object, its VHE gamma-ray emission was detected
by MAGIC during 2005 August to December with an integral flux of $(0.6\pm0.2)\times10^{-11}$cm$^{-2}$s$^{-1}$,
corresponding to $3\%$ crab flux (Albert et al. 2007a). The photon index from 150 to 900 GeV is rather steep
with $\Gamma=-3.6\pm0.5$, and the light curve shows no significant variability. The simultaneous observation for
MAGIC in optical band was performed by KVA. The broadband SED together with EGRET data (open blue squares) in
1995 is shown in Figure 1 (blue squares). During an optical outburst in July 1997, BL Lacertae was detected in
X-ray band by RXTE and in gamma-ray band by EGRET, implying that the source was bright and variable in both
bands (Madejski et al. 1999). The spectra in X-ray and gamma-ray bands are hard. The broadband observational SED
for this outburst is also presented in Figure 1 (green circles). Madejski et al. (1999) considered that the
X-rays are produced by synchrotron radiation while the gamma-rays are produced by Comptonization of the broad
emission line flux. The multiwavelength data during Fermi/LAT observation are also considered and taken from
Tavecchio et al. (2010b), but no simultaneous data in TeV regime are obtained.

\emph{1ES 2344+514.} The third BL Lac object detected with VHE emission by Whipple. The detection of VHE
emission mostly came from an apparent flare on 1995 December and the average flux above 350 GeV was
$(6.6\pm1.9)\times10^{-11}$ cm$^{-2}$ s$^{-1}$, $63\%$ of the crab flux (Catanese et al. 1998). The observation
of MAGIC between 2005 August 3 and 2006 January 1 presented a steep spectrum with photon index $\Gamma=-2.95$
and a flux 6 times below the 1995 flare, indicating that the source was in low state (Albert et al. 2007b). No
evidence for variability was found during the MAGIC observations. The simultaneous optical observation with
MAGIC was performed by KVA. The broadband SED is shown as red squares in Figure 1. The data of Swift satellite
observation on 19 April 2005 (Tramacere et al. 2007) and the bow-tie of Fermi/LAT observation (Abdo et al.
2009) are also presented to constrain the radiation model. A broadband SED for a high state
quasi-simultaneously obtained by VERITAS and Swift on December 2007 is also considered in this work, and the
data are from the web \emph{http://veritas.sao.arizona.edu/content/view/174/72/}. The measured flux above 300
GeV was $(6.76\pm0.62)\times10^{-11}$ cm$^{-2}$ s$^{-1}$, corresponding to $48\%$ of crab flux. The highest
X-ray emission ever observed was measured by Swift/XRT, and was correlated with the VHE gamma-ray emission.

\emph{1ES 1959+650.} A HBL object with a strong TeV outburst in May 2002 detected by VERITAS (Holder et al.
2003), HEGRA (Horns et al. 2002) and CAT (Aharonian et al. 2003). During the outburst, the flux level reached to
3 times the Crab flux and there was evidences for strong variability. The multiwavelength campaign in radio,
optical, X-ray and TeV gamma-ray bands was performed from 2002 May 18 to August 14 (Krawczynski et al. 2004). A
time-averaged spectrum corresponding to the TeV gamma-ray high state is considered in this work (blue squares in
Figure 1). The observations show that X-ray flux and gamma-ray flux had tentative correlation, but no
correlations of optical variability with X-ray and gamma-ray were found. Another multiwavelength campaign was
performed in 2006 May, when the source exhibited a high state in optical and X-ray bands and was at the lowest
level in the VHE band (Tagliaferri et al. 2008). There were variabilities in optical and X-ray bands. The
derived broadband SED during this campaign and spectrum butterfly of Fermi/LAT observation are shown as red
squares in Figure 1.

\emph{PKS 2005-489.} A HBL object detected with VHE gamma-ray emission by HESS in 2003 and 2004 (Aharonian et
al. 2005). A significant signal of VHE emission was detected in 2004 with the integral flux above 200 GeV of
$\sim6.9\times10^{-12}$ cm$^{-2}$ s$^{-1}$, corresponding to $2.5\%$ crab flux. The flux level in 2003 was lower
than that measured in 2004, indicating that the activity in 2004 increased, but no significant variability on
time scale less than a year was found. The multiwavelength observation by HESS, XMM-Newton and RXTE satellites
from 2004 to 2007 was performed (HESS Collaboration 2010a) and the observation indicated that the large flux
variations in the X-ray band are coupled with weak or no variations in the VHE band. We consider a broadband SED
for the low X-ray flux, which is taken as the low state and shown as red symbols in Figure 1. A very high state
of X-ray emission for this source in 1998 was detected by BeppoSAX and RossiXTE (Tagliaferri et al. 2001). HESS
and Fermi/LAT simultaneously observed the source in 2009, when the X-ray flux was comparable to the flux level
in 1998 (Kaufmann et al. 2010). Therefore, we compile the data of the two observations together and take it as
the high state (blue symbols in Figure 1).

\emph{S5 0716+714.} It is a LBL object. The MAGIC observations were performed in November 2007 and in April 2008
and detected its TeV emission (Teshima et al. 2008; Anderhub et al. 2009b). The integral flux above 400 GeV was
$\sim7.5\times10^{-12}$ cm$^{-2}$ s$^{-1}$, corresponding to $9\%$ crab flux. The optical emission of S5
0716+714 was simultaneously observed by KVA when the source was in a high state at optical band, and most of the
gamma-ray emission signal came from the phase of the optical high state of the object, suggesting a possible
correlation between the VHE emission and optical emission (Anderhub et al. 2009b). The data quasi-simultaneously
obtained in April 2008 by KVA, Swift and MAGIC are considered and shown as blue symbols. The data of Fermi/LAT
observation with quasi-simultaneous Swift observation from Tavecchio et al. (2010b) are also considered and
shown as red symbols in this work.

\emph{1ES 0347-121.} A BL Lac object detected by HESS between August and December 2006 with an integral flux
corresponding to $2\%$ of crab flux (Aharonian et al. 2007b). The photon spectrum from 250 GeV to 3 TeV can be
described by a power law with a photon index $\Gamma\sim3.1$. During the VHE observation, no significant
variability was detected. The broadband SED (blue squares in Figure 1) was compiled with the data from the
simultaneous observations of HESS, Swift and ATOM  (Aharonian et al. 2007b). Another quasi-simultaneous
multiwavelength campaign during Fermi/LAT observation (Abdo et al. 2010b) is also considered and shown in Figure
1 (red squares).

\emph{PG 1553+113.} Its VHE emission with a very soft spectrum (photon index of $\Gamma=4.0\pm0.6$) was detected
in 2005 by HESS, and no evidence of variability was found (Aharonian et al. 2006a). The VHE emission
subsequently confirmed by MAGIC (Albert et al. 2007c), and the integral flux levels of HESS and MAGIC are
consistent. PG 1553+113 is in the Fermi LAT bright AGN source list, but was not detected by EGRET because it was
in a low state during the observation. This source is a HBL object and a bright X-ray source with many
observations, but no strong or fast variability was detected in the X-ray band (Reimer et al. 2008). The
redshift of PG1553+113 is unknown and the VHE observation indicates that the redshift is greater than 0.25. In
this work, we take $z=0.3$. The observations by Fermi/LAT from 4 August 2008 to 22 February 2009 show that it
was a steady source with a hard spectrum in Fermi/LAT energy band (Abdo et al. 2010c). The data of broadband SED
(black squares) are from Abdo et al. (2010c), including quasi-simultaneous observations by KVA, Suzaku, MAGIC
and HESS in July 2006 and the observation of Fermi/LAT.

\emph{3C 66A.} A IBL object, and the observation of VHE was performed from September 2007 through January 2008
by VERITAS and it was confirmed to be a TeV source with an integral flux above 200 GeV $6\%$ of crab flux
(Acciari et al. 2009b). The observed spectrum can be characterized by a soft power-law with photon index
$\Gamma=4.1$ and a variability on the time-scale of days was found. The simultaneous GeV-TeV observations by
Fermi/LAT and VERITAS were performed in October 2008 (Reyes et al. 2009). These data and the follow-up
observations in low energy bands are combined to create a broadband SED.

\emph{Mkn 180.} A HBL object detected with VHE gamma-ray emission by MAGIC during an optical outburst in 2006.
The integral flux above 200 GeV is $(2.3\pm0.7)\times10^{-11}$ cm$^{-2}$ s$^{-1}$ and corresponds to $11\%$ crab
flux. The observed spectrum was rather soft with a photon index of $\Gamma\sim3.3\pm0.7$, and no variability was
found (Albert et al. 2006a). Only the data of KVA and UMRAO (University of Michigan Radio Observatory)
observation are simultaneous with the MAGIC observation, but the historical data (open squares) and the bow-tie
of Fermi/LAT observation (from Abdo et al. 2009) are also given in the SED in our work.

\emph{H 2356-309.} A HBL object detected with VHE emission by HESS from June to December 2004, with a integral
flux above 200 GeV of
$4.1\pm0.5\times10^{-12}$ cm$^{-2}$ s$^{-1}$ (Aharonian et al. 2006b). A simultaneous observation with HESS in 2004
at lower energy bands was performed by ROTSE-III (optical) and RXTE (X-rays). A broadband SED obtained
simultaneously by XMM-Newton and HESS on June 2005 (HESS Collaboration et al. 2010b) is presented in this work,
in which the flux levels in X-ray and TeV bands are comparable with that measured in 2004. In the VHE band,
significant small-amplitude variations on time scales of
months and years were detected. The X-ray measurements show that it was in a low state in this band.
Unfortunately, the observations of Fermi/LAT only give an upper limit (Tavecchio et al. 2010b).

\emph{1ES 1218+30.4.} A HBL object confirmed to be a TeV source by MAGIC in 2005 January, but no variability on
timescales of days was found within the statistical errors (Albert et al. 2006b). In optical band, KVA observed
the source simultaneously with MAGIC. From 2008 December to 2009 May, VERITAS monitored the source and revealed
a prominent flare reaching $\sim20\%$ of the Crab flux (Acciari et al. 2010a). The light curve of TeV emission
for this source showed day-scale variability. The observational flux of VERITAS is comparable to MAGIC
(Weidinger \& Spanier 2010). Swift observed this source between March and December 2005, and was
quasi-simultaneous with the observation of MAGIC (R\"{u}ger et al. 2010). In this work, the broadband SED
includes the observations of VERITAS, Fermi/LAT (from Abdo et al. 2009), swift and KVA.

\emph{1ES 1011+496.} A HBL object observed by MAGIC from 2007 March to May after an optical outburst in March
2007, and obtained an integral flux above 200 GeV of $1.58\pm0.32\times10^{-11}$ cm$^{-2}$ s$^{-1}$. The
variation of VHE emission comparing with that in 2006 March-April, implies that the state of VHE emission may be
related to the optical emission state (Albert et al. 2007d). The broadband SED includes the quasi-simultaneous
observations of Swift and Fermi/LAT (from Tavecchio et al. 2010b) and the observation of MAGIC in 2007.

\emph{1ES 0806+524.} A HBL object detected by VERITAS in VHE gamma-ray regime between November 2006 and April
2008, and no significant variability on months time-scale was found (Acciari et al. 2009c). The observed photon
spectrum from November 2007 to April 2008 can be fitted by a power law with $\Gamma\sim-3.6$ between 300 to 700
GeV. The integral flux above 300 GeV is $\sim2.2\times10^{-12}$ cm$^{-2}$ s$^{-1}$, corresponding to $1.8\%$ of
crab flux. The data obtained quasi-simultaneously by Swift and VERITAS observations (Acciari et al. 2009c) and
the spectrum butterfly of Fermi/LAT observations (Abdo et al. 2009) are considered in this work.

\emph{RGB J0710+591.} A well known HBL object not detected by EGRET. It was observed in the VHE waveband by
VERITAS between December 2008 and March 2009, and confirmed to be a TeV source (Ong et al. 2009, Acciari et al.
2010b). The observed spectrum from 0.31 to 4.6 TeV can be fit by a power law with a photon spectral index
$\sim-2.69$, and the integral flux above 300 GeV is $3.9\pm0.8\times10^{-12}$ cm$^{-2}$ s$^{-1}$, corresponding
to $3\%$ of crab flux. The VERITAS observation was complemented by contemporaneous observations from Fermi/LAT,
Swift and Michigan-Dartmouth-MIT observatory (Acciari et al. 2010b).

\emph{PKS 1424+240.} A HBL object with an unknown redshift, and its VHE emission was detected by VERITAS with a
flux normalization at 200 GeV of $\sim5.1\times10^{-11}$ TeV$^{-1}$ cm$^{-2}$ s$^{-1}$ (Acciari et al. 2010c).
The photon spectrum above 140 GeV can be described well by a power law with $\Gamma\sim3.8$. During the period
from February 2009 to June 2009, the flux of VHE emission was steady and the contemporaneous observation of
Fermi/LAT also did not detect any variability (Acciari et al. 2010c). The broadband SED is established by
simultaneous observations of VERITAS, Fermi/LAT, Swift and MDM. Considering the EBL absorption, a redshift upper
limit of 0.66 is inferred (Acciari et al. 2010c) and $z=0.5$ is taken in this work.

\emph{RGB J0152+017.} A HBL object detected by HESS in late October and November 2007 (Aharonian et al. 2008).
The observed spectrum is well fit by a power law with $\Gamma\sim2.95$, and the integral flux above 300 GeV
corresponds to $\sim2\%$ of crab flux. The broadband SED also includes the simultaneous observations of Swift
and RXTE (Aharonian et al. 2008), and the upper limit detected by Fermi/LAT (from Tavecchio et al. 2010b).

\emph{1ES 0229+200} A HBL object observed by HESS in 2005/2006, and confirmed to be a TeV source (Aharonian et
al. 2007c). The integral flux above 580 GeV is $\sim9.4\times10^{-13}$ cm$^{-2}$ s$^{-1}$, corresponding to
$\sim1.8\%$ of crab flux, and the observed spectrum is characterized by a hard power law with $\Gamma\sim2.5$
from 500 GeV to 15 TeV. During the observation, no significant variability on any scale was detected. Except for
the data of the HESS observation, the broadband SED considered in this work also includes the data of Swift
observation in August 2008 and an upper limit detected Fermi/LAT (Tavecchio et al. 2010b).

\emph{PKS 0548-322.} A HBL object observed between October 2004 and January 2008 with the HESS, and confirmed to
be a TeV source (Superina et al. 2008, Aharonian et al. 2010). The integral flux above 200 GeV is $1.3\%$ of the
crab flux and the observed spectrum is characterized by a power-law with a photon index $\Gamma\sim2.86$.
Contemporaneous UV and X-ray observations in November 2006 were made by Swift, but it was not be detected by
Fermi/LAT (Tavecchio et al. 2010b). No significant variability was detected by HESS and Swift. In this work, the
broadband SED includes the data of the Swift and HESS observations, together with the upper limit of Fermi/LAT
observation.

\emph{H 1426+428.} A HBL object with the strongest TeV emission detected in 2000 and 2001 by Whipple with an
integral flux of $(2.04\pm0.35)\times10^{-11}$ cm$^{-2}$ s$^{-1}$ above 280 GeV (Horan et al. 2002). The object
was monitored by Whipple from 1995 to 1998 during a general blazar survey, but no statistical signal was
detected. No simultaneous broadband data in Figure 1 are found and the data of broadband SED are from Wolter et
al. (2008). The spectrum butterfly of the Fermi/LAT observations (Abdo et al. 2009) is also taken into account.

\begin{deluxetable}{llllllllllllll}
\tabletypesize{\footnotesize} \rotate \tablecolumns{13}\tablewidth{45pc} \tablecaption{Observations and SED fit
results for the sources in our sample}\tablenum{1} \tablehead{\colhead{Source} &
\colhead{State\tablenotemark{a}} & \colhead{z\tablenotemark{b}}  & \colhead{$p_{1}$}
& \colhead{$p_{2}$} & \colhead{$R$}  & \colhead{$B$} & \colhead{$\delta$} & \colhead{$\gamma_{\rm min}$}& \colhead{$F_{\rm 1TeV}$} & \colhead{$L_{\rm bol}$} & \colhead{$P_{\rm jet}$} & \colhead{$M_{\rm BH}\tablenotemark{\rm c}$} & \colhead{Ref}\\
\colhead{}& \colhead{} & \colhead{} & \colhead{} & \colhead{} & \colhead{($10^{15}$cm)} & \colhead{(G)} &
\colhead{} & \colhead{} & \colhead{(Jy)} & \colhead{(erg/s)} &
\colhead{(erg/s)} & \colhead{$\log M_{\bigodot}$} & \colhead{}\\
\colhead{(1)}& \colhead{(2)} & \colhead{(3)} & \colhead{(4)}& \colhead{(5)} & \colhead{(6)} & \colhead{(7)} &
\colhead{(8)} &\colhead{(9)} &\colhead{ (10)} & \colhead{(11)} & \colhead{(12)}& \colhead{(13)}& \colhead{(14)}}
\startdata
W Com&L& &1.94&4.24&2.6&0.23&19&250&1.4E-15&1.5E45&6.1E46&\nodata&1\\
&H& &2&3.56&2.6&0.25&28&200&7.8E-15&8.6E44&3.2E46&\nodata& \\
Mkn 421&L&0.031&2.36&4.4&10&0.14&27&180&2.9E-14&2.5E44&6.4E45&8.3&2\\
&H& &2.36&3.6&10&0.13&27&20&1.8E-13&7.3E45&2.5E46&\nodata& \\
Mkn 501&L&0.034&2.2&3.72&24&0.06&19&150&7.3E-15&1.7E44&1.8E45&9.2&3\\
&H& &1.86&4.6&2.6&0.13&19&2&1.5E-13&1.9E45&2.2E46&\nodata& \\
PKS 2155-304&L&0.116&2.2&4&14&0.6&22&500&3.3E-15&2.4E44&5.9E46&8.7\tablenotemark{Re}&4\\
&H& &2.4&3.8&14&0.13&29&200&3.6E-13&5E45&4.8E47&\nodata& \\
1ES 1101-232&L&0.186&1.86&3.42&6&1&22&2&5.7E-16&5E44&1.4E46&\nodata&5\\
&H& &2.06&3.54&6&0.29&24&2&1.7E-15&6.3E45&2.3E46&\nodata& \\
BL Lacertae&97&0.069&1.6&4&2.6&0.35&22&30&8.9E-16&4.2E45&3.2E46&8.2&6\\
&L& &1.64&4&2.6&0.83&20&30&9.2E-17&1.7E45&1.2E46&\nodata& \\
&H& &2.1&3.8&2.6&0.35&22&30&7.3E-16&5.2E45&1E46&\nodata& \\
1ES 2344+514&L&0.044&1.78&3.84&2.6&0.45&18&2&1.2E-15&2.2E44&9.9E44&8.8&7\\
&H& &2.24&3.4&2.6&0.19&18&2&1.8E-14&1.8E46&3.5E45&\nodata& \\
1ES 1959+650&L&0.048&2.3&3.5&2.6&1&28&2&2.9E-15&6.6E45&6.9E45&8.1&8\\
&H& &2.4&3.4&2.6&0.4&21&2&6.7E-14&6.4E46&2.5E46&\nodata& \\
PKS 2005-489&L&0.071&1.8&4.8&26&0.1&15&200&1.8E-15&2.9E44&1.1E46&9&9\\
&H& &2.5&3.2&26&0.45&22&150&2.5E-15&5.1E44&2.6E46&\nodata& \\
S5 0716+714&L&0.26&1.9&4.16&26&0.5&22&150&1.5E-16&1.3E45&1.6E47&8.6\tablenotemark{Re}&10\\
&H& &1.96&3.9&26&0.2&22&200&6.4E-15&3E45&4.6E47&\nodata& \\
1ES 0347-121&L&0.188&1.22&4.6&33&0.02&14&100&3.6E-17&8.1E45&5E46&8.7&11\\
&H& &2.42&3.5&7&0.6&15&2&2.1E-15&7E46&2.3E46&\nodata& \\
PG1553+113&&0.3&1.46&3.68&100&0.18&20&200&5.9E-15&8.5E44&3.4E47&\nodata&12\\
3C 66A&&0.44&2&4.5&26&0.36&22&200&2.2E-15&1.9E45&9.6E47&\nodata&13\\
Mkn 180&&0.045&1.62&3.4&9&0.45&18&2&5.6E-16&1.2E44&1.9E45&8.2&14\\
H2356-309&&0.165&2.1&3.4&2.6&0.38&27&2&8.4E-16&5.5E45&6.5E45&8.6&15\\
1ES 1218+30.4&&0.182&1.86&3.6&5.6&0.43&25&300&2.1E-15&7E43&2.3E46&8.6&16\\
1ES 1011+496&&0.212&2.1&4.8&26&0.22&22&2&1.2E-14&3.2E46&1.3E47&8.3\tablenotemark{Wu}&17\\
1ES 0806+524&&0.138&2.2&3.8&2.6&1.06&22&2&2.4E-16&9.9E45&7E45&8.9\tablenotemark{Wu}&18\\
RGB J0710+591&&0.125&2.24&3.12&20&0.11&19&1000&2E-15&4.6E43&1.2E46&8.3&19\\
PKS 1424+240&&0.5&1.9&5.4&50&0.37&25&100&6.9E-16&1.8E45&7.4E47&\nodata&20\\
RGB J0152+017&&0.08&2.1&3.4&2.6&0.12&22&2&1.2E-15&4.8E45&1E45&\nodata&21\\
1ES 0229+200&&0.14&2.08&3.16&2.6&0.8&21&2&2.3E-15&1.8E45&2.8E46&9.2&22\\
PKS 0548-322&&0.069&2.1&3.8&2.6&0.5&20&2&5.4E-16&9E44&1.5E45&8.2&23\\
H1426+428&&0.129&2.4&3.2&2.6&0.14&20&200&2.3E-14&7.4E44&2.8E46&9.1&24\\

\enddata
\tablenotetext{a}{The state of each source in TeV band. ``H" indicates ``high state", ``L" indicate ``low
state", and ``97" indicates the observation of the 1997 flare state.}
\tablenotetext{b}{z: redshift.} \tablenotetext{\rm c}{BH mass for 18 sources, among which 14 are from Woo \&
Urry (2002), two are from Wu et al. (2002), and two are from other references. The superscript ``Re" denotes the
reference given in column (14) and the superscript ``Wu" denote the reference Wu et al. (2002).}
\tablecomments{Columns: (4) (5) The energy indices of electrons below and above the break; (6) Size of the
emitting region; (7) The inferred magnetic field strength; (8) The beaming factors $\delta$; (9) The minimum
Lorenz factor of electrons; (10) Flux density at 1 TeV; (11) The bolometric luminosity of each source; (12) The
total jet power; (13) The mass of the black hole in the center of each host galaxy; (14) The references.}

\tablerefs{ (1) Tagliaferri et al. 2000; Acciari et al. 2008; Acciari et al. 2009a; Tavecchio et al. 2010b; (2)
BLazejowski et al. 2005; Paneque et al. 2009; (3) Tavecchio et al. 2001; Anderhub et al. 2009a; (4) Aharonian et
al. 2009a; Aharonian et al. 2009b; Rieger \& Volpe 2010; (5) Costamante 2007; Reimer et al. 2008; (6) Albert et
al. 2007a; Ravasio et al. 2002; Tavecchio et al. 2010b; (7) Albert et al. 2007b; Tramacere et al. 2007; (8)
Krawczynski et al. 2004; Tagliaferri et al. 2008; (9) HESS Collaboration 2010a; Tagliaferri et al. 2001;
Kaufmann et al. 2010; (10) Anderhub et al. 2009b; Vittorini et al. 2009; (11) Aharonian et al. 2007b; Abdo et
al. 2010b; (12) Abdo et al. 2010c; (13) Reyes et al. 2009; (14) Albert et al. 2006a; Abdo et al. 2009; (15)
HESS Collaboration et al. 2010b; (16) R\"{u}ger et al. 2010; Abdo et al. 2009; (17) Tavecchio et al. 2010b;
(18) Acciari et al. 2009c; Abdo et al. 2009; Tavecchio et al. 2010b; (19) Acciari et al. 2010b; (20) Acciari et
al. 2010c; (21) Aharonian et al. 2008; (22) Tavecchio et al. 2010b;  (23) Aharonian et al. 2010; (24) Wolter et
al. 2008 ; Abdo et al. 2009}
\end{deluxetable}

\clearpage

\begin{figure*}
\includegraphics[angle=0,scale=0.30]{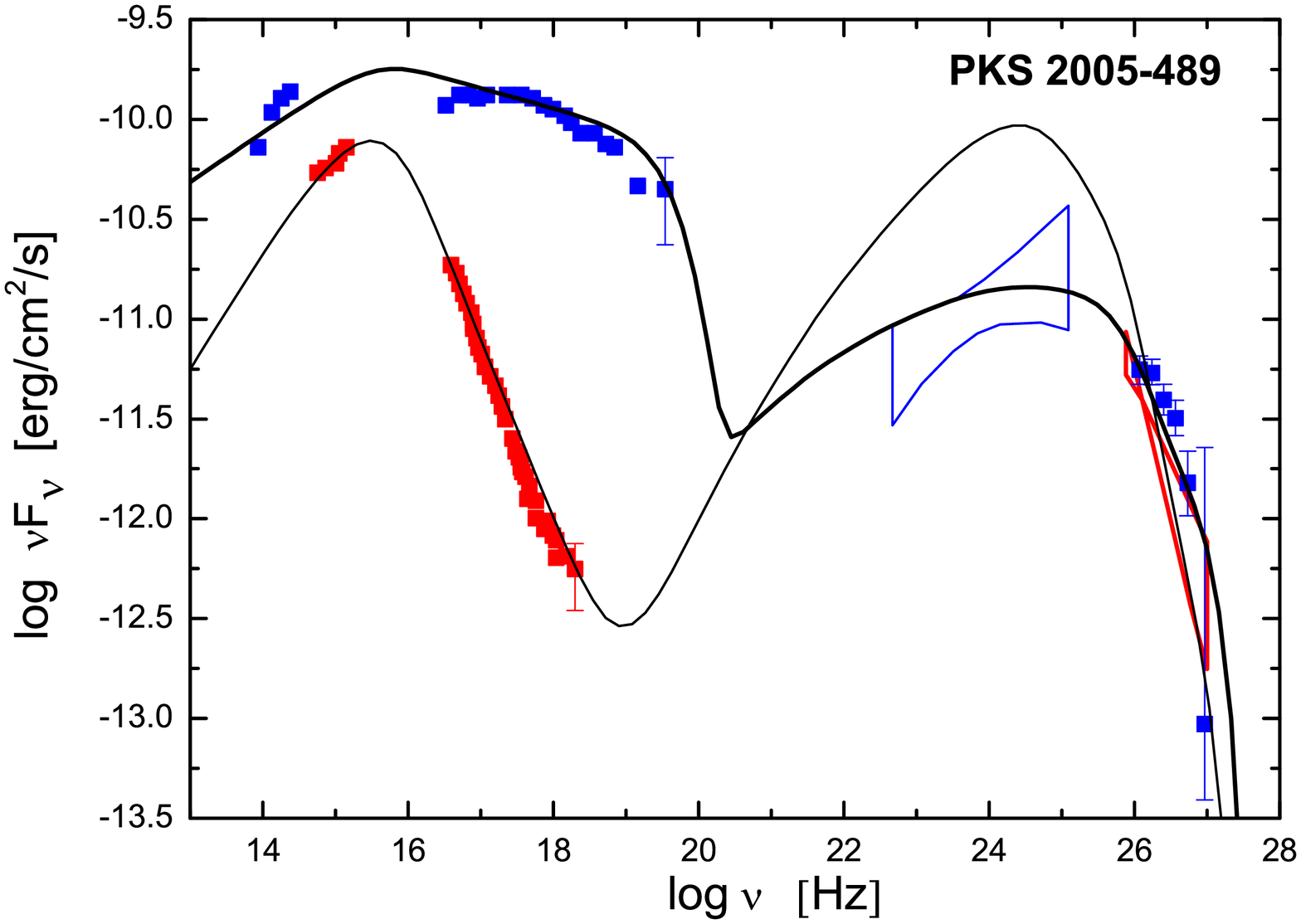}
\includegraphics[angle=0,scale=0.30]{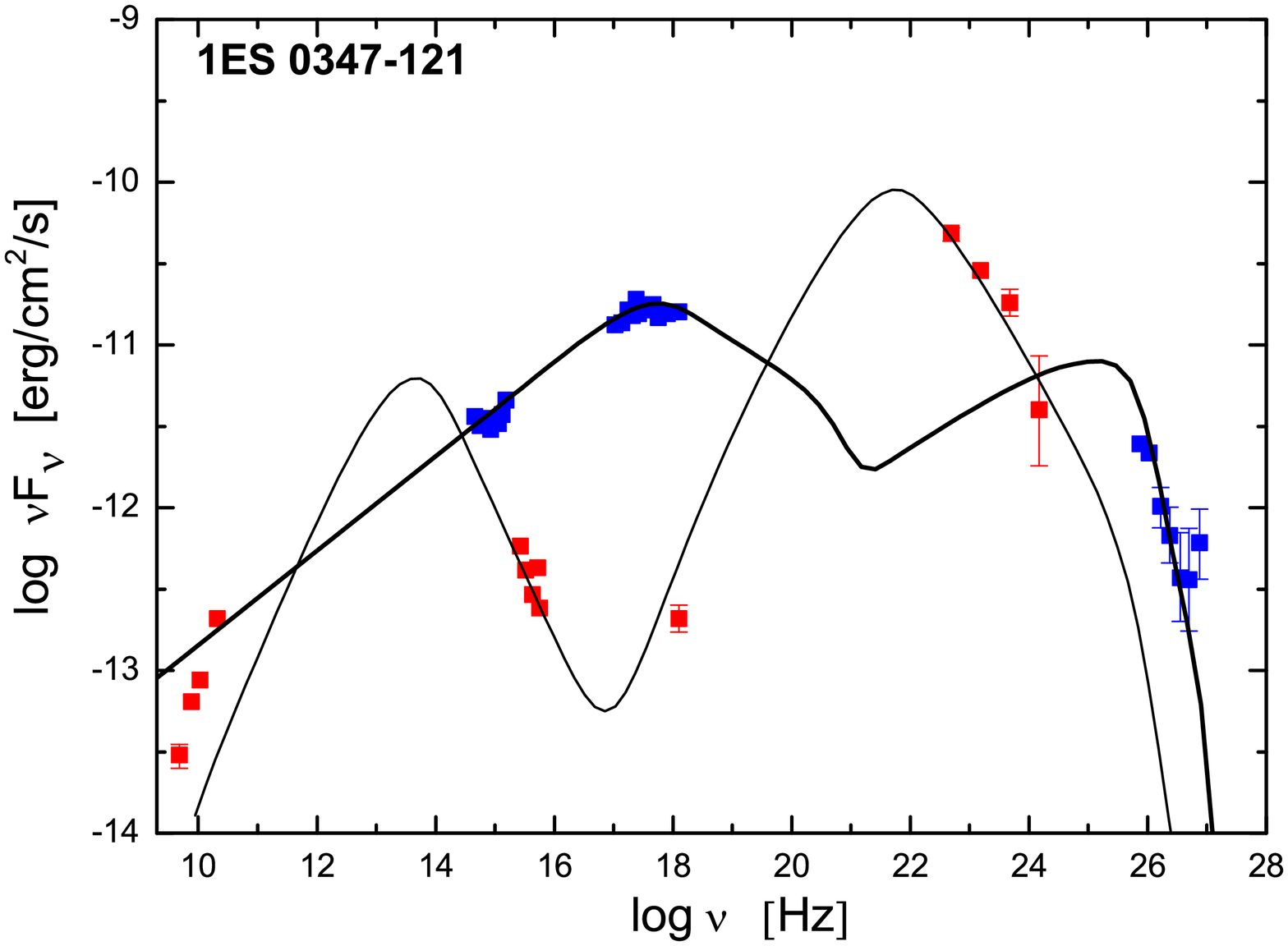}
\includegraphics[angle=0,scale=0.30]{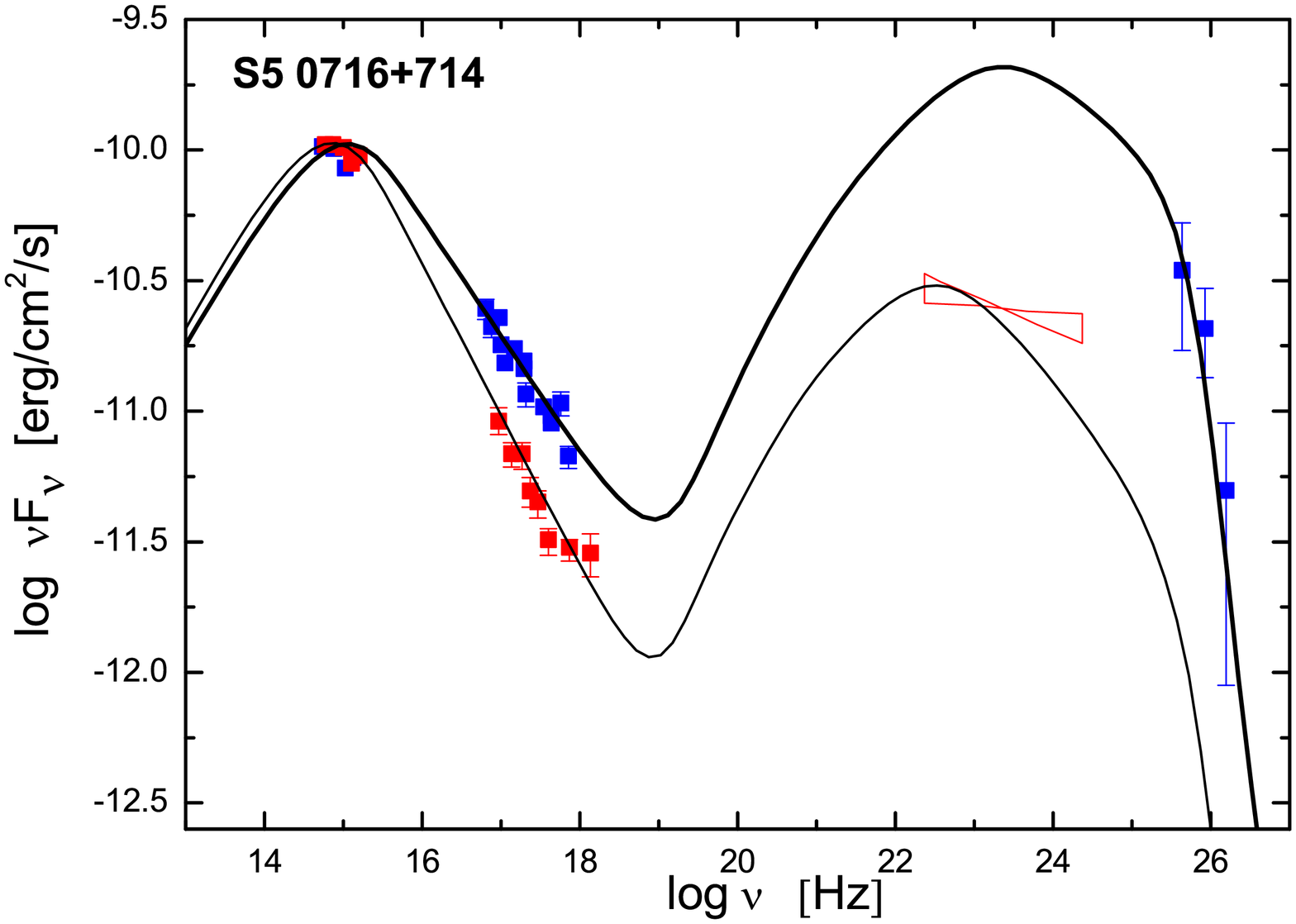}
\includegraphics[angle=0,scale=0.30]{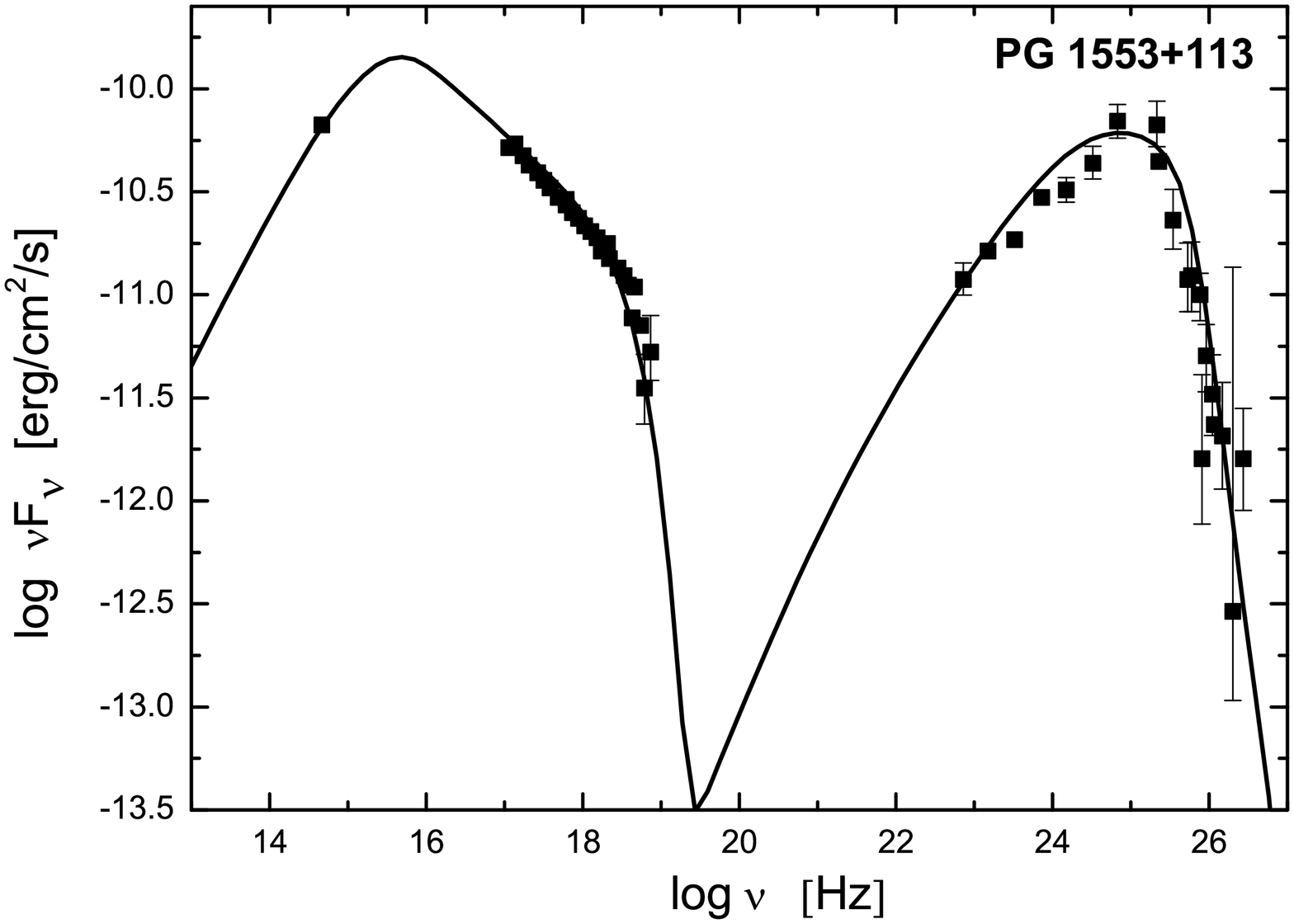}
\includegraphics[angle=0,scale=0.30]{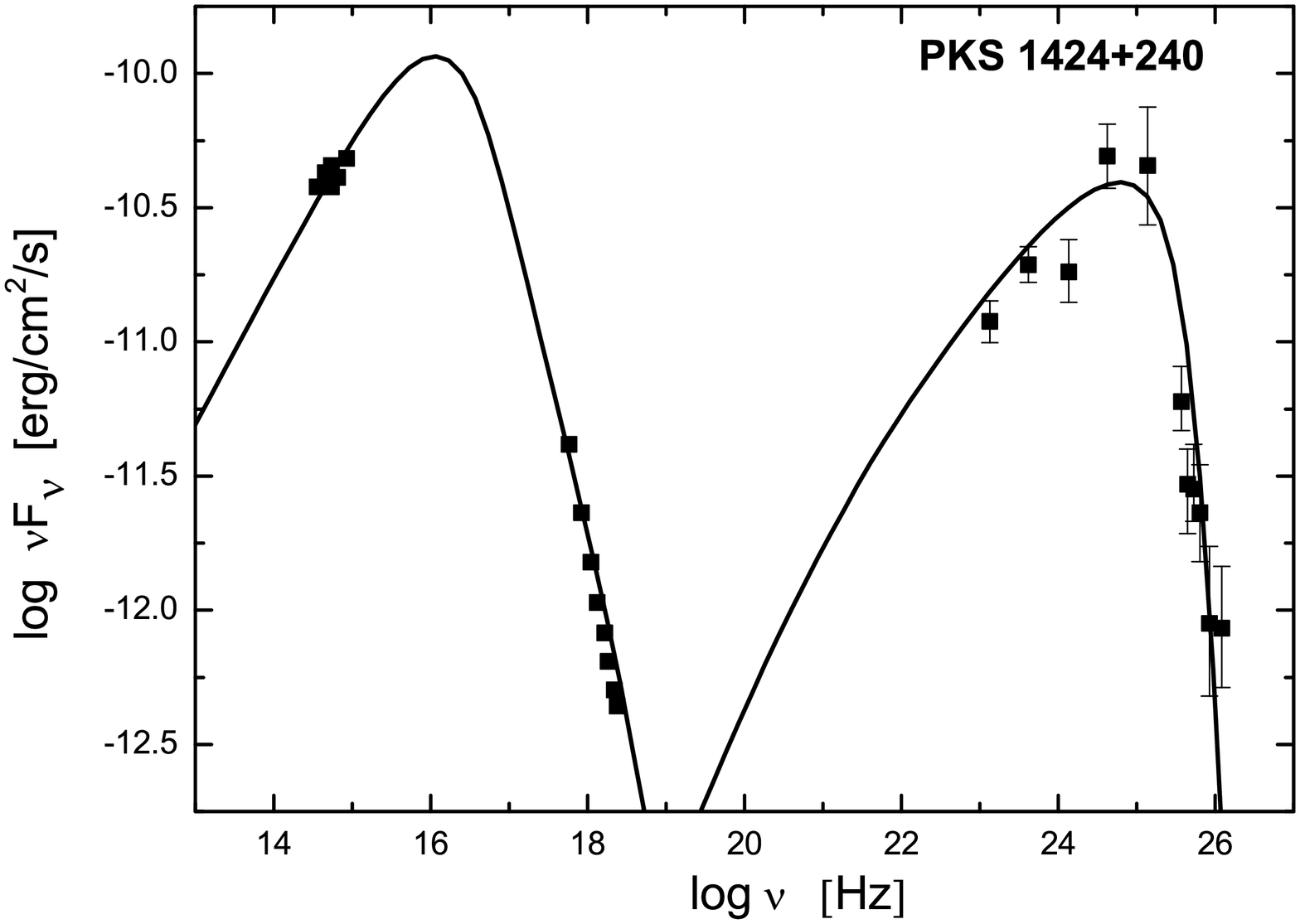}
\includegraphics[angle=0,scale=0.30]{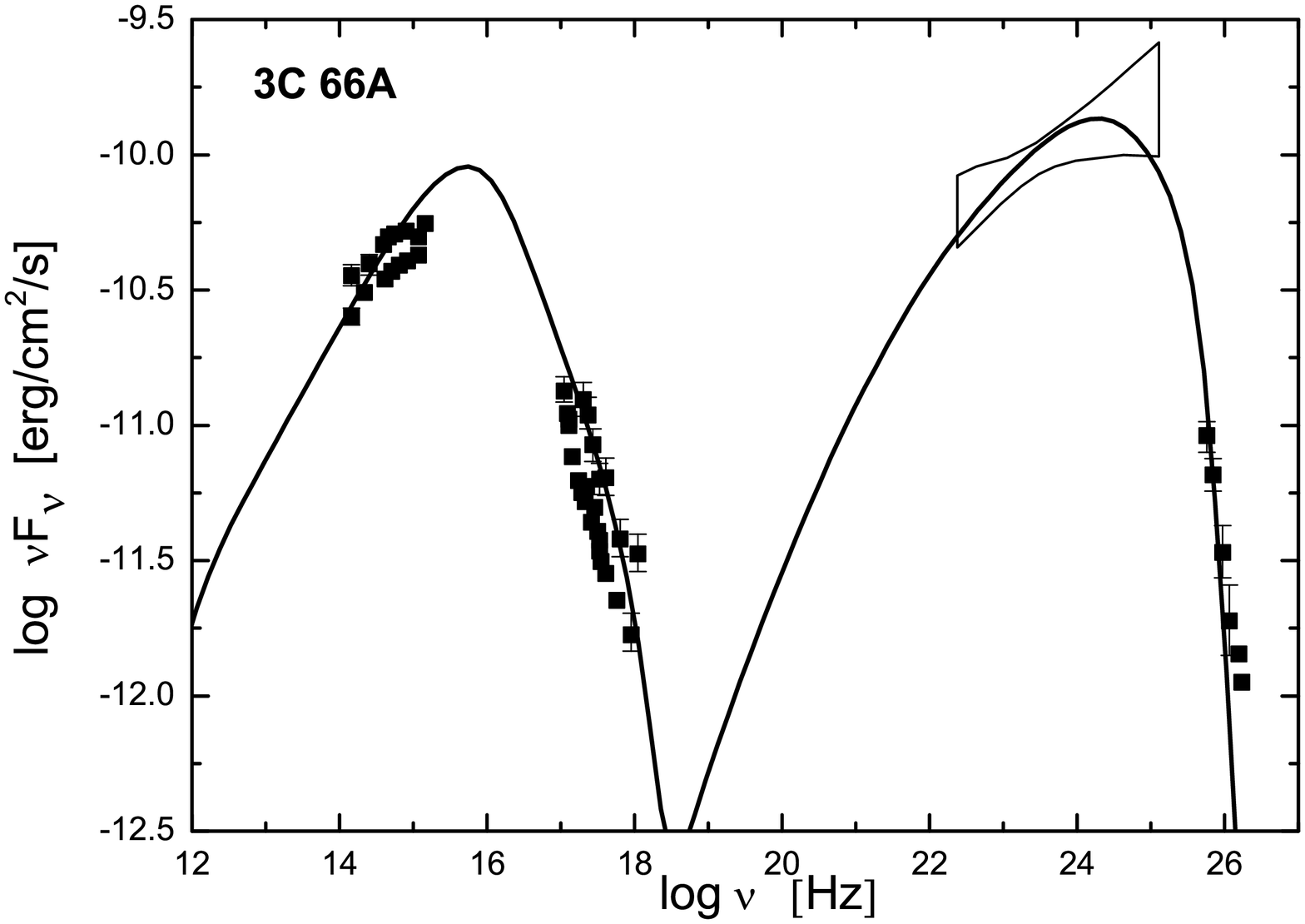}
\includegraphics[angle=0,scale=0.30]{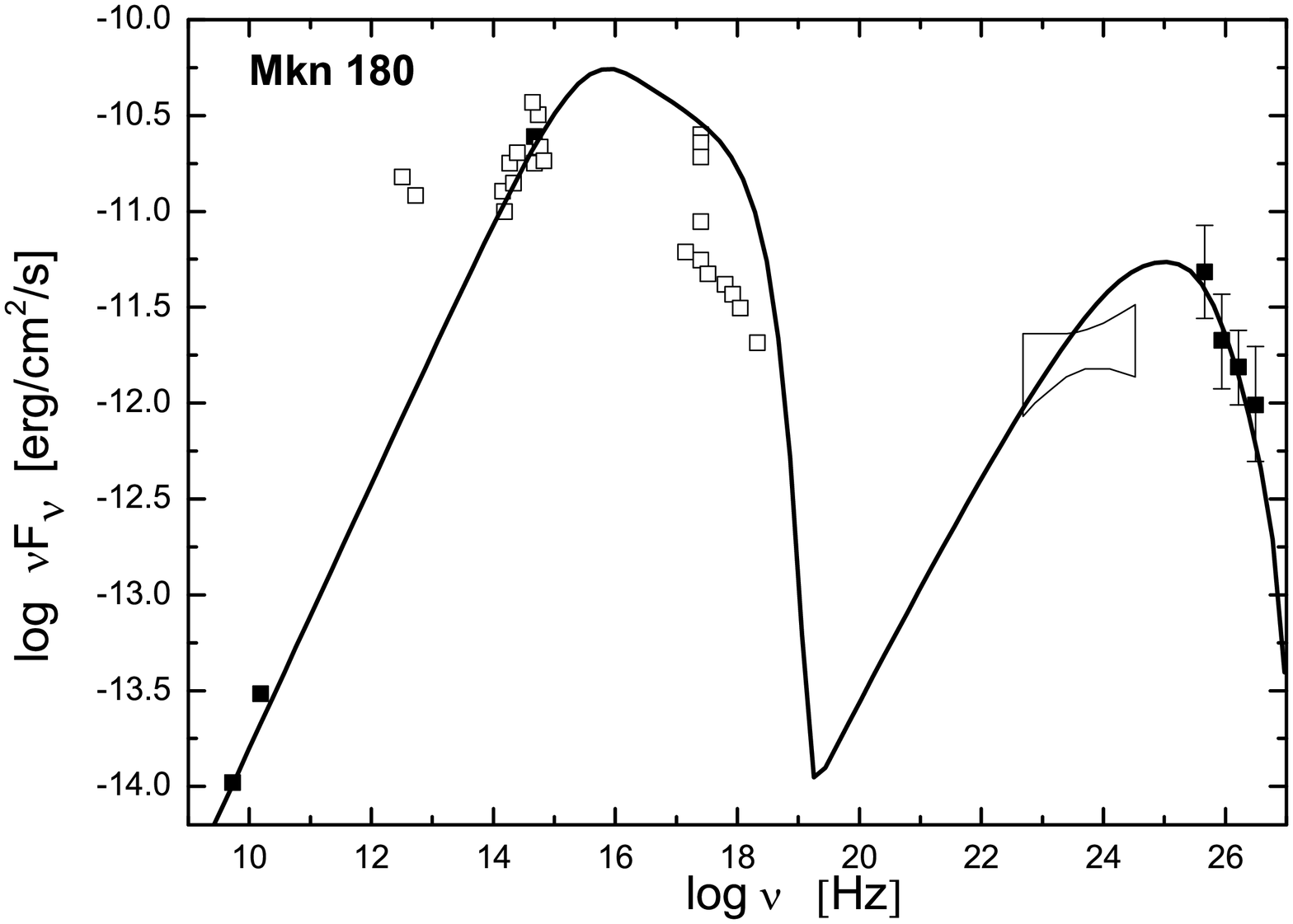}
\includegraphics[angle=0,scale=0.30]{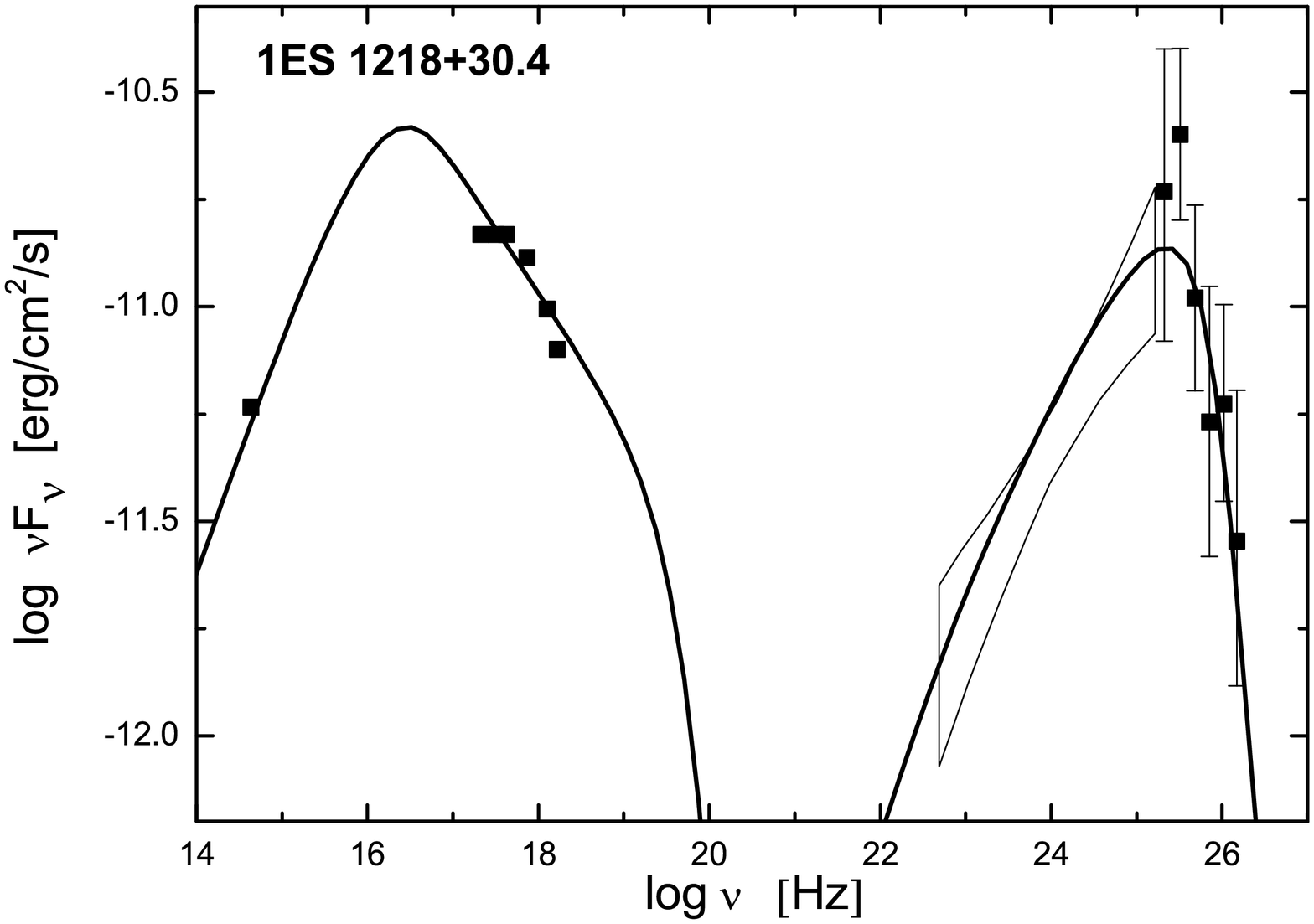}
\includegraphics[angle=0,scale=0.30]{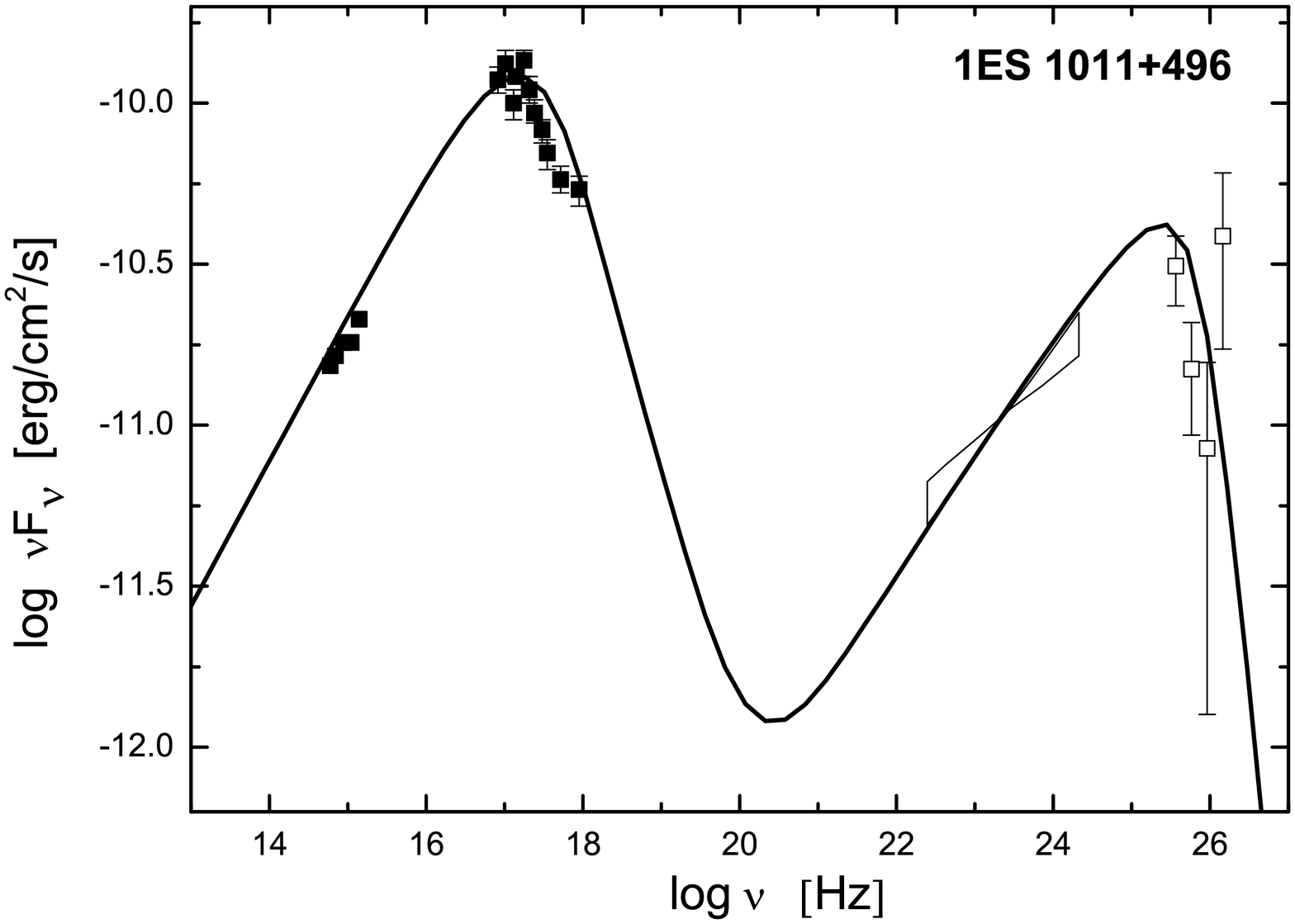}
\hfill
\includegraphics[angle=0,scale=0.30]{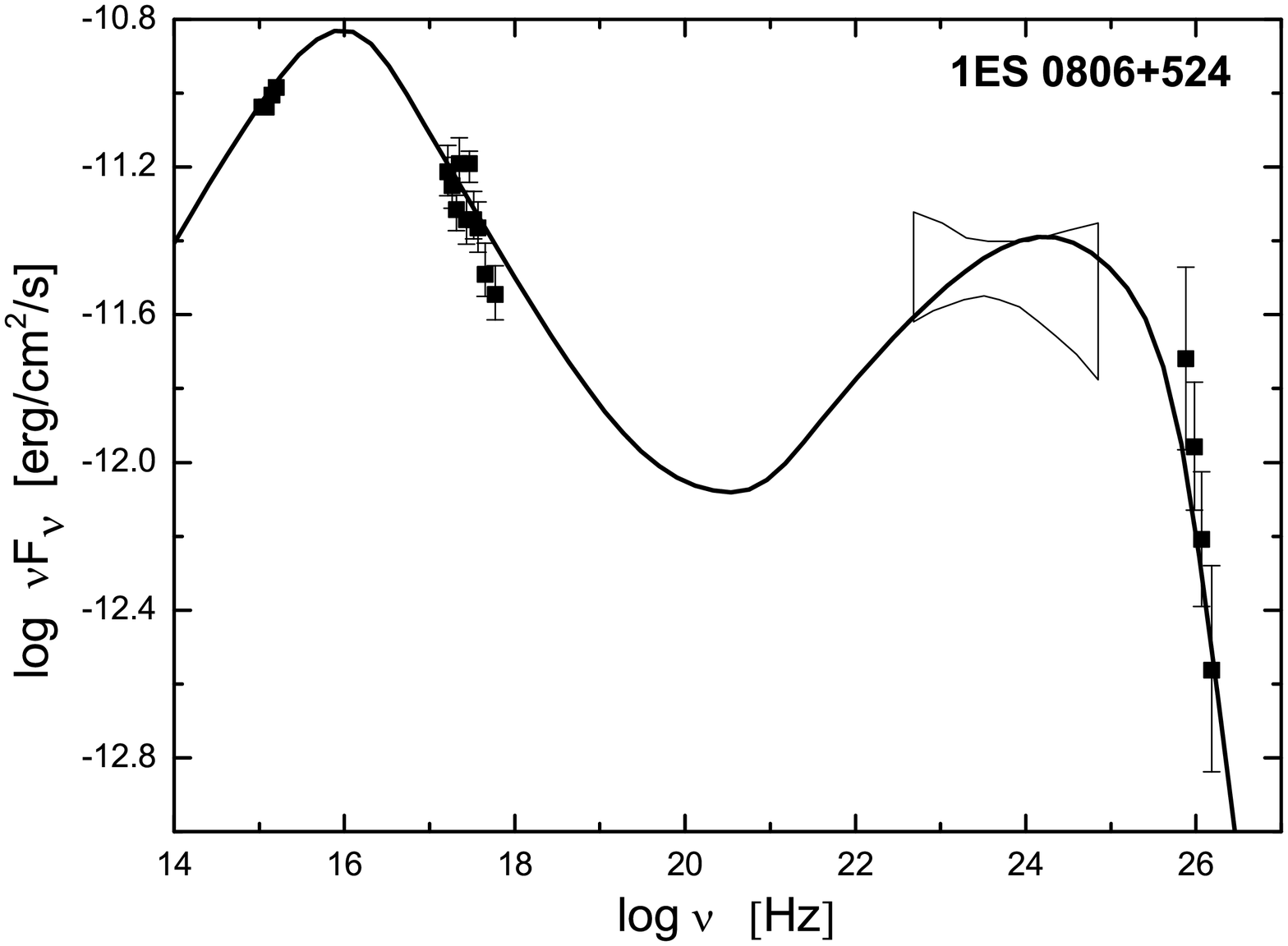}
\center{Fig. 1---  continued}
\end{figure*}

\begin{figure*}

\includegraphics[angle=0,scale=0.30]{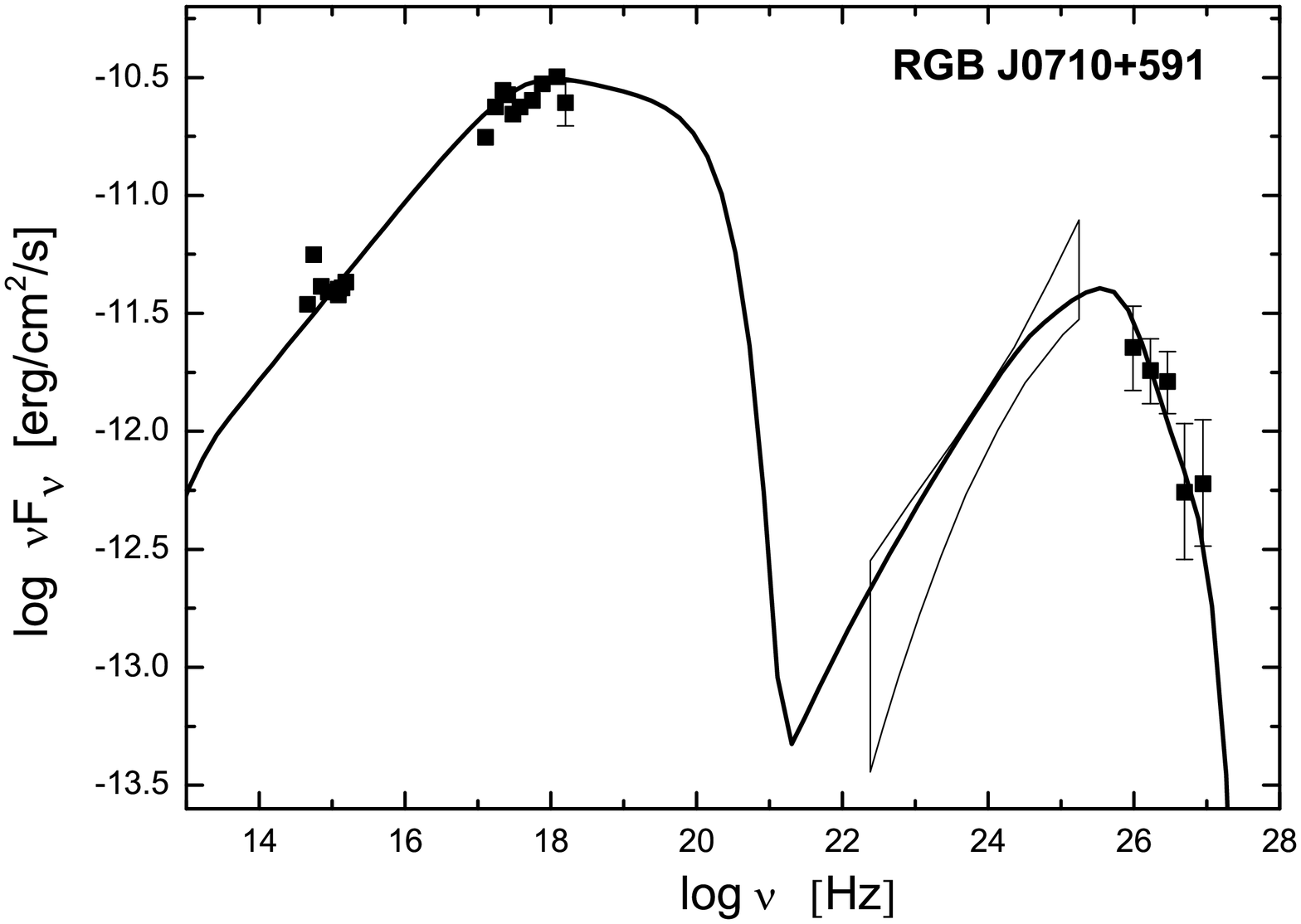}
\includegraphics[angle=0,scale=0.30]{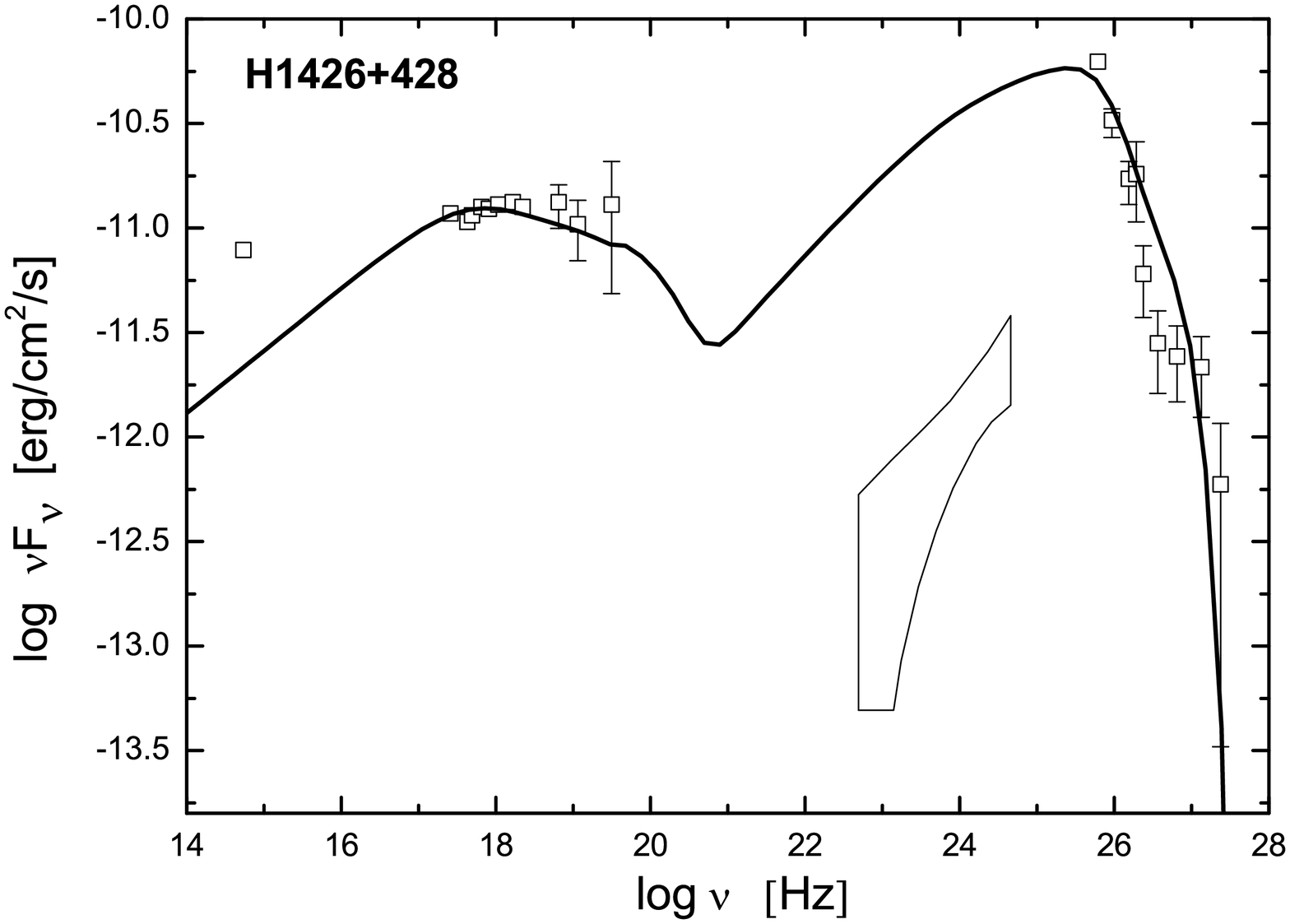}
\includegraphics[angle=0,scale=0.30]{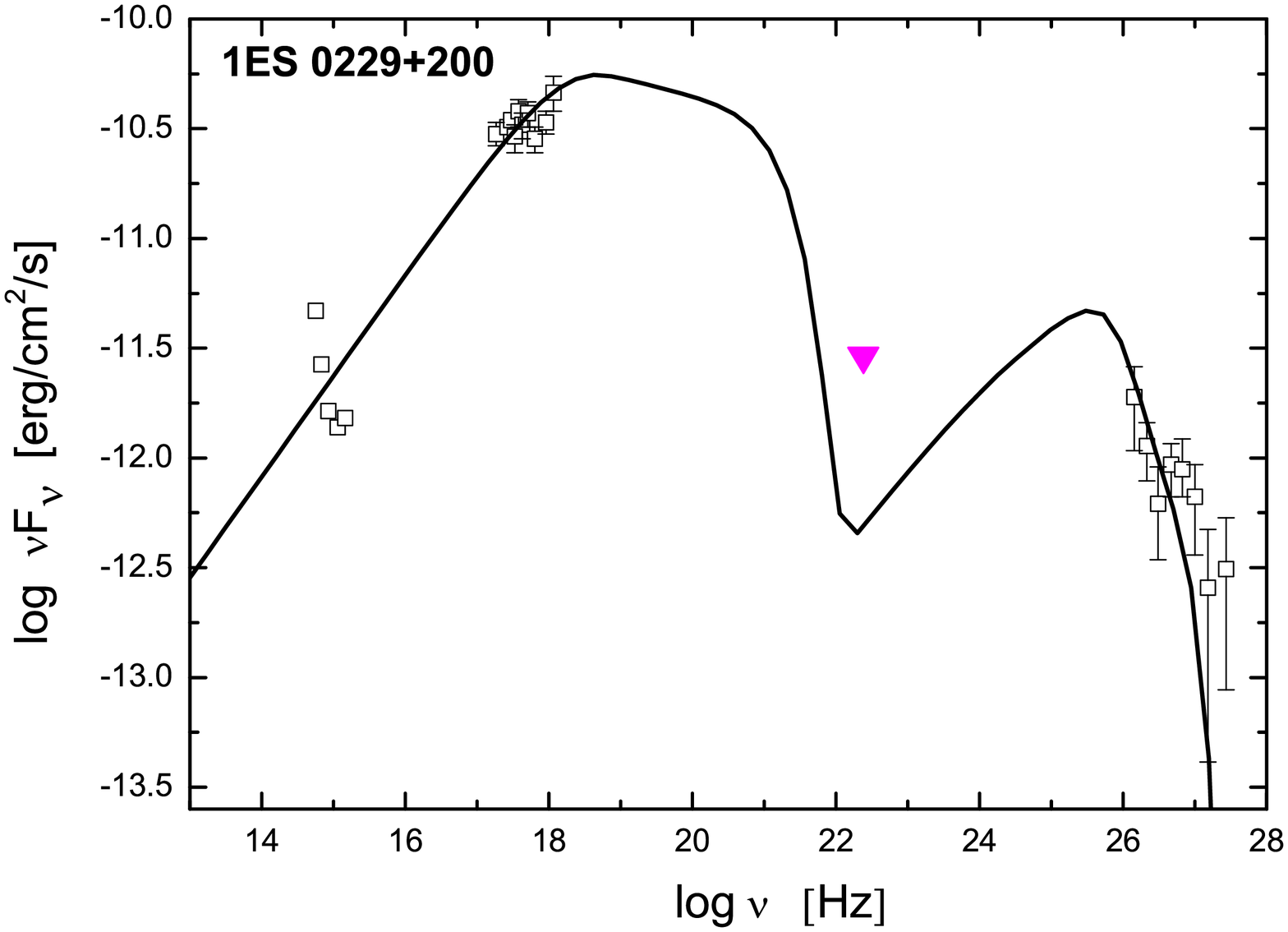}
\includegraphics[angle=0,scale=0.30]{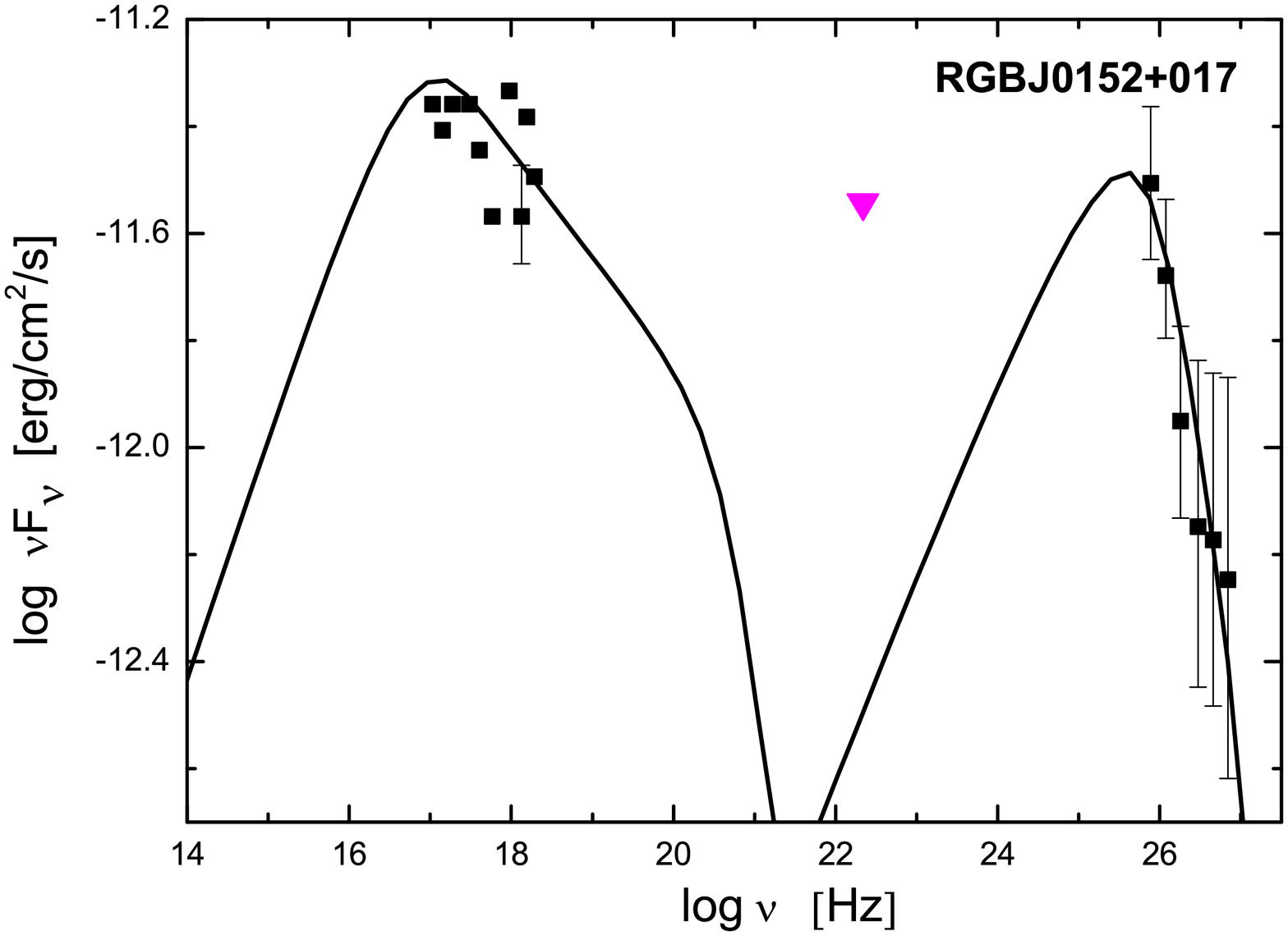}
\includegraphics[angle=0,scale=0.30]{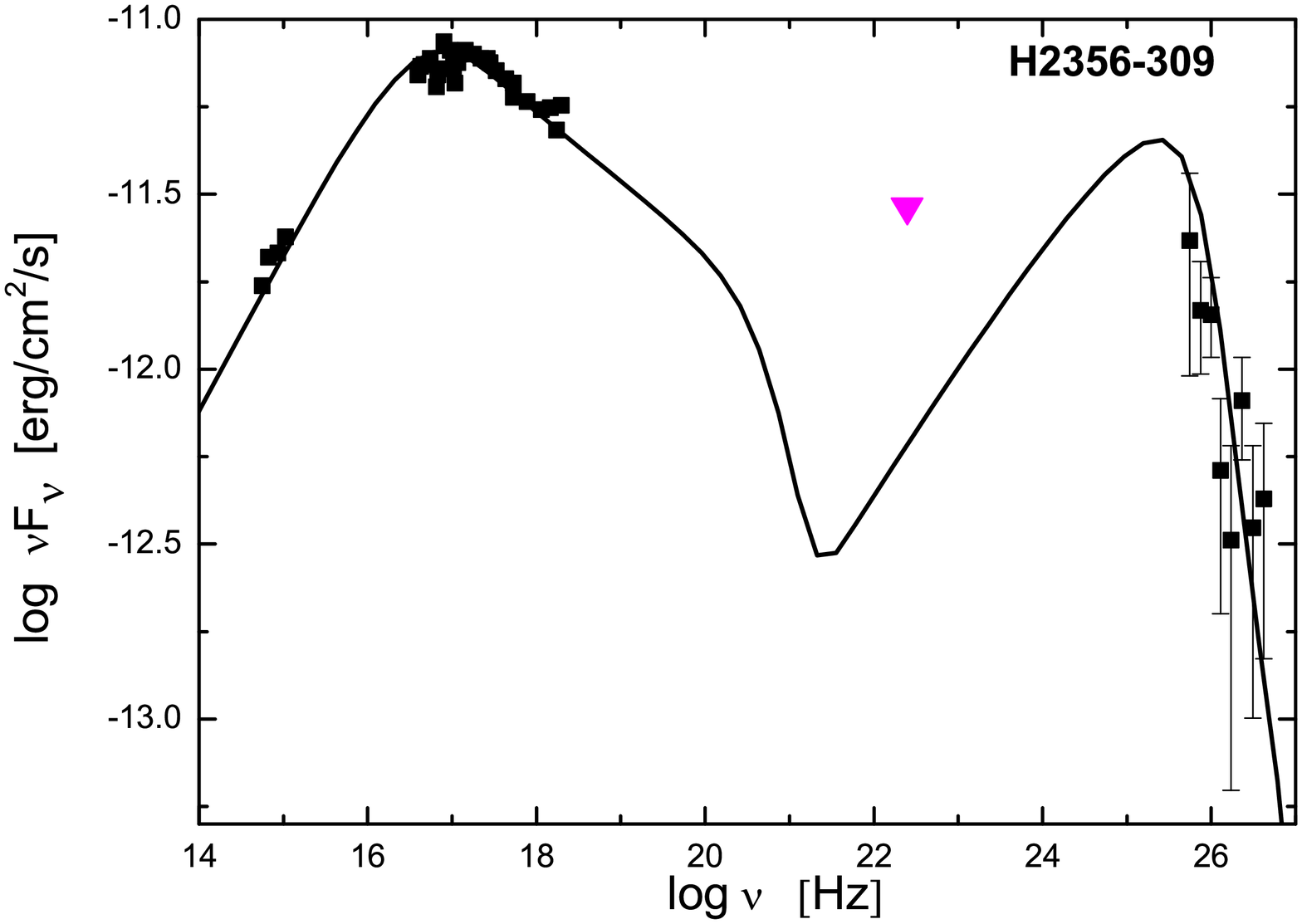}
\hfill
\includegraphics[angle=0,scale=0.30]{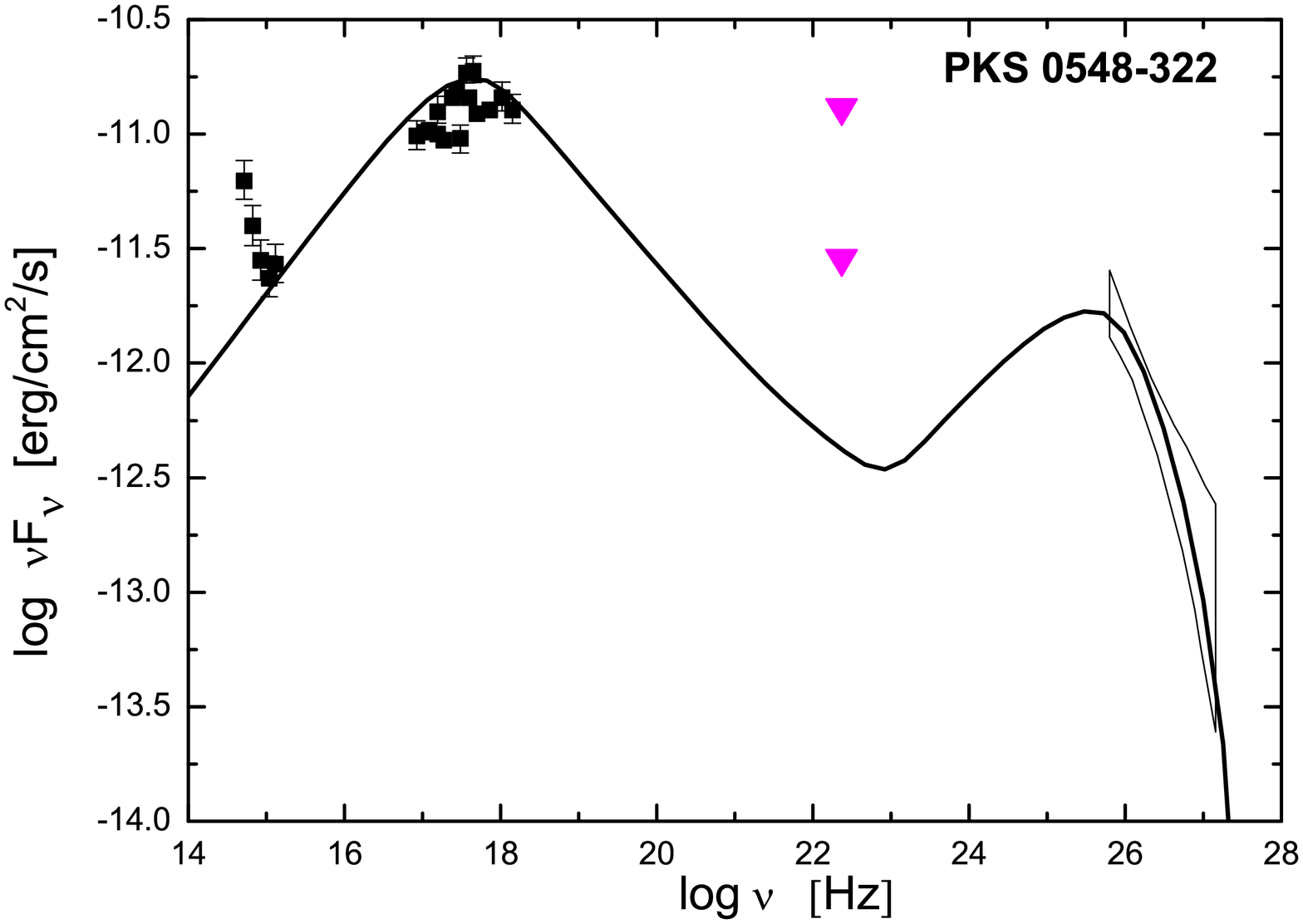}
\center{Fig. 1---  continued}
\end{figure*}

\begin{figure*}
\includegraphics[angle=0,scale=0.3]{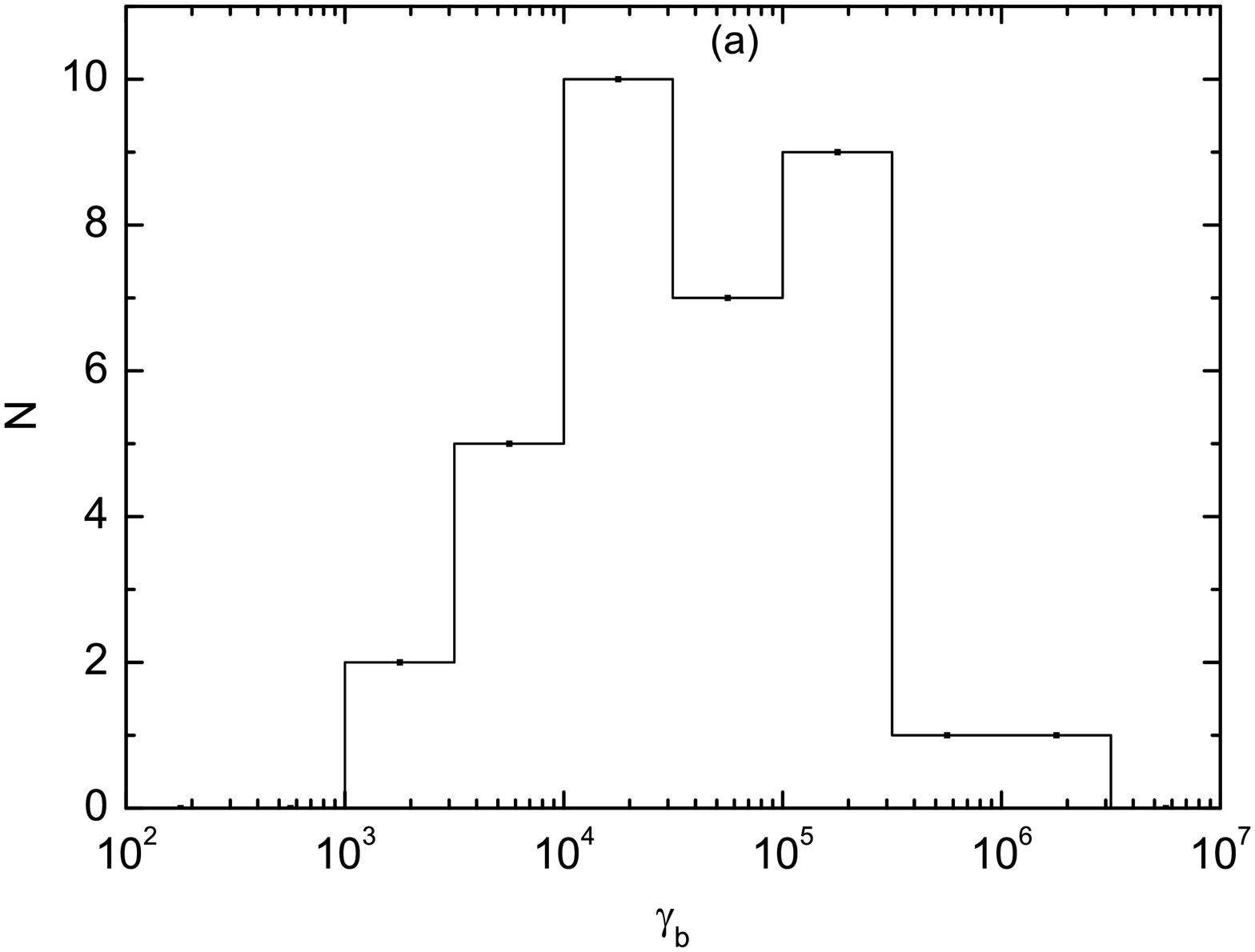}
\includegraphics[angle=0,scale=0.3]{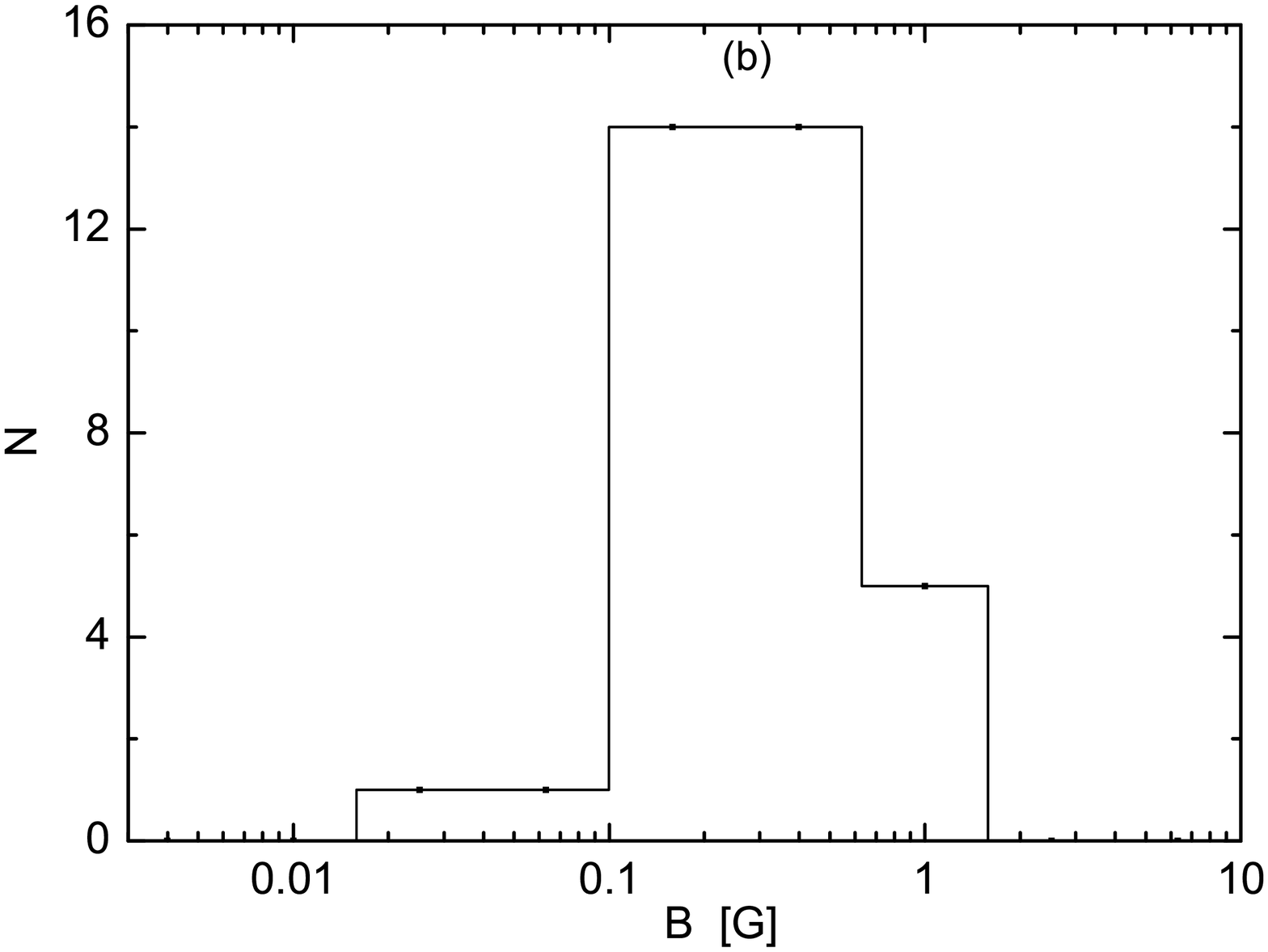}
\caption{Distributions of the break Lorenz factor $\gamma_{\rm b}$ for electrons ({\em Panel a}) and the
magnetic field strength ({\em Panel b}).}\label{Fig:2}
\end{figure*}

\begin{figure*}
\includegraphics[angle=0,scale=0.22]{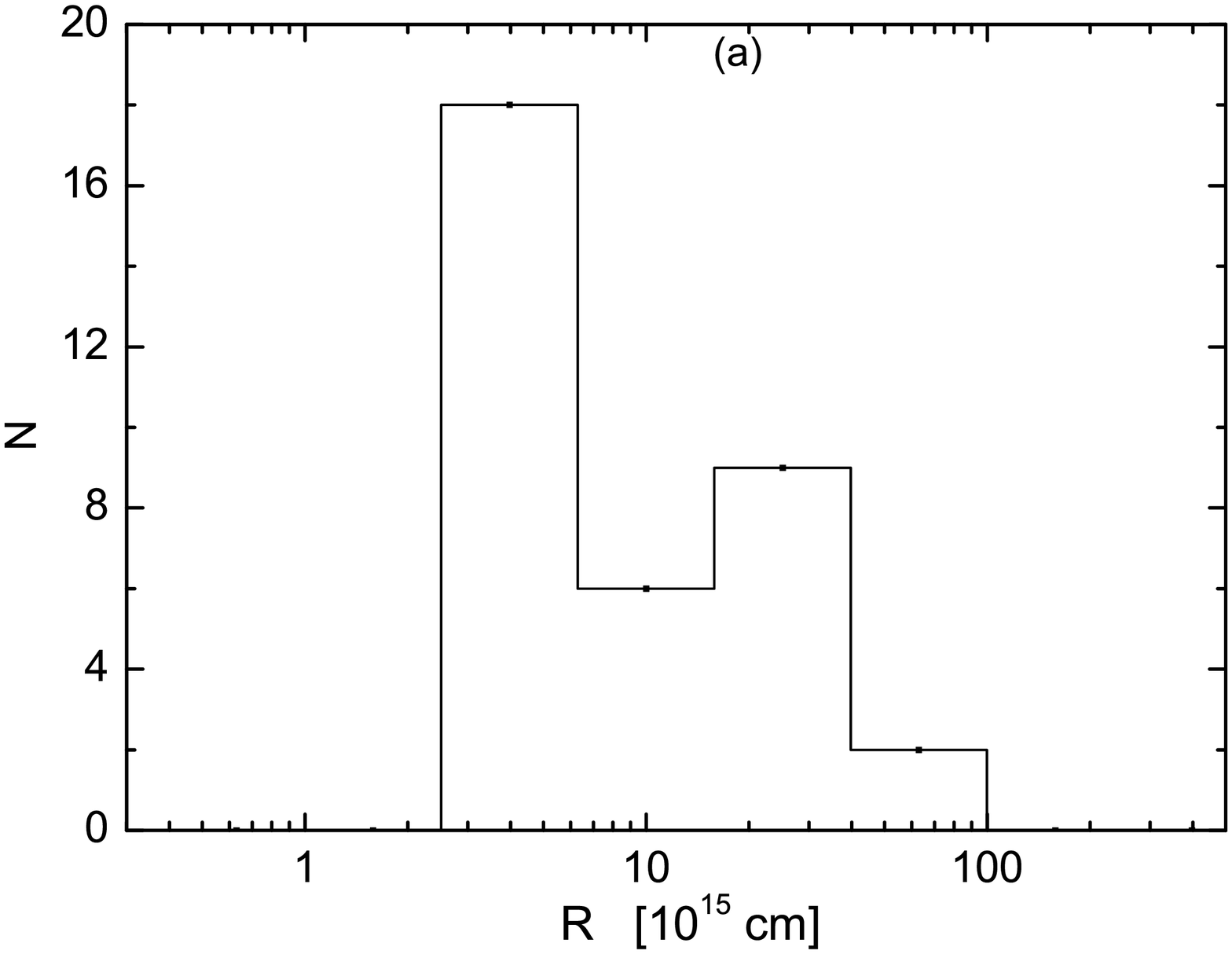}
\includegraphics[angle=0,scale=0.22]{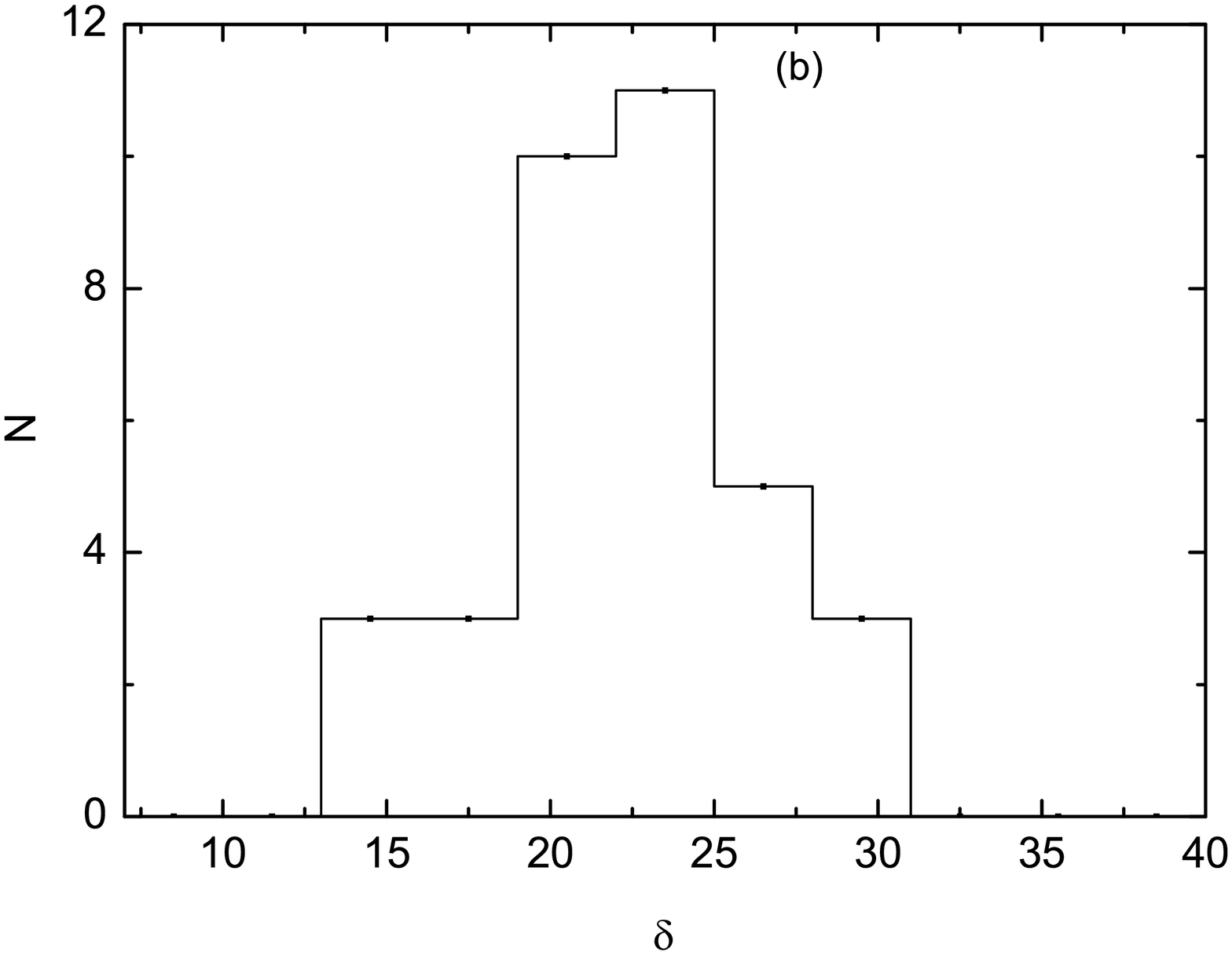}
\includegraphics[angle=0,scale=0.22]{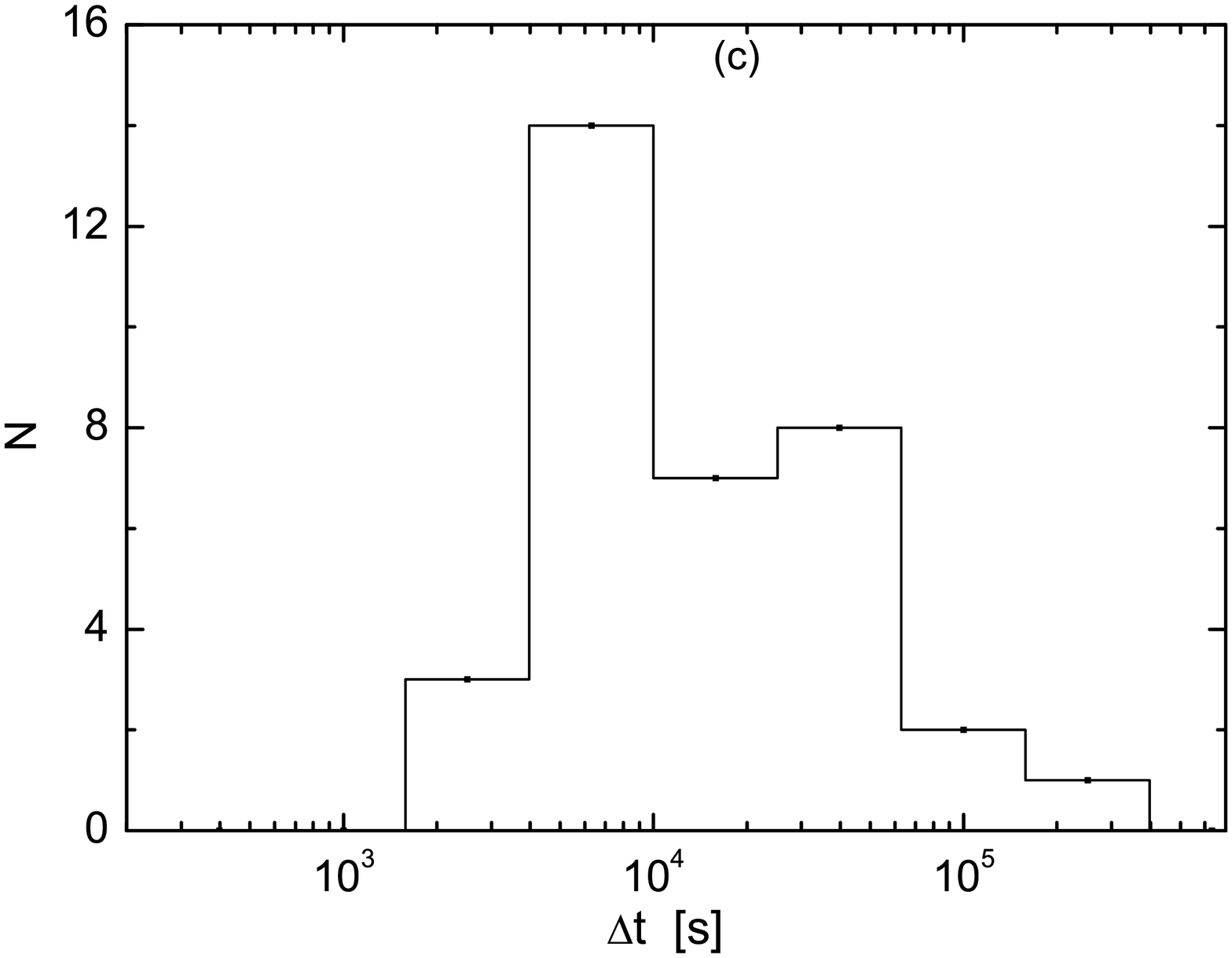}
\caption{Distributions of the size for the radiating region ({\em Panel a}), the beaming factor ({\em Panel b}),
and the minimum variation timescale ({\em Panel c}) for the sources in our sample.}\label{Fig:3}
\end{figure*}

\begin{figure*}
\includegraphics[width=3.in,height=2.5in]{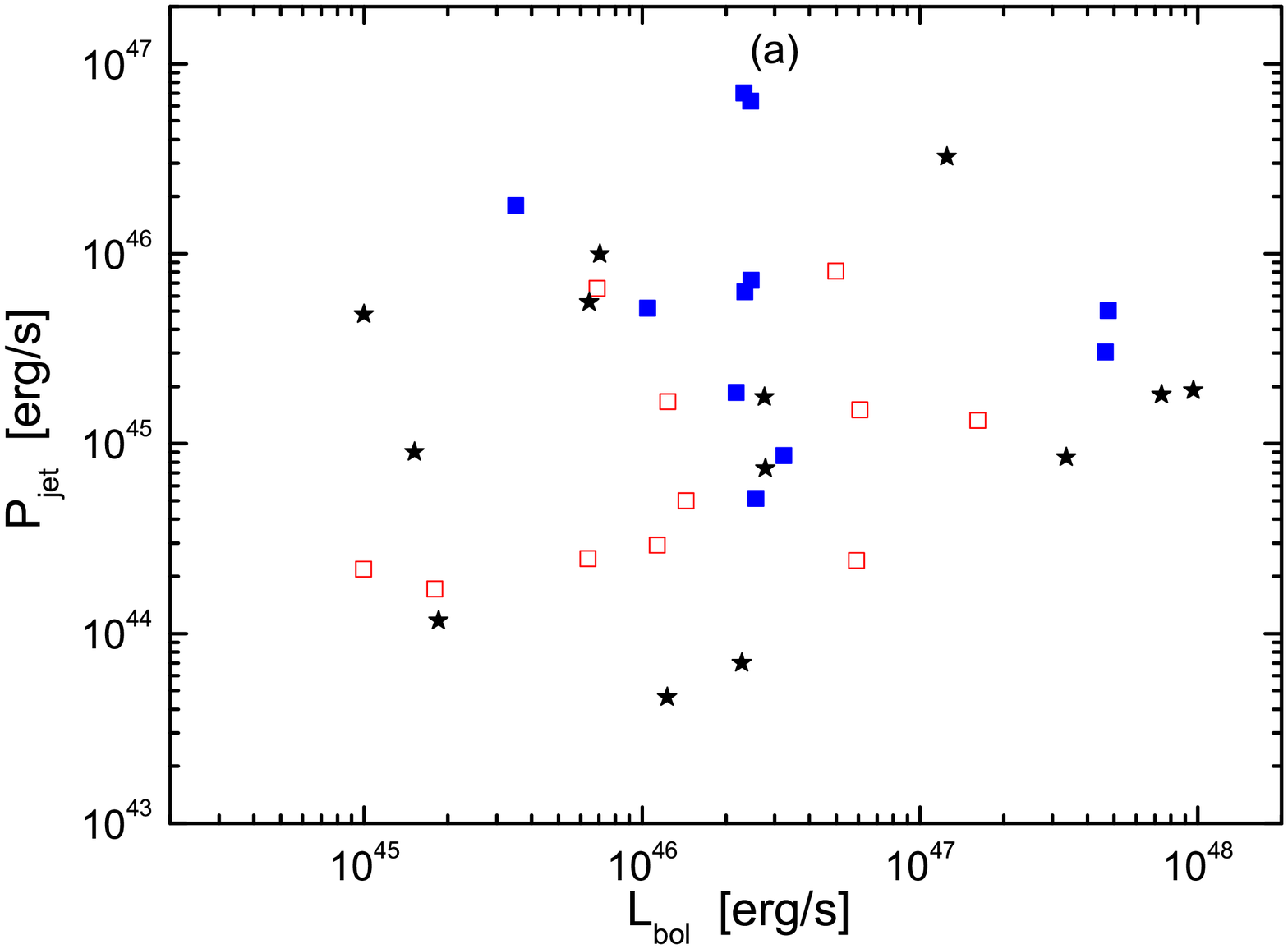}
\includegraphics[width=3.in,height=2.5in]{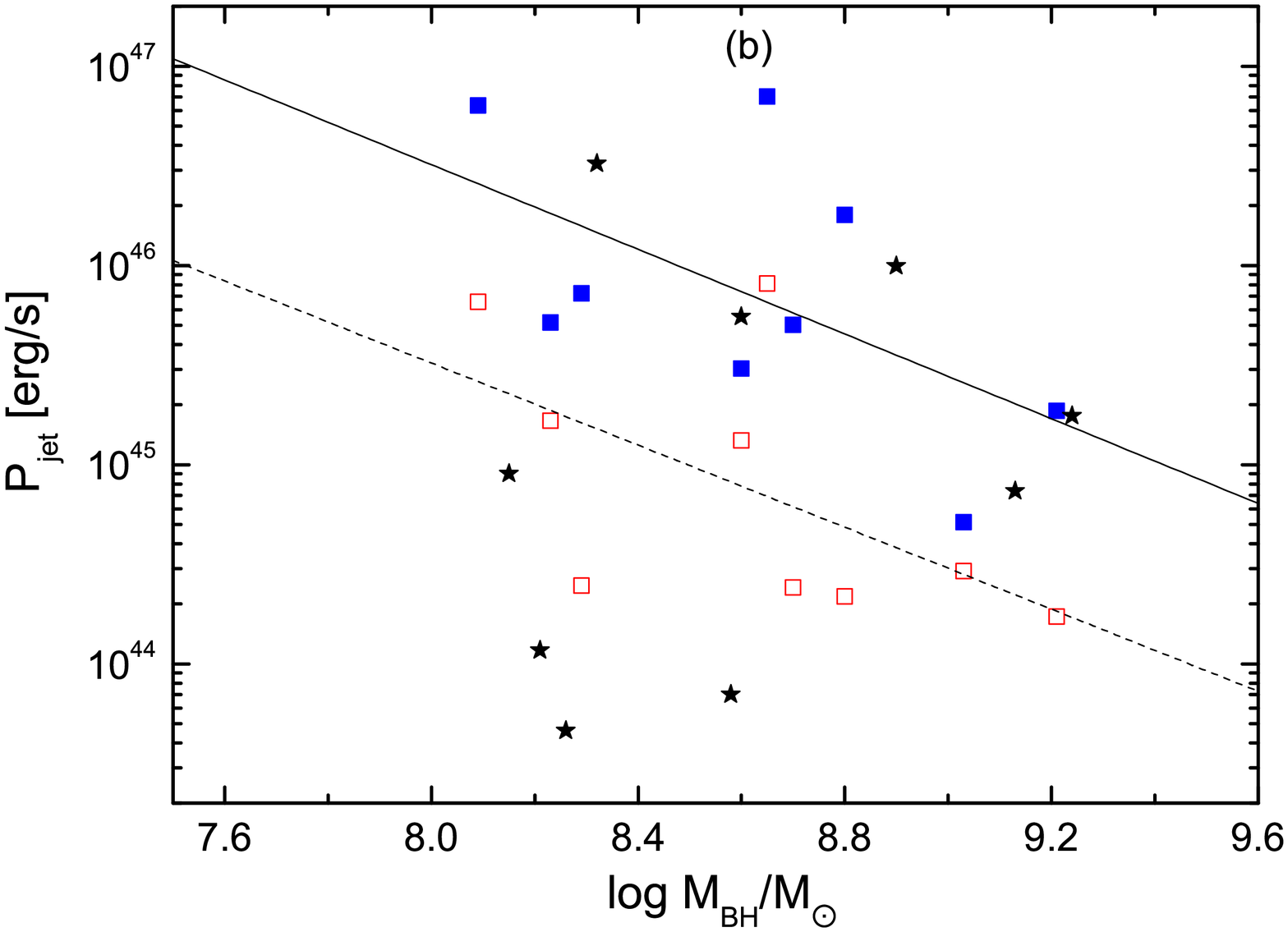}
\caption{{\em Panel a}---Correlation of the jet power with the bolometric luminosity. {\em Panel b}---Jet power
as a function of BH mass. The best fit lines are $\log P_{\rm jet}=(53.7\pm4.7)-(1.03\pm0.55)\log M_{\rm BH}$
for the low state data ({\em dashed line}) and $\log P_{\rm jet}=(55\pm5.1)-(1.06\pm0.59)\log M_{\rm BH}$ for
the high state data ({\em solid line}). {\em Blue and red} squares are for the sources in the high and low
states, respectively, {\em black} stars are for the sources with only one SED available.}\label{Fig:4}
\end{figure*}

\begin{figure*}
\includegraphics[angle=0,scale=0.24]{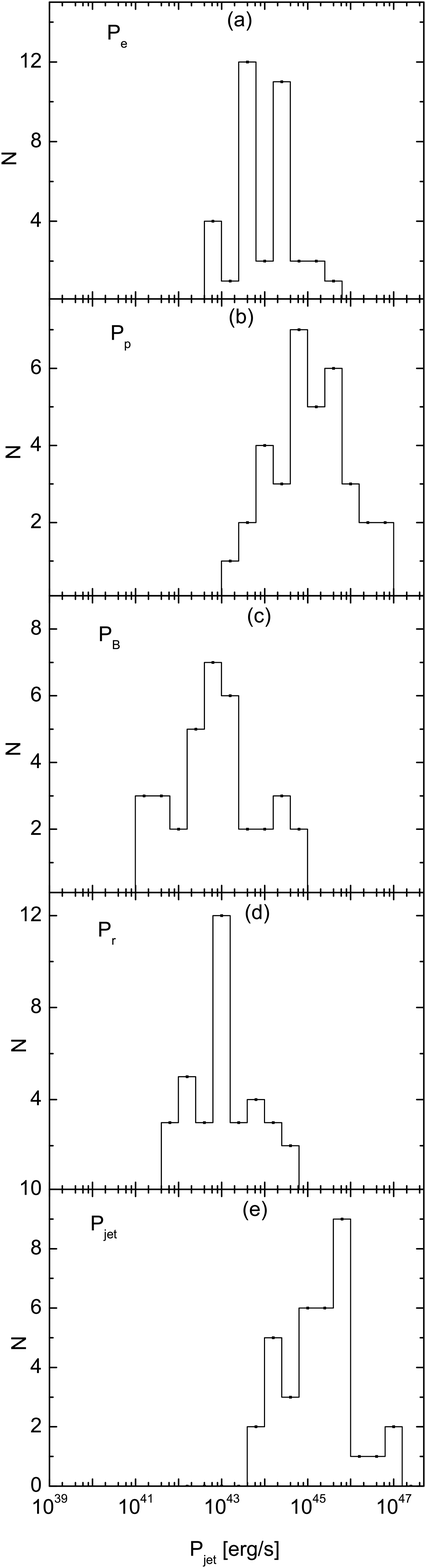}
\includegraphics[angle=0,scale=0.24]{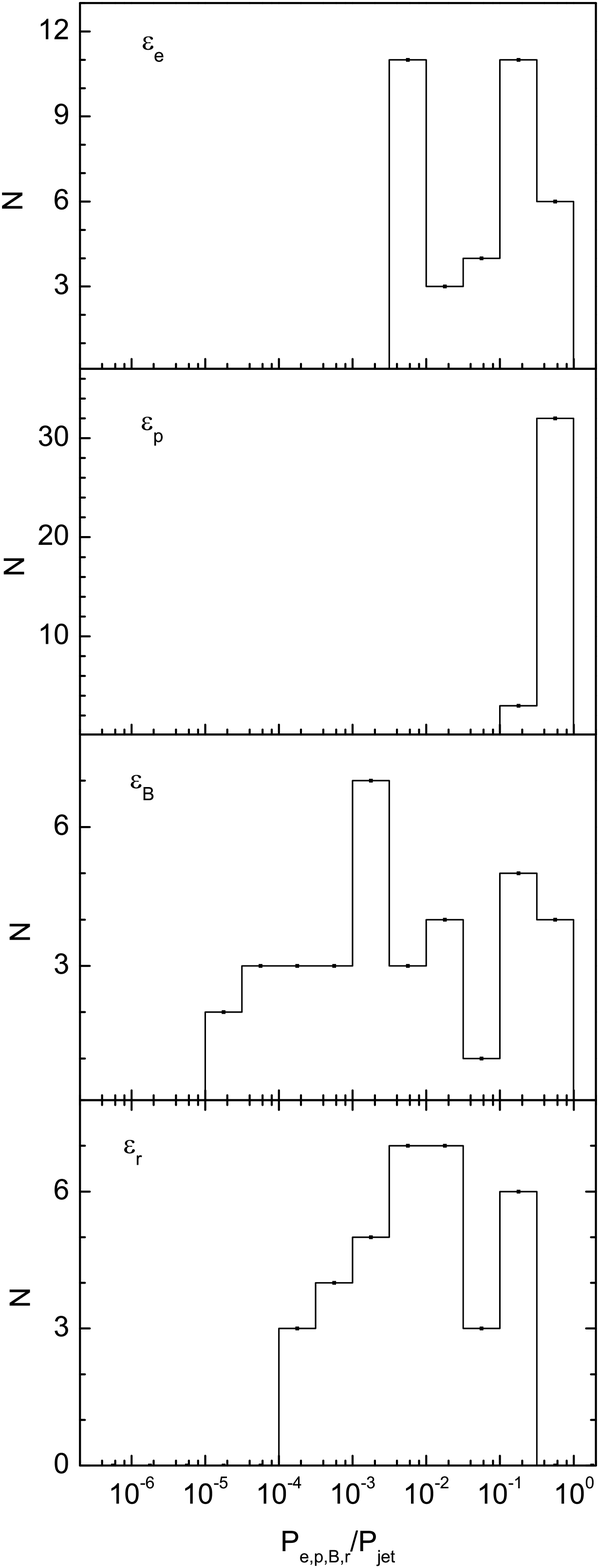}
\caption{Distributions of the powers associated with relativistic electrons $P_{\rm e}$, cold protons $P_{p}$,
Poynting flux $P_B$, radiation component $P_{\rm r}$, and the total power $P_{\rm jet}$ of the jets ({\em Panel
a}), together with the distributions of the ratios of these powers to the total jet power ({\em Panel
b}).}\label{Fig:5}
\end{figure*}

\begin{figure*}
\includegraphics[angle=0,scale=0.22]{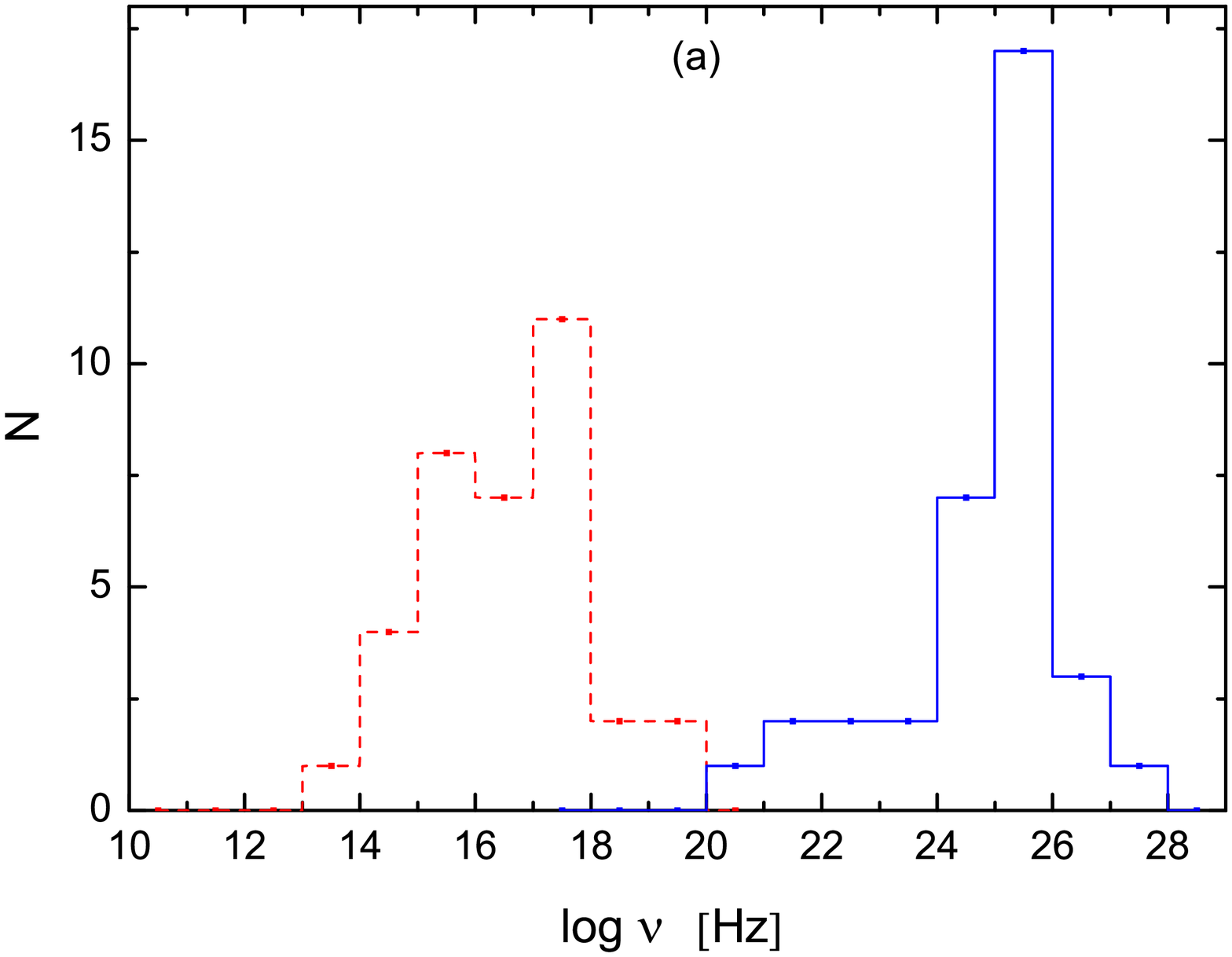}
\includegraphics[angle=0,scale=0.22]{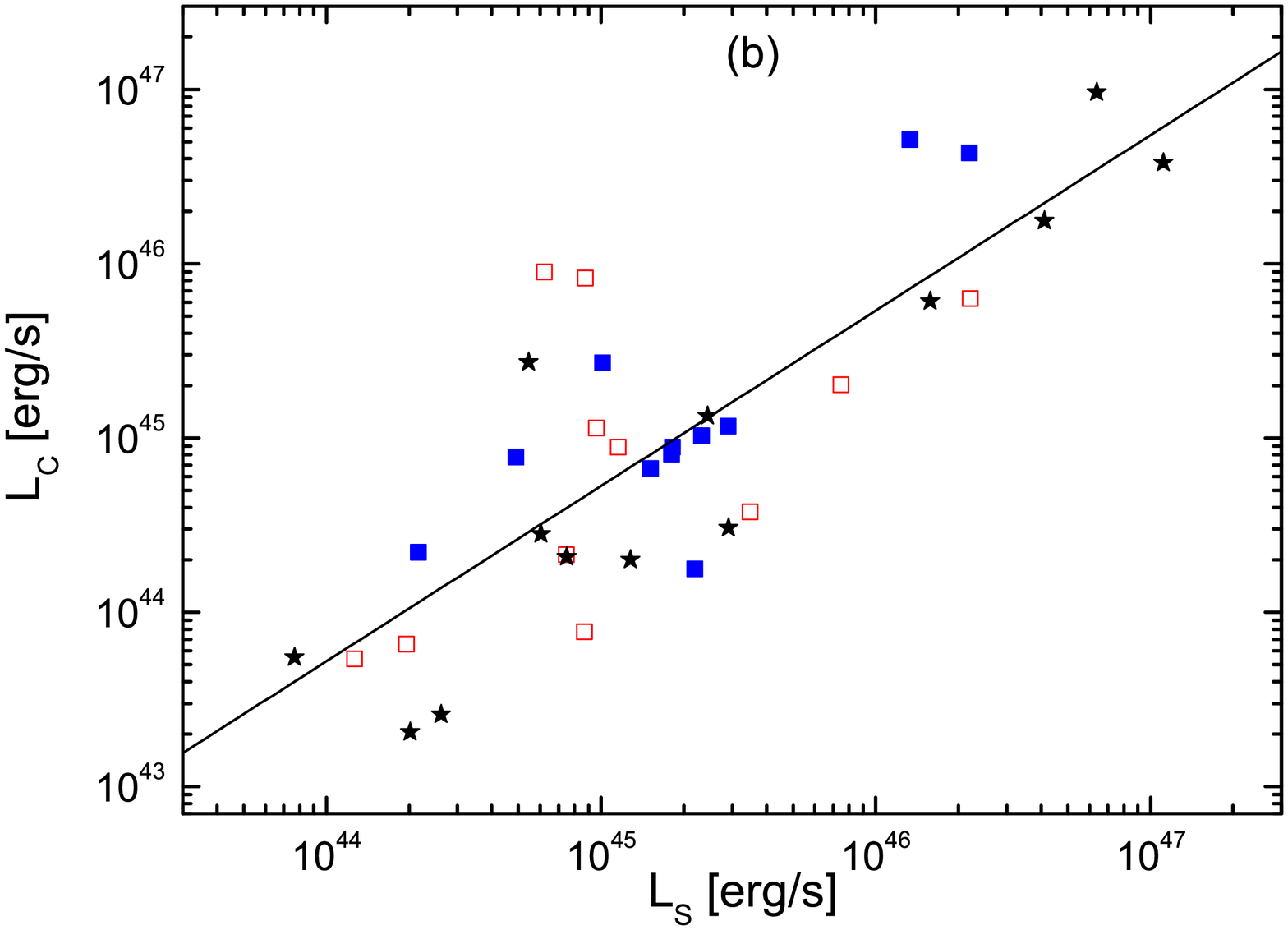}
\includegraphics[angle=0,scale=0.22]{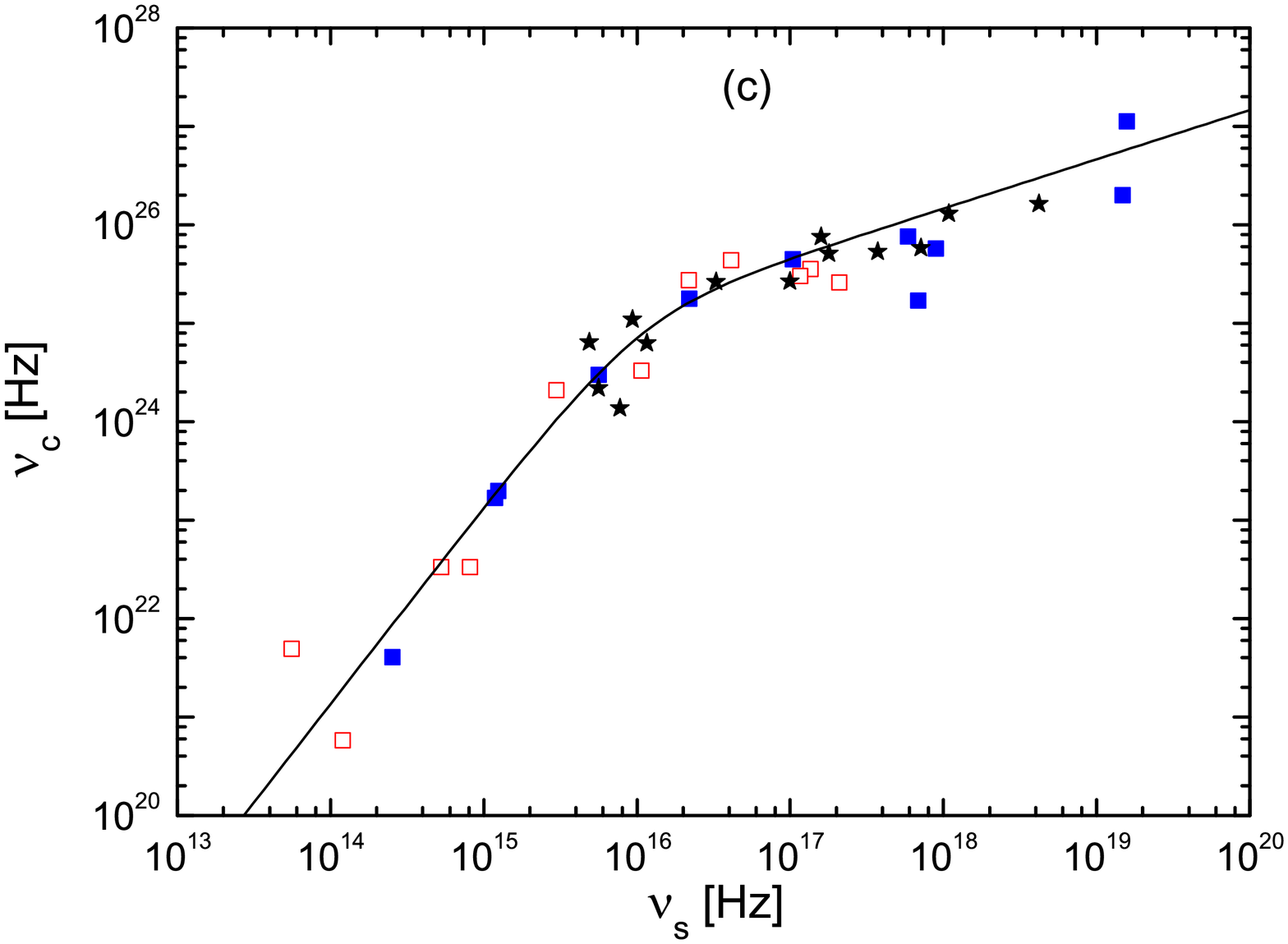}
\caption{{\em Panel a}--- Distributions of the synchrotron radiation peak frequency $\nu_{\rm s}$ ({\em red
dashed line}) and the inverse Compton scattering peak frequency $\nu_{\rm c}$ ({\em blue solid} line). {\em
Panel b}---Peak luminosity of the synchrotron radiation component as a function of the peak luminosity of the
SSC component. The symbol styles are the same as in Figure 4. The {\em solid line} is the best linear fit to all
data points, which is $\log L_{\rm c}=(0.55\pm5.75)+(1\pm0.13)\log L_{\rm s}$. {\em Panel c}---Correlation between
$\nu_{\rm s}$ and $\nu_{\rm c}$. The symbol styles are the same as in Figure 4. The {\em solid line} is the fit
to all the data points with a smoothly broken power-law. The slopes before and after the break ($\nu_{\rm
s}=1.05\times 10^{16}$) Hz are $s_1=2$ and $s_2=0.5$.} \label{Fig:6}
\end{figure*}

\begin{figure*}
\includegraphics[width=3.in,height=2.5in]{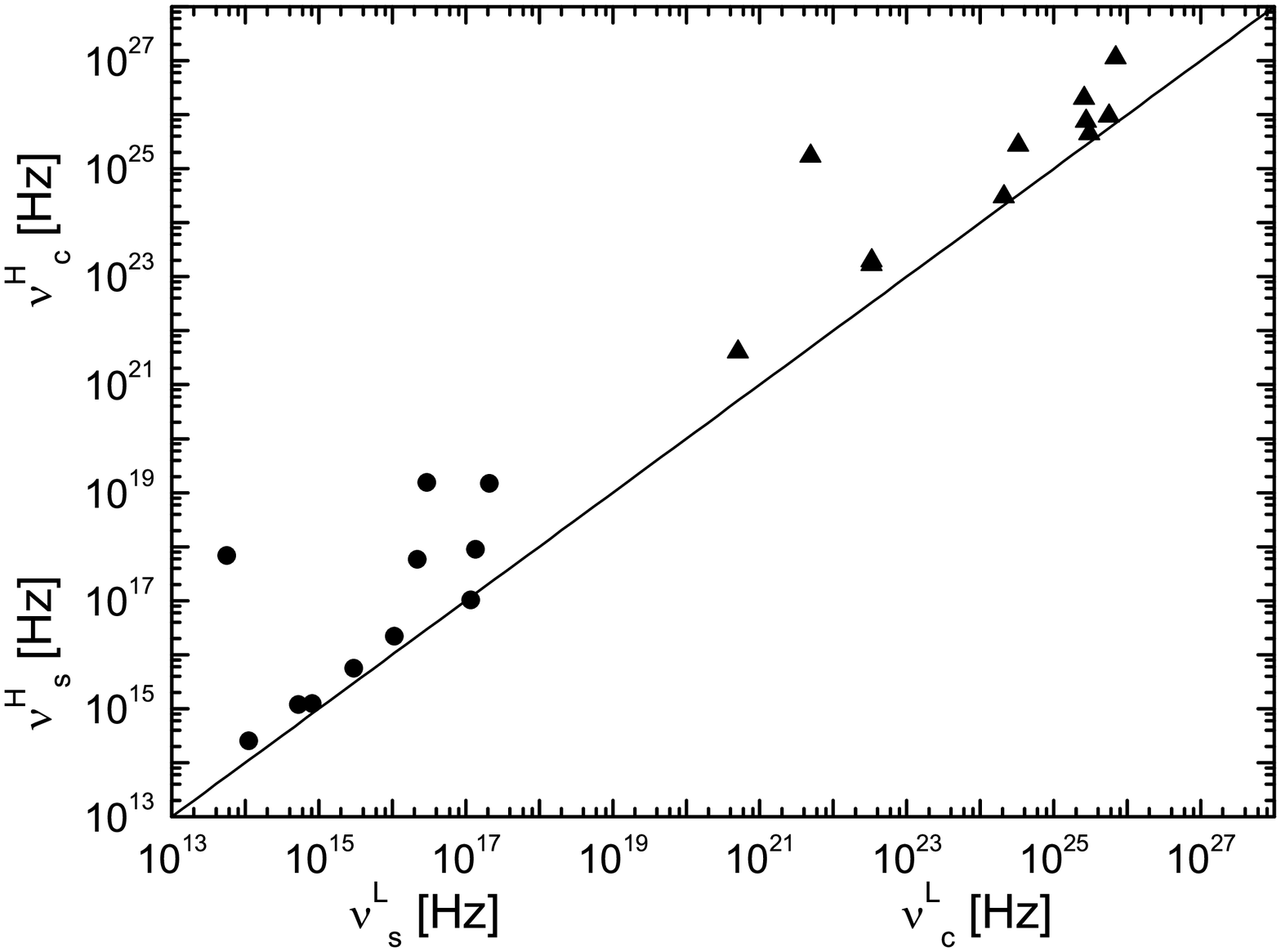}
\caption{{\em Panel a}---Comparison of the peak frequencies $\nu_{\rm s}$ (and $\nu_{\rm c}$) between the high
and low states. Circles are for $\nu_{\rm s}$, and triangles for $\nu_{\rm c}$. The line is the equality
line.}\label{Fig:7}
\end{figure*}

\begin{figure*}
\includegraphics[width=2.5 in,height=8.5 in]{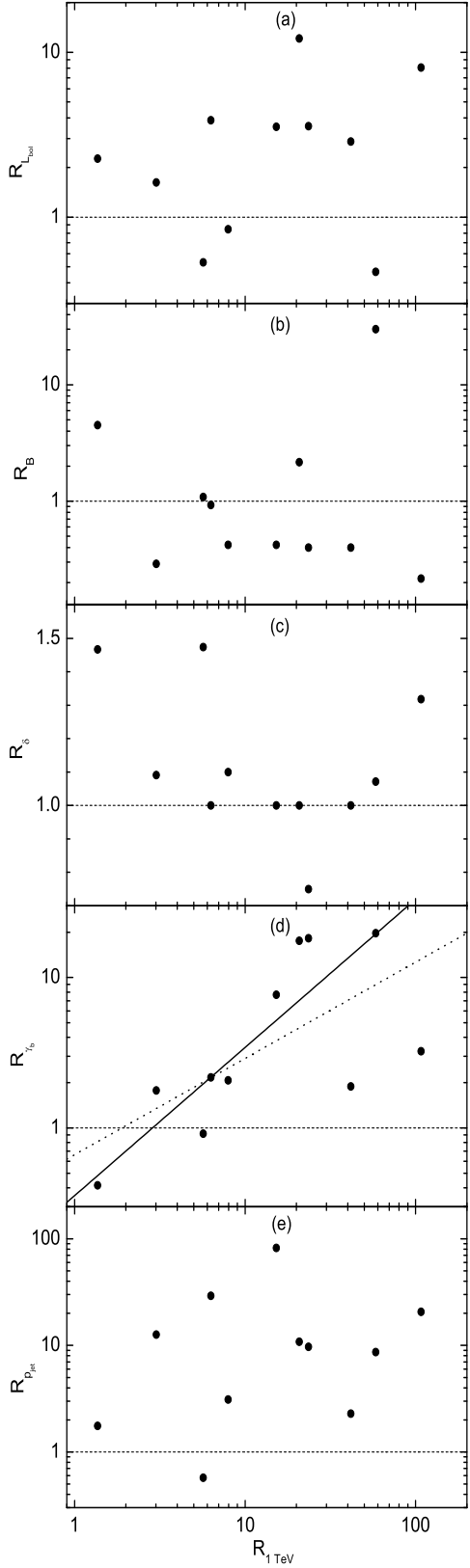}
\caption{Ratios of $R_x=x^H/x^L$ as a function of the ratio $R_{\rm 1\ TeV}=F^{H}_{\rm 1\ TeV}/F^{L}_{\rm 1\
TeV}$, where x=[$L_{\rm bol}$, $B$, $\delta$, $\gamma_{\rm b}$, $P_{\rm jet}$],  $F$ is the flux density at 1
TeV, and the letters ``H" and ``L" mark the high and low states, respectively. The fit lines for $R_{\gamma_{\rm
b}}-R_{\rm 1\ TeV}$ are shown in {\em Panel d}, {\em solid line} for robust fit $\log R_{\gamma_{\rm b}}=-0.45+0.985\log R_{\rm
1\ TeV}$ and {\em dashed line} for best fit $\log R_{\gamma_{\rm b}}=(-0.18\pm0.31)+(0.64\pm0.25)\log R_{\rm 1\
TeV}$.}\label{Fig:8}
\end{figure*}

\begin{figure*}
\includegraphics[width=6.in,height=2.5in]{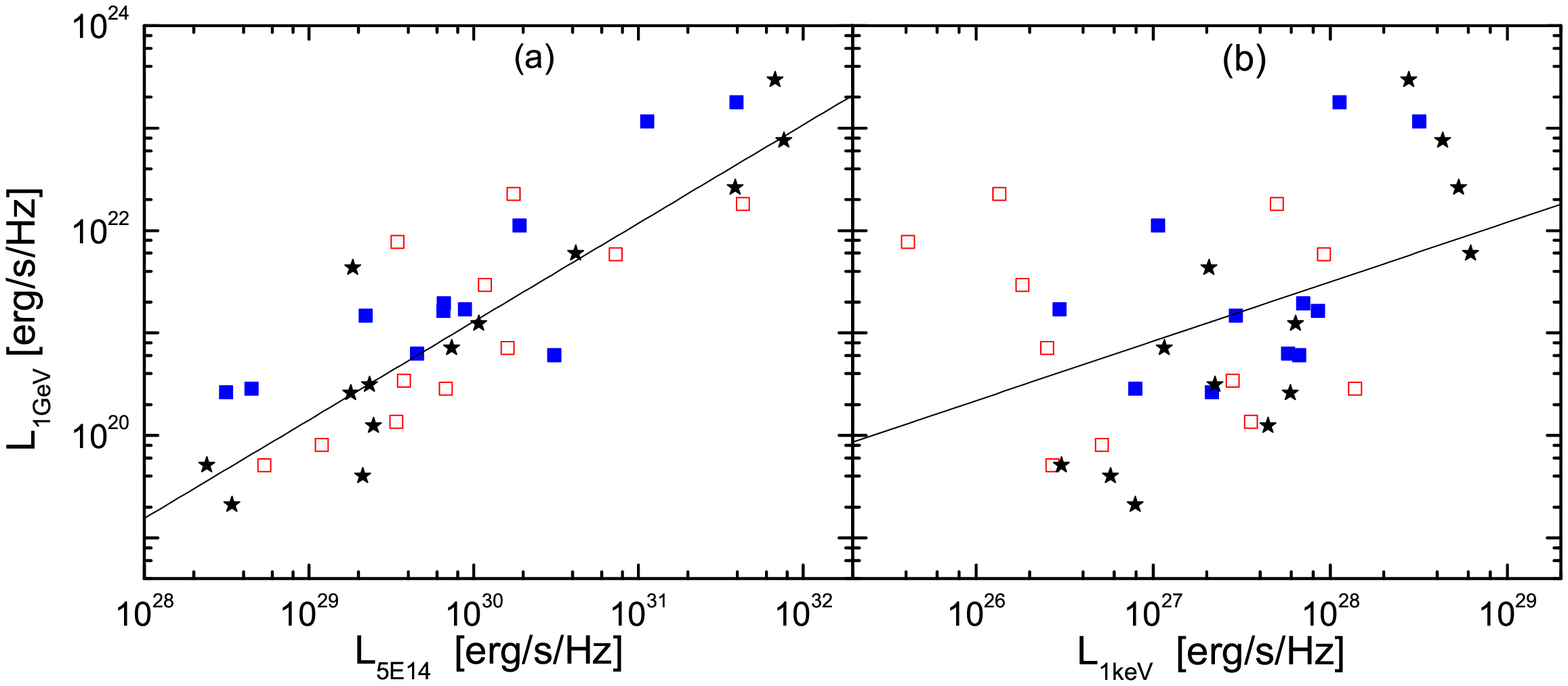}
\caption{Mono-luminosity at 1 GeV as the functions of that at $5\times10^{14}$ Hz ({\em Panel a}) and 1 keV
({\em Panel b}); the symbols are the same as that in Figure 4. The best fit lines are $L_{\rm 1\
GeV}=(-7.69\pm2.95)+(0.96\pm0.1)\log L_{\rm 5E14}$ for {\em Panel a} and $\log L_{\rm 1\
GeV}=(5.26\pm5.9)+(0.58\pm0.22)\log L_{\rm 1\ keV}$ for {\em Panel b}.}\label{Fig:9}
\end{figure*}

\begin{figure*}
\includegraphics[width=3.in,height=2.5in]{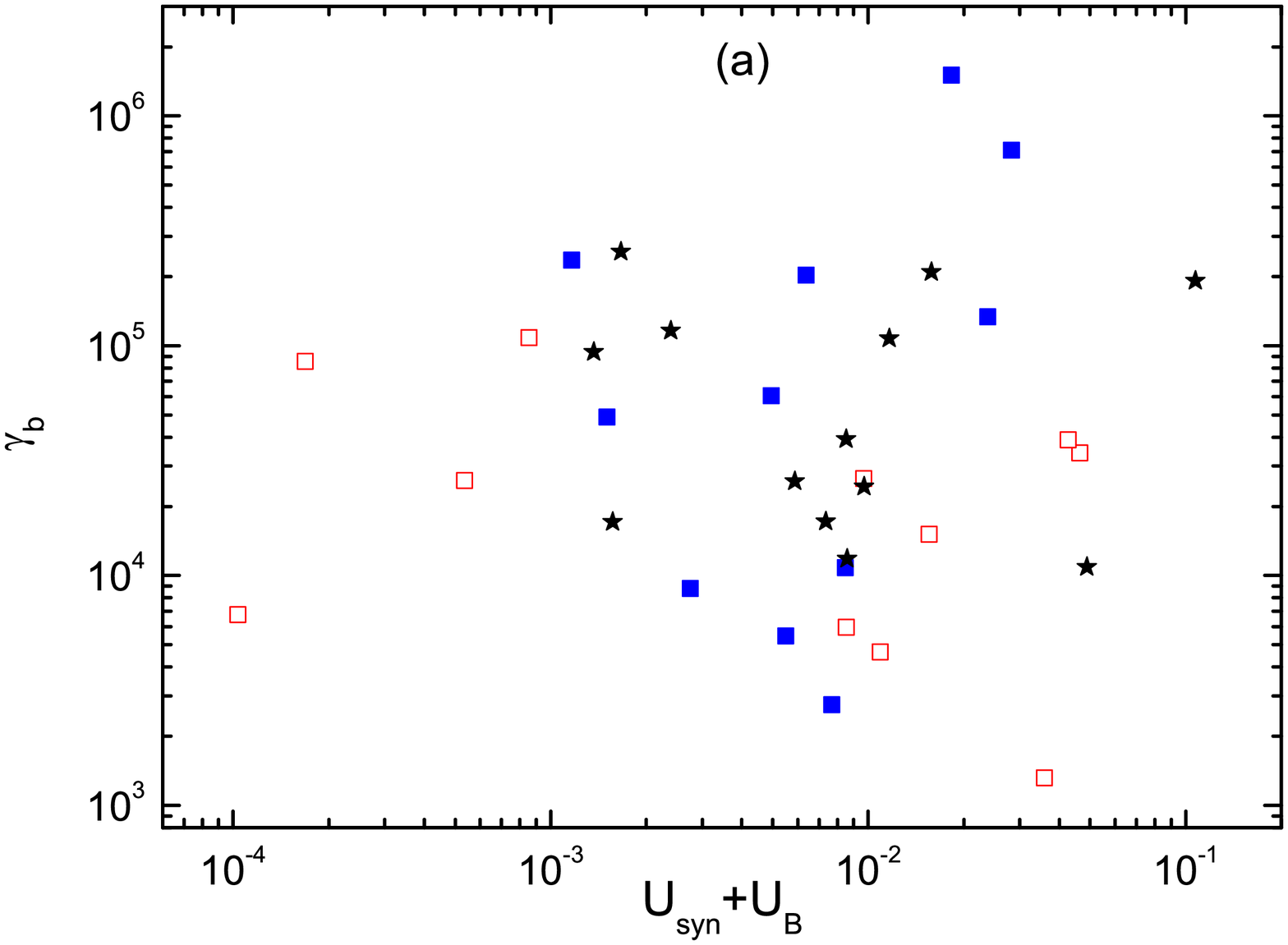}
\includegraphics[width=3.in,height=4.in]{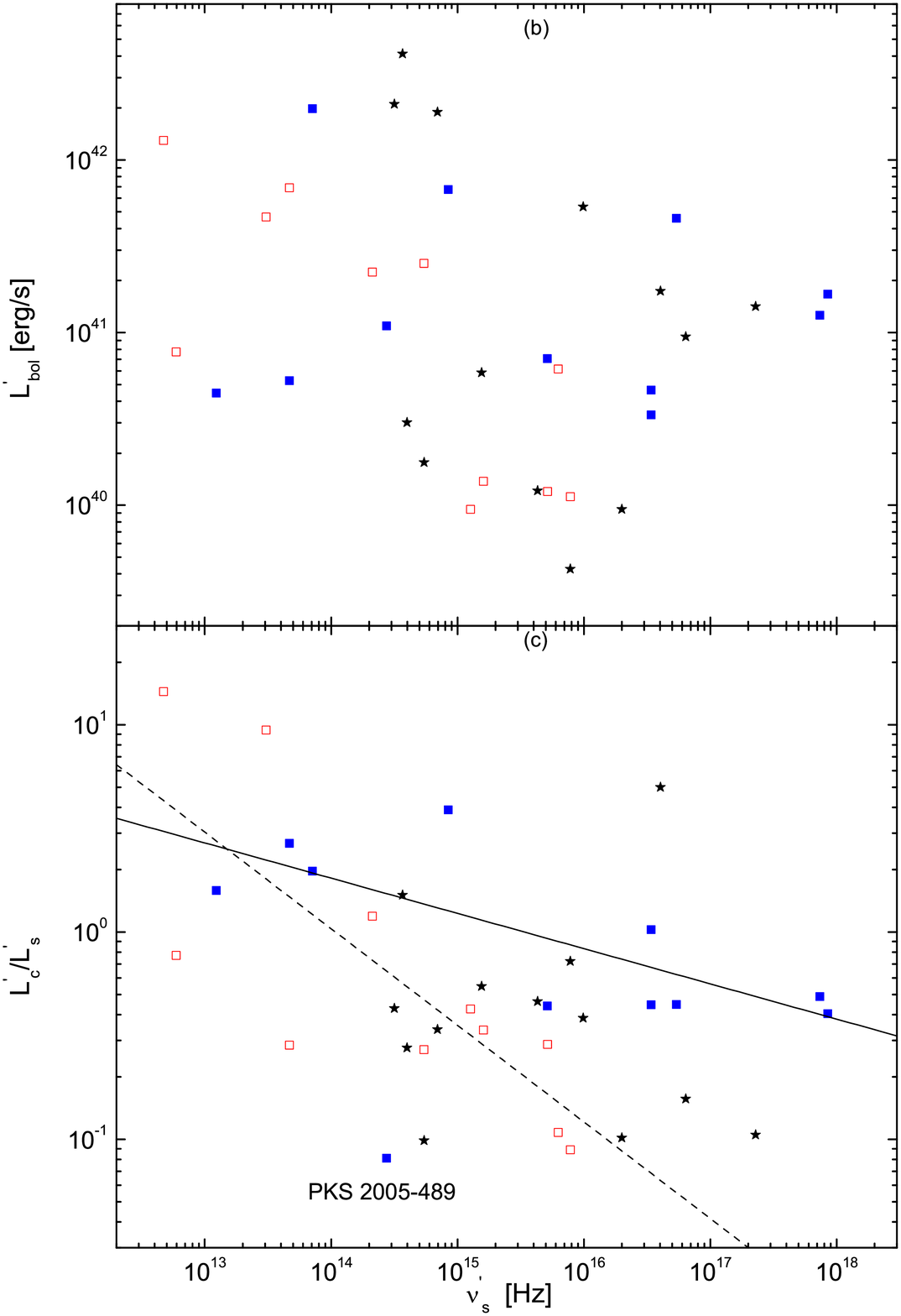}
\caption{{\em Panel a}---$\gamma_{\rm b}$ as a function of energy density $U_{\rm syn} + U_B$. {\em Panel b} and
{\em Panel c}---Bolometric luminosity $L_{\rm bol}^{'}$ and the ratio of the peak luminosities $L_{\rm
c}^{'}/L_{\rm s}^{'}$ as a function of synchrotron radiation peak frequency $\nu_{\rm s}^{'}$. The symbols are
the same as in Figure 4. The best fit lines are $\log L_{\rm c}^{'}/L_{\rm s}^{'}=(6.54\pm1.84)-(0.47\pm0.13)\log {\nu_{\rm
s}^{'}}$ for the low state data ({\em dashed line}) and $\log L_{\rm c}^{'}/L_{\rm
s}^{'}=(2.64\pm0.75)-(0.17\pm0.05)\log {\nu_{\rm s}^{'}}$ for the high state data ({\em solid line}).
}\label{Fig:10}
\end{figure*}

\acknowledgments

This work is supported by the National Natural Science Foundation of China (Grants 11078008, 11025313, 10873002,
10821061£¬10733010, 10725313, 10973034), the National Basic Research Program (973 Programme) of China (Grant
2009CB824800), China Postdoctoral Science Foundation, Guangxi Science Foundation (2011GXNSFB018063,
2010GXNSFC013011), and Guangxi SHI-BAI-QIAN project (Grant 2007201).

\label{lastpage}


\clearpage
\begin{thebibliography}{99}

\bibitem[Abdo et al.(2009)]{2009ApJ...707.1310A} Abdo, A.~A., et al.\ 2009,
\apj, 707, 1310

\bibitem[Abdo et al.(2010)]{2010ApJ...715..429A} Abdo, A.~A., et al.\ 2010a,
\apj, 715, 429
\bibitem[Abdo et al.(2010)]{2010ApJ...716...30A} Abdo, A.~A., et al.\ 2010b,
\apj, 716, 30
\bibitem[Abdo et al.(2010)]{2010ApJ...708.1310A} Abdo, A.~A., et al.\ 2010c,
\apj, 708, 1310
\bibitem[Acciari et al.(2008)]{2008ApJ...684L..73A} Acciari, V.~A., et al.\
2008, \apjl, 684, L73

\bibitem[Acciari et al.(2009)]{2009ApJ...707..612A} Acciari, V.~A., et al.\
2009a, \apj, 707, 612

\bibitem[Acciari et al.(2009)]{2009ApJ...693L.104A} Acciari, V.~A., et al.\
2009b, \apjl, 693, L104

\bibitem[Acciari et al.(2009)]{2009ApJ...690L.126A} Acciari, V., et al.\
2009c, \apjl, 690, L126

\bibitem[Acciari et al.(2010)]{2010ApJ...709L.163A} Acciari, V.~A., et al.\
2010a, \apjl, 709, L163

\bibitem[Acciari et al.(2010)]{2010ApJ...715L..49A} Acciari, V.~A., et al.\
2010b, \apjl, 715, L49

\bibitem[Acciari et al.(2010)]{2010ApJ...708L.100A} Acciari, V.~A., et al.\
2010c, \apjl, 708, L100

\bibitem[Aharonian et al.(1994)]{1994ApJ...423L...5A} Aharonian, F.~A.,
Coppi, P.~S., \& Voelk, H.~J.\ 1994, \apjl, 423, L5

\bibitem[Aharonian et
al.(2003)]{2003A&A...406L...9A} Aharonian, F., et al.\ 2003, \aap, 406, L9

\bibitem[Aharonian et
al.(2005)]{2005A&A...436L..17A} Aharonian, F., et al.\ 2005, \aap, 436, L17

\bibitem[Aharonian et
al.(2006)]{2006A&A...448L..19A} Aharonian, F., et al.\ 2006a, \aap, 448, L19

\bibitem[Aharonian et
al.(2006)]{2006A&A...455..461A} Aharonian, F., et al.\ 2006b, \aap, 455, 461

\bibitem[Aharonian et
al.(2007)]{2007A&A...470..475A} Aharonian, F., et al.\ 2007a, \aap, 470, 475

\bibitem[Aharonian et
al.(2007)]{2007A&A...473L..25A} Aharonian, F., et al.\ 2007b, \aap, 473, L25

\bibitem[Aharonian et
al.(2007)]{2007A&A...475L...9A} Aharonian, F., et al.\ 2007c, \aap, 475, L9

\bibitem[Aharonian et
al.(2008)]{2008A&A...481L.103A} Aharonian, F., et al.\ 2008, \aap, 481, L103
\bibitem[Aharonian et
al.(2009)]{2009A&A...502..749A} Aharonian, F., et al.\ 2009a, \aap, 502, 749

\bibitem[Aharonian et al.(2009)]{2009ApJ...696L.150A} Aharonian, F., et
al.\ 2009b, \apjl, 696, L150

\bibitem[H.~E.~S.~S.~collaboration : F.~Aharonian et
al.(2010)]{2010arXiv1006.5289H} Aharonian, F., et al.\ 2010, arXiv:1006.5289

\bibitem[Albert et al.(2006)]{2006ApJ...648L.105A} Albert, J., et al.\
2006a, \apjl, 648, L105

\bibitem[Albert et al.(2006)]{2006ApJ...642L.119A} Albert, J., et al.\
2006b, \apjl, 642, L119

\bibitem[Albert et al.(2007)]{2007ApJ...666L..17A} Albert, J., et al.\
2007a, \apjl, 666, L17

\bibitem[Albert et al.(2007)]{2007ApJ...662..892A} Albert, J., et al.\
2007b, \apj, 662, 892

\bibitem[Albert et al.(2007)]{2007ApJ...654L.119A} Albert, J., et al.\
2007c, \apjl, 654, L119

\bibitem[Albert et al.(2007)]{2007ApJ...667L..21A} Albert, J., et al.\
2007d, \apjl, 667, L21

\bibitem[Anderhub et al.(2009)]{2009ApJ...705.1624A} Anderhub, H., et al.\
2009a, \apj, 705, 1624

\bibitem[Anderhub et al.(2009)]{2009ApJ...704L.129A} Anderhub, H., et al.\
2009b, \apjl, 704, L129

\bibitem[Bai et
al.(1998)]{1998A&AS..132...83B} Bai, J.~M., Xie, G.~Z., Li, K.~H., Zhang, X., \& Liu, W.~W.\ 1998, \aaps, 132,
83

\bibitem[B{\l}a{\.z}ejowski et al.(2005)]{2005ApJ...630..130B}
Blazejowski, M., et al.\ 2005, \apj, 630, 130

\bibitem[B{\l}a{\.z}ejowski et al.(2000)]{2000ApJ...545..107B}
Blazejowski, M., Sikora, M., Moderski, R., \& Madejski, G.~M.\ 2000, \apj, 545, 107

\bibitem[Bloom
\& Marscher(1996)]{1996ApJ...461..657B} Bloom, S.~D., \& Marscher, A.~P.\ 1996, \apj, 461, 657

\bibitem[Buckley et al.(1996)]{1996ApJ...472L...9B} Buckley, J.~H., et al.\
1996, \apjl, 472, L9

\bibitem[Cao
\& Li(2008)]{2008MNRAS.390..561C} Cao, X., \& Li, F.\ 2008, \mnras, 390, 561

\bibitem[Catanese et al.(1998)]{1998ApJ...501..616C} Catanese, M., et al.\
1998, \apj, 501, 616

\bibitem[Chadwick et al.(1999)]{1999ApJ...513..161C} Chadwick, P.~M., et
al.\ 1999, \apj, 513, 161

\bibitem[Chen
\& Bai(2011)]{2011ApJ...735..108C} Chen, L., \& Bai, J.~M.\ 2011, \apj, 735, 108

\bibitem[Costamante(2007)]{2007Ap&SS.309..487C} Costamante, L.\ 2007, \apss, 309, 487

\bibitem[Dai et al.(2002)]{2002ApJ...580L...7D} Dai, Z.~G., Zhang, B., Gou,
L.~J., M{\'e}sz{\'a}ros, P., \& Waxman, E.\ 2002, \apjl, 580, L7

\bibitem[Dermer et
al.(1992)]{1992A&A...256L..27D} Dermer, C.~D., Schlickeiser, R., \& Mastichiadis, A.\ 1992, \aap, 256, L27

\bibitem[Dolag et al.(2009)]{2009ApJ...703.1078D} Dolag, K.,
Kachelrie{\ss}, M., Ostapchenko, S., \& Tom{\`a}s, R.\ 2009, \apj, 703, 1078

\bibitem[Dondi
\& Ghisellini(1995)]{1995MNRAS.273..583D} Dondi, L., \& Ghisellini, G.\ 1995, \mnras, 273, 583

\bibitem[Falcke et
al.(2004)]{2004A&A...414..895F} Falcke, H., K{\"o}rding, E., \& Markoff, S.\ 2004, \aap, 414, 895

\bibitem[Fanidakis et al.(2011)]{2011MNRAS.410...53F} Fanidakis, N., Baugh,
C.~M., Benson, A.~J., Bower, R.~G., Cole, S., Done, C., \& Frenk, C.~S.\ 2011, \mnras, 410, 53

\bibitem[Fossati et al.(1998)]{1998MNRAS.299..433F} Fossati, G., Maraschi,
L., Celotti, A., Comastri, A., \& Ghisellini, G.\ 1998, \mnras, 299, 433

\bibitem[Franceschini et
al.(2008)]{2008A&A...487..837F} Franceschini, A., Rodighiero, G., \& Vaccari, M.\ 2008, \aap, 487, 837

\bibitem[Ghisellini et
al.(1996)]{1996A&AS..120C.503G} Ghisellini, G., Maraschi, L., \& Dondi, L.\ 1996, \aaps, 120, 503

\bibitem[Ghisellini et al.(1998)]{1998MNRAS.301..451G} Ghisellini, G.,
Celotti, A., Fossati, G., Maraschi, L.,
\& Comastri, A.\ 1998, \mnras, 301, 451

\bibitem[Ghisellini
\& Tavecchio(2008)]{2008MNRAS.387.1669G} Ghisellini, G., \& Tavecchio, F.\ 2008, \mnras, 387, 1669

\bibitem[Ghisellini et al.(2009)]{2009MNRAS.399.2041G} Ghisellini, G.,
Tavecchio, F., \& Ghirlanda, G.\ 2009, \mnras, 399, 2041

\bibitem[Ghisellini et al.(2010)]{2010MNRAS.402..497G} Ghisellini, G.,
Tavecchio, F., Foschini, L., Ghirlanda, G., Maraschi, L., \& Celotti, A.\ 2010, \mnras, 402, 497

\bibitem[HESS Collaboration et
al.(2010)]{2010A&A...511A..52H} HESS Collaboration, et al.\ 2010a, \aap, 511, A52

\bibitem[HESS Collaboration et
al.(2010)]{2010A&A...516A..56H} HESS Collaboration, et al.\ 2010b, \aap, 516, A56

\bibitem[Holder et al.(2003)]{2003ApJ...583L...9H} Holder, J., et al.\
2003, \apjl, 583, L9

\bibitem[Horan et al.(2002)]{2002ApJ...571..753H} Horan, D., et al.\ 2002,
\apj, 571, 753

\bibitem[Horns
\& Konopelko(2002)]{2002ATel...96....1H} Horns, D., \& Konopelko, K.\ 2002, The Astronomer's Telegram, 96, 1

\bibitem[Kaufmann et al.(2010)]{2010HEAD...11.3325K} Kaufmann, S., et al.\
2010, Bulletin of the American Astronomical Society, 42, 709

\bibitem[King
\& Pringle(2006)]{2006MNRAS.373L..90K} King, A.~R., \& Pringle, J.~E.\ 2006, \mnras, 373, L90

\bibitem[King et al.(2008)]{2008MNRAS.385.1621K} King, A.~R., Pringle,
J.~E., \& Hofmann, J.~A.\ 2008, \mnras, 385, 1621

\bibitem[King et al.(2011)]{2011ApJ...729...19K} King, A.~L., et al.\ 2011,
\apj, 729, 19

\bibitem[Krawczynski et al.(2004)]{2004ApJ...601..151K} Krawczynski, H., et
al.\ 2004, \apj, 601, 151

\bibitem[Lagos et al.(2009)]{2009MNRAS.395..625L} Lagos, C.~D.~P., Padilla,
N.~D., \& Cora, S.~A.\ 2009, \mnras, 395, 625

\bibitem[Liang
\& Liu(2003)]{2003MNRAS.340..632L} Liang, E.~W., \& Liu, H.~T.\ 2003, \mnras, 340, 632

\bibitem[Li et al.(2010)]{2010ApJ...710..878L} Li, Y.-R., Wang, J.-M.,
Yuan, Y.-F., Hu, C., \& Zhang, S.\ 2010, \apj, 710, 878

\bibitem[Madejski et al.(1999)]{1999ApJ...521..145M} Madejski, G.~M.,
Sikora, M., Jaffe, T., B{\l}a{\.z}ejowski, M., Jahoda, K., \& Moderski, R.\ 1999, \apj, 521, 145

\bibitem[Maraschi et al.(1992)]{1992ApJ...397L...5M} Maraschi, L.,
Ghisellini, G., \& Celotti, A.\ 1992, \apjl, 397, L5

\bibitem[Marscher et al.(2002)]{2002Natur.417..625M} Marscher, A.~P.,
Jorstad, S.~G., G{\'o}mez, J.-L., Aller, M.~F., Ter{\"a}sranta, H., Lister, M.~L., \& Stirling, A.~M.\ 2002,
\nat, 417, 625

\bibitem[Mannheim
\& Biermann(1992)]{1992A&A...253L..21M} Mannheim, K., \& Biermann, P.~L.\ 1992, \aap, 253, L21

\bibitem[Merloni et al.(2003)]{2003MNRAS.345.1057M} Merloni, A., Heinz, S.,
\& di Matteo, T.\ 2003, \mnras, 345, 1057

\bibitem[Neronov
\& Semikoz(2009)]{2009PhRvD..80l3012N} Neronov, A., \& Semikoz, D.~V.\ 2009, \prd, 80, 123012

\bibitem[Ong(2009)]{2009ATel.1941....1C} Ong, R.\ 2009, The Astronomer's
Telegram, 1941, 1

\bibitem[Paneque(2009)]{Paneque et al. 2009} Paneque, D., et al. 2009, in preparation

\bibitem[Punch et al.(1992)]{1992Natur.358..477P} Punch, M., et al.\ 1992,
\nat, 358, 477

\bibitem[Quinn et al.(1996)]{1996ApJ...456L..83Q} Quinn, J., et al.\ 1996,
\apjl, 456, L83

\bibitem[R{\"u}ger et al.(2010)]{2010MNRAS.401..973R} R{\"u}ger, M.,
Spanier, F., \& Mannheim, K.\ 2010, \mnras, 401, 973

\bibitem[Ravasio et
al.(2002)]{2002A&A...383..763R} Ravasio, M., et al.\ 2002, \aap, 383, 763

\bibitem[Reimer et al.(2008)]{2008ApJ...682..775R} Reimer, A., Costamante,
L., Madejski, G., Reimer, O., \& Dorner, D.\ 2008, \apj, 682, 775

\bibitem[Reyes et al.(2009)]{2009arXiv0907.5175R} Reyes, L.~C., for the
Fermi LAT collaboration, \& the VERITAS collaboration 2009, arXiv:0907.5175

\bibitem[Rieger
\& Volpe(2010)]{2010A&A...520A..23R} Rieger, F.~M., \& Volpe, F.\ 2010, \aap, 520, A23

\bibitem[Sambruna et al.(2000)]{2000ApJ...538..127S} Sambruna, R.~M., et
al.\ 2000, \apj, 538, 127

\bibitem[Sikora et al.(1994)]{1994ApJ...421..153S} Sikora, M., Begelman,
M.~C., \& Rees, M.~J.\ 1994, \apj, 421, 153

\bibitem[Sikora et al.(2007)]{2007ApJ...658..815S} Sikora, M., Stawarz,
{\L}., \& Lasota, J.-P.\ 2007, \apj, 658, 815

\bibitem[Superina et al.(2008)]{2008ICRC....3..913S} Superina, G., Benbow,
W., Boutelier, T., \& et al.\ 2008, International Cosmic Ray Conference, 3, 913

\bibitem[Tagliaferri et
al.(2000)]{2000A&A...354..431T} Tagliaferri, G., et al.\ 2000, \aap, 354, 431

\bibitem[Tagliaferri et
al.(2001)]{2001A&A...368...38T} Tagliaferri, G., et al.\ 2001, \aap, 368, 38

\bibitem[Tagliaferri et al.(2008)]{2008ApJ...679.1029T} Tagliaferri, G., et
al.\ 2008, \apj, 679, 1029

\bibitem[Takahashi et al.(2000)]{2000ApJ...542L.105T} Takahashi, T., et
al.\ 2000, \apjl, 542, L105

\bibitem[Tavecchio et al.(1998)]{1998ApJ...509..608T} Tavecchio, F.,
Maraschi, L., \& Ghisellini, G.\ 1998, \apj, 509, 608


\bibitem[Tavecchio et al.(2000)]{2000ApJ...544L..23T} Tavecchio, F.,
Maraschi, L., Sambruna, R.~M., \& Urry, C.~M.\ 2000, \apjl, 544, L23

\bibitem[Tavecchio et al.(2001)]{2001ApJ...554..725T} Tavecchio, F., et
al.\ 2001, \apj, 554, 725

\bibitem[Tavecchio et al.(2009)]{2009MNRAS.399L..59T} Tavecchio, F.,
Ghisellini, G., Ghirlanda, G., Costamante, L., \& Franceschini, A.\ 2009, \mnras, 399, L59

\bibitem[Tavecchio et al.(2010)]{2010MNRAS.406L..70T} Tavecchio, F.,
Ghisellini, G., Foschini, L., Bonnoli, G., Ghirlanda, G., \& Coppi, P.\ 2010a, \mnras, 406, L70

\bibitem[Tavecchio et al.(2010)]{2010MNRAS.401.1570T} Tavecchio, F.,
Ghisellini, G., Ghirlanda, G., Foschini, L., \& Maraschi, L.\ 2010b, \mnras, 401, 1570

\bibitem[Teshima
\& The MAGIC Collaboration(2008)]{2008ATel.1500....1T} Teshima, M., \& The MAGIC Collaboration 2008, The
Astronomer's Telegram, 1500, 1

\bibitem[Tramacere et
al.(2007)]{2007A&A...467..501T} Tramacere, A., et al.\ 2007, \aap, 467, 501

\bibitem[Ulvestad
\& Ho(2001)]{2001ApJ...562L.133U} Ulvestad, J.~S., \& Ho, L.~C.\ 2001, \apjl, 562, L133

\bibitem[Urry(1999)]{1999APh....11..159U} Urry, C.~M.\ 1999, Astroparticle
Physics, 11, 159

\bibitem[Vermeulen et al.(1995)]{1995ApJ...452L...5V} Vermeulen, R.~C.,
Ogle, P.~M., Tran, H.~D., Browne, I.~W.~A., Cohen, M.~H., Readhead, A.~C.~S., Taylor, G.~B., \& Goodrich, R.~W.\
1995, \apjl, 452, L5

\bibitem[Vittorini et al.(2009)]{2009ApJ...706.1433V} Vittorini, V., et
al.\ 2009, \apj, 706, 1433

\bibitem[Wang et al.(2009)]{2009ApJ...697L.141W} Wang, J.-M., et al.\ 2009,
\apjl, 697, L141

\bibitem[Weidinger
\& Spanier(2010)]{2010A&A...515A..18W} Weidinger, M., \& Spanier, F.\ 2010, \aap, 515, A18

\bibitem[Wolter et al.(2008)]{2008ASPC..386..302W} Wolter, A., Beckmann,
V., Ghisellini, G., Tavecchio, F., \& Maraschi, L.\ 2008, Extragalactic Jets: Theory and Observation from Radio
to Gamma Ray, 386, 302

\bibitem[Woo
\& Urry(2002)]{2002ApJ...579..530W} Woo, J.-H., \& Urry, C.~M.\ 2002, \apj, 579, 530

\bibitem[Wu et
al.(2002)]{2002A&A...389..742W} Wu, X.-B., Liu, F.~K., \& Zhang, T.~Z.\ 2002, \aap, 389, 742

\bibitem[Wu et al.(2011)]{2011ApJ...735...50W} Wu, Q., Cao, X.,
\& Wang, D.-X.\ 2011, \apj, 735, 50

\bibitem[Xie et al.(2001)]{2001ApJ...548..200X} Xie, G.~Z., Li, K.~H., Bai,
J.~M., Dai, B.~Z., Liu, W.~W., Zhang, X., \& Xing, S.~Y.\ 2001, \apj, 548, 200

\bibitem[Yang et al.(2008)]{2008ApJ...682..767Y} Yang, C.~Y., Fang, J.,
Lin, G.~F., \& Zhang, L.\ 2008, \apj, 682, 767

\bibitem[Zhang et al.(2000)]{2000Sci...287.1239Z} Zhang, S.~N., Cui, W.,
Chen, W., Yao, Y., Zhang, X., Sun, X., Wu, X.-B., \& Xu, H.\ 2000, Science, 287, 1239

\bibitem[Zhang et al.(2010)]{2010ApJ...710.1017Z} Zhang, J., Bai, J.~M.,
Chen, L., \& Liang, E.\ 2010, \apj, 710, 1017

\bibitem[Zhang(2009)]{2009RAA.....9..777Z} Zhang, J.\ 2009, Research in
Astronomy and Astrophysics, 9, 777

\end{thebibliography}
\end{document}